\documentclass[twocolumn]{aastex631}

\usepackage{graphicx}
\usepackage{epstopdf}
\usepackage{subfigure} 
\usepackage{color}
\usepackage{float}
\usepackage[figuresright]{rotating}
\usepackage{footnote}
\usepackage{amsmath}
\usepackage{subfigure} 
\usepackage{appendix}
\usepackage{longtable}
\usepackage{threeparttable}
\usepackage{booktabs}
\usepackage{hyperref}
\usepackage{CJKutf8}

\newcommand{\teff}{$T_{\rm eff}$}

\begin{document}
\begin{CJK}{UTF8}{gbsn}
\title{The Distribution of Semi-Detached Binaries. I.An Efficient Pipeline}

\author[0000-0003-4829-6245]{Jianping Xiong}
\affiliation{Yunnan Observatories, Chinese Academy of Sciences, 396 YangFangWang, Guandu District, Kunming, 650216, Peopleʼs Republic of China}

\author[0000-0002-2427-161X]{Xu Ding}
\affiliation{Yunnan Observatories, Chinese Academy of Sciences, 396 YangFangWang, Guandu District, Kunming, 650216, Peopleʼs Republic of China}

\author[0000-0002-3651-5482]{Jiadong Li}\affiliation{Max-Planck-Institut für Astronomie, Königstuhl 17, D-69117 Heidelberg, Germany}

\author[0000-0002-6398-0195]{Hongwei Ge}
\affiliation{Yunnan Observatories, Chinese Academy of Sciences, 396 YangFangWang, Guandu District, Kunming, 650216, Peopleʼs Republic of China}
\affiliation{Key Laboratory for Structure and Evolution of Celestial Objects, Chinese Academy of Sciences, P.O. Box 110, Kunming 650216, People's Republic of China}
\affiliation{International Centre of Supernovae, Yunnan Key Laboratory, Kunming 650216, People's Republic of China}

\author{Qiyuan Cheng}
\affiliation{Yunnan Observatories, Chinese Academy of Sciences, 396 YangFangWang, Guandu District, Kunming, 650216, Peopleʼs Republic of China}
\affiliation{University of Chinese Academy of Sciences\\
Beijing 100049, China}

\author{Kaifan Ji}
\affiliation{Yunnan Observatories, Chinese Academy of Sciences, 396 YangFangWang, Guandu District, Kunming, 650216, Peopleʼs Republic of China}

\author[0000-0001-9204-7778]{Zhanwen Han}
\affiliation{Yunnan Observatories, Chinese Academy of Sciences, 396 YangFangWang, Guandu District, Kunming, 650216, Peopleʼs Republic of China}
\affiliation{Key Laboratory for Structure and Evolution of Celestial Objects, Chinese Academy of Sciences, P.O. Box 110, Kunming 650216, People's Republic of China}
\affiliation{International Centre of Supernovae, Yunnan Key Laboratory, Kunming 650216, People's Republic of China}

\author[0000-0001-5284-8001]{Xuefei Chen}
\affiliation{Yunnan Observatories, Chinese Academy of Sciences, 396 YangFangWang, Guandu District, Kunming, 650216, Peopleʼs Republic of China}
\affiliation{Key Laboratory for Structure and Evolution of Celestial Objects, Chinese Academy of Sciences, P.O. Box 110, Kunming 650216, People's Republic of China}
\affiliation{International Centre of Supernovae, Yunnan Key Laboratory, Kunming 650216, People's Republic of China}
\correspondingauthor{Jianping Xiong, Xuefei Chen}
\email{xiongjianping@ynao.ac.cn, cxf@ynao.ac.cn}



\begin{abstract}
Semi-detached binaries are in the stage of mass transfer and play a crucial role in studying mass transfer physics between interacting binaries. Large-scale time-domain surveys provide massive light curves of binary systems, while Gaia offers high-precision astrometric data. In this paper, we develop, validate, and apply a pipeline that combines the MCMC method with a forward model and DBSCAN clustering to search for semi-detached binary and estimate its inclination, relative radius, mass ratio, and temperature ratio using light curve. We train our model on the mock light curves from PHOEBE, which provides broad coverage of light curve simulations for semi-detached binaries. Applying our pipeline to TESS sectors 1-26, we have identified $77$ semi-detached binary candidates. Utilizing the distance from Gaia, we determine their masses and radii with median fractional uncertainties of $\sim$26\% and $\sim$7\%, respectively. With the added $77$ candidates, the catalog of semi-detached binaries with orbital parameters has been expanded by approximately $20\%$. The comparison and statistical results show that our semi-detached binary candidates align well with the compiled samples and the PARSEC model in \teff-$L$ and $M$-$R$ relations. Combined with the literature samples, comparative analysis with stability criteria for conserved mass transfer indicates that $\sim$ 97.4\% of samples are undergoing nuclear-timescale mass transfer, and two samples (GO Cyg and TIC 454222105) are located within the limits of stability criteria for dynamical- and thermal-timescale mass transfer, which are currently undergoing thermal-timescale mass transfer. Additionally, one system (IR Lyn) is very close to the upper limit of delayed dynamical-timescale mass transfer.
\end{abstract}



\section{Introduction} \label{sec:intro}
Eclipsing binaries (EBs) play a crucial role in modern astronomy. More than 50\% stars with masses of 1 $M_{\odot}$ or higher are discovered to exist in binary (multiple) systems \citep{2012Sci...337..444S,2017ApJS..230...15M}. Studying binaries provides insights into the star formation and evolution \citep{2010MNRAS.403...45S}, stellar model calibration \citep{2020A&A...637A..60T,2023AJ....165...30X}, accretion physics \citep{2010ASPC..435..287B} and mass transfer physic\citep{2015ApJ...812...40G,2020ApJ...899..132G}. The interactions between binary systems give rise to various intriguing objects, including compact binaries, supernovae, gamma-ray bursts, X-ray binaries, pulsars, cataclysmic variables, etc \citep{2012Sci...337..444S,2020RAA....20..161H}. Semi-detached binaries (SDs) are in a state of mass transfer and therefore serve as excellent laboratories for studying the mass transfer of binaries. Analyzing a large sample of SDs containing accurate parameters will enhance our understanding of statistical properties and mass transfer processes in SDs.

With photometric observations, numerous SDs have been discovered; for instance, \citet{2004A&A...417..263B} compiled a catalog of 411 SDs, \citet{2006MNRAS.368.1311P} revealed a list of 2949 SDs from the All Sky Automated Survey (ASAS), and \citet{2018ApJS..238....4P} identified 449 SDs by analyzing the Catalina Sky Surveys (CSS) data. Moreover, recent large-scale time-domain surveys such as Kepler/K2, TESS (Transiting Exoplanet Survey Satellite), ZTF(Zwicky Transient Facility), and ASAS-SN (All-Sky Automated Survey for Supernovae) have greatly contributed to the discovery of millions of eclipsing binaries \citep{2016AJ....151...68K,2020ApJS..249...18C,2021AA...652A.120I,2022ApJS..258...16P,2023MNRAS.519.5271C}. However, only a small fraction of SDs have measurements of complete parameters, where \citet{2004yCat.5115....0S} have compiled approximately 232 SDs samples with stellar parameters, of which 96 had spectroscopic and photometric observations, \citet{2006MNRAS.373..435I} also listed the absolute parameters of 61 SDs, and recently, \citet{2020MNRAS.491.5489M} added 31 new samples to the previous catalog \citep{2004yCat.5115....0S} with available light and radial-velocity curve solutions, and \citet{2022RAA....22k5015M} has compiled physical parameters of 48 individually studied Near-Contact Binaries. 

TESS is a pioneering space mission that can capture nearly the entire sky \citep{2015JATIS...1a4003R}, and it has monitored $\sim$200,000 bright stars (V=5$\sim$12\texttt{mag}) in its first 2-year prime mission. Notably, in the TESS survey, by using the light curves from sectors 1-26, \citet{2021AA...652A.120I} and \citet{2022ApJS..258...16P} have reported the confirmation of 3155 early-type eclipsing binaries (EBs) and 4584 EBs, respectively. These light curves can provide valuable information on various properties of binaries, such as mass ratio ($q$), period ($P$), temperature ratio ($T\rm _{2}/T\rm _{1}$), luminosity ratio ($L_{2}/L_{1}$), inclination ($i$) and relative radii ($R_{(1,2)}/a$), where $a$ is the semi-major axis. This information will greatly contribute to our understanding of semi-detached binaries and the mass transfer in binary systems.

PHOEBE\footnote{http://phoebe-project.org} (PHysics Of Eclipsing BinariEs \citealt{2005ApJ...628..426P,2016ApJS..227...29P,2020ApJS..247...63J}) is now a widely used tool for analyzing and fitting light curves of binaries, which is built upon the WD code \citep{1971ApJ...166..605W}. However, when dealing with a large number of data points, the light curve fitting process in PHOEBE is time-consuming. This is primarily due to the integration of internal physical models. Therefore, motivated by the need for measuring stellar properties in a large number of semi-detached binaries, there is an increasing demand for a pipeline that can effectively process the data and extract relevant information for studying these systems.

Recently, machine learning techniques have been widely applied to the processing and analysis of massive astronomical observations, including the measurement of stellar parameters \citep{2018MNRAS.475.2978F,2020ApJ...891...23W}, target detection \citep{2019MNRAS.487.2874I,2022ApJ...932..118C}, and classification \citep{2020ApJS..249...18C,2022MNRAS.514.2793B}. In the specific context of binary systems analysis, machine learning methods have also been used. For example, \citet{2022ApJS..258...26Z} employed the CNN network to identify the binaries in the LAMOST survey. Additionally, \citet{2022AJ....164..200D} proposed a fast approach for deriving parameters of contact binaries using a combination of a neural network (NN) and the MCMC (Markov chain Monte Carlo) algorithm. This method significantly enhances the speed of light curve fitting for contact binaries.

Furthermore, a large number of stars with accurate measurements of astronomical and atmospheric parameters have recently been released from Gaia DR3\citep{2023A&A...674A...1G, 2023A&A...674A..37G,2023MNRAS.524.1855Z}. By combining these parameters and light curves, a catalog containing the complete parameters for semi-detached binaries can be constructed. Consequently, in this paper, we develop a fast pipeline using a machine learning method to accurately fit the light curves of semi-detached binaries. As a result of our work, we will analyze the semi-detached binaries in the TESS survey and present a catalog containing the complete parameters for these systems. The result will serve as a valuable resource for further research and analysis of the physical and evolutionary properties of semi-detached binaries.

The paper is organized as follows: We describe the light curve fitting model for semi-detached binaries in Section~\ref{sec:model}. The pipeline for deriving parameters from the light curve is presented in Section~\ref{sec: method}. Then, we analyze the semi-detached binaries from the TESS survey in Section~\ref{sec:tess}. The results and statistical analysis are indicated in Section~\ref{sec:result}. Finally, we summarize in Section~\ref{sec:Conclusion}.

\section{Light curve fitting model establishment} \label{sec:model}

\subsection{Training dataset} \label{sec:traindata}
\begin{figure*}[t]
  \centering
  \includegraphics[scale=0.5,width=\textwidth]{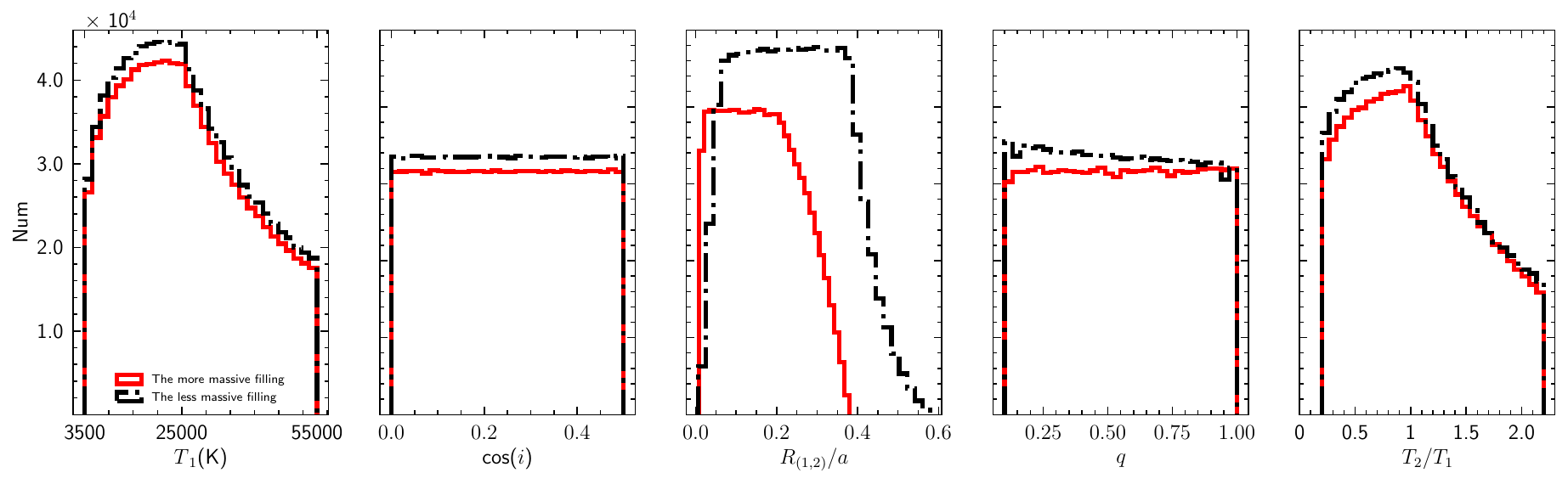}
  \centering
  \caption{The input parameters' distributions of the training sample for the model with a more massive component filling its Roche lobe (red solid histograms) and the model with a less massive component filling its Roche lobe (black dash-dotted histograms). From left to right, the distributions of the effective temperature of primary (more massive) star ($T\rm_{1}$), cosine of the inclination (cos($i$)), relative radii of unfilled component ($R\rm_{(1,2)}/a$), mass ratio ($q$=$M_{2}$/$M_{1}$) and temperature ratio ($T\rm_{2}$/$T\rm_{1}$) are shown. The non-uniform distributions of $T\rm_{1}$, $R\rm_{(1,2)}/a$ and $T\rm_{2}$/$T\rm_{1}$ are primarily caused by the constraints of the physical models for semi-detached binaries.}\label{fig:fig1}
\end{figure*}
Semi-detached binaries are the binaries in which one component completely fills its Roche lobe while the other component remains inside its Roche lobe. There are two types of semi-detached binaries according to the mass, the more massive component and the less massive one, of the Roche lobe filling stars. In some specific parameters space, the light curves of a more massive or a less massive component overfill its Roche lobe will be similar. So, we need to construct two individual models for fitting the light curves of these two types of semi-detached binaries.

For the two models, the following free parameters are considered, and the more massive star is defined as the primary star (star 1):
\begin{itemize}
    \item mass ratio ($q$=$M_{2}$/$M_{1}$) within the range of [0.1, 1].
    \item inclination (\rm cos($i$)) in the range of [0,0.5] that corresponds to inclinations in [60$^{\circ}$, 90$^{\circ}$].
    \item effective temperature of primary star ($T\rm_{1}$) within the range of [3500 K, 55000 K]
    \item effective temperature ratio ($T\rm _{2}$/$T\rm _{1}$) in the range of [0.2, 2.25]
    \item relative radii of unfilled component($R\mathrm{_{1}}/a$ or $R\mathrm{ _{2}}/a$) are generated in the range of [0.01, $R\rm _{L_{1}}$] and [0.01, $R\rm _{L_{2}}$], where $a$ is the semi-major axis, $R\rm _{L_{1}}$ and $R\rm _{L_{2}}$ are the Roche lobe radii of two components. And the Roche lobe radii are constrained by mass ratio as \citep{1983ApJ...268..368E} given:
    \begin{equation}\label{eq:RL1}
        \frac{R_{\mathrm{L_{1}}}}{a} = \frac{0.49q^{-2/3}}{0.6q^{-2/3}+\mathrm{ln}(1+q^{-1/3})}.
    \end{equation}
    In Eq.\ref{eq:RL1}, the mass ratio $q$ is defined as $M_{2}$/$M_{1}$.
    \item gravity darkening coefficients (\emph{g}$_{(1,2)}$) and reflection coefficients (Albedos, \emph{A}$_{(1,2)}$) are set to the adopted values corresponding to the temperatures of the components, following the law\citep{1924MNRAS..84..665V,1967ZA.....65...89L,1969AcA....19..245R}.
\end{itemize}

Based on the specified parameters' ranges, we uniformly and randomly sample the parameters within their given ranges to generate training samples by using the PHOEBE package. We generate 100 sample points of light curves within the phase range of 0 to 1. The passband used for PHOEBE is set as ``TESS:T'', which represents the passband of TESS mission. Then we use the Eq. \ref{Equ:normalized} to convert the flux values obtained from PHOEBE to magnitudes.
\begin{equation}\label{Equ:normalized}
\begin{aligned}
    m_{i}&=-2.5\times \mathrm{log\rm_{10}}(f_{i})\\
    m^{'}_{i}&=m_{i}-\frac{\sum^{n}_{i}m_{i}}{n}
    \end{aligned}
\end{equation}
In Eq.\ref{Equ:normalized}, $f\rm_{i}$ is the mock light curve from PHOEBE, $n$ represents the number of sampling points, and $m^{'}\rm_{i}$ is the normalized light curve in ``\texttt{mag}'' unit. For each model, we generate approximately 1 million light curves for model construction. Finally, the dataset is divided into training and validation sets in a ratio of 8:2 for model training and evaluation. Fig.\ref{fig:fig1} shows the distribution histograms of the training dataset, and we found that the distributions of $R\rm_{(1,2)}/a$, and effective temperatures ($T\rm_{1}$ and $T\rm_{2}$/$T\rm_{1}$) are non-uniform. This non-uniformity is primarily due to the constraints imposed by the physical models and the composition conditions of semi-detached binaries. Therefore, during the model training process, we randomly sample the training set during each iteration to mitigate the influence of bias caused by the dataset's distribution on model training. This approach aims to enhance the model's coverage of the entire dataset, facilitating more effective training. By employing random sampling, we ensure that each iteration's batch or sample represents the dataset as a whole, minimizing any potential biases towards specific subsets. Random sampling also helps prevent the model from memorizing the order or specific characteristics of the training examples. Consequently, the model can learn more generalized patterns and improve its ability to handle diverse data.

\begin{figure*}[t]
  \centering
  \includegraphics[scale=0.15,width=\textwidth]{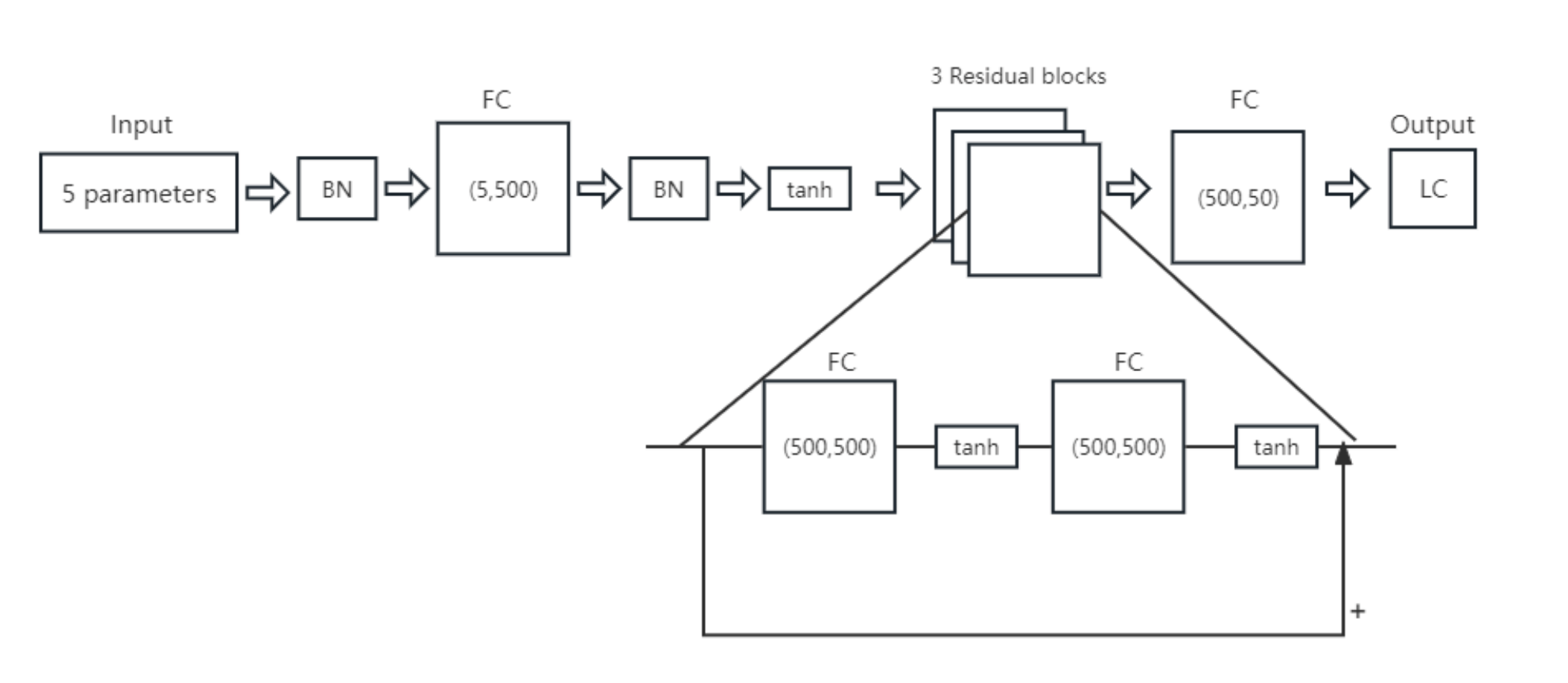}
  \centering
  \caption{The architecture of the network}, which comprises 1 input layer, 8 hidden layers, and 1 output layer. The input layer contains 5 neurons, representing the effective temperature of primary (more massive) star ($T\rm_{1}$), inclination ($i$), relative radii of unfilled component ($R\rm_{(1,2)}/a$), mass ratio ($q$=$M_{2}$/$M_{1}$) and temperature ratio ($T\rm_{2}$/$T\rm_{1}$). Each hidden layer is a fully connected layer (FC), and the number of neurons in each layer is indicated by the numbers. The output is the corresponding light curve.\label{fig:net}
\end{figure*}

\subsection{Model establishment} 
In this study, we utilize a Multi-Layer Perceptron (MLP) network to establish the mapping relationship between parameters and light curves. We apply the MLP network to the mock light curves constructed in Section \ref{sec:traindata}. The complete network architecture is illustrated in Fig.\ref{fig:net}. This MLP network comprises 1 input layer, 8 hidden layers, and 1 output layer. The input layer is composed of 5 neurons, representing the effective temperature of primary (more massive) star ($T\rm_{1}$), inclination ($i$), relative radii of unfilled component ($R\rm_{(1,2)}/a$), mass ratio ($q$=$M_{2}$/$M_{1}$) and temperature ratio ($T\rm_{2}$/$T\rm_{1}$). The output layer predicts the corresponding light curve based on these input parameters. Each hidden layer in the network is a fully connected layer, and the number of neurons in each layer is indicated in Fig.\ref{fig:net}. In order to reduce the complexity of our model, we train the model using only 50 sample points from the 0-0.5 phase range of the light curve. This choice was motivated by the circular and symmetrical characteristics of our simulated light curves, resulting in the model generating a light curve with 50 data points. During the light curve fitting process, we will concatenate the 50-point light curve to create a 100-point curve. To enhance the model’s expressive capacity, we incorporate residual blocks and utilize the hyperbolic tangent (\texttt{tanh}) activation function. The inclusion of residual blocks helps address the vanishing gradient problem, allowing the model to learn more effectively. Finally, during the training process, we employ the backpropagation (BP) algorithm to train the model \citep{1986Natur.323..533R}. For optimization, we utilize the \texttt{Adam} optimizer and \texttt{L2} loss function. These choices aid in optimizing the model's performance and improving its ability to accurately predict the corresponding light curves based on the given input parameters.

\subsection{Model verification}
\begin{figure*}[t]
  \centering
    \subfigure{
   \includegraphics[scale=0.5]{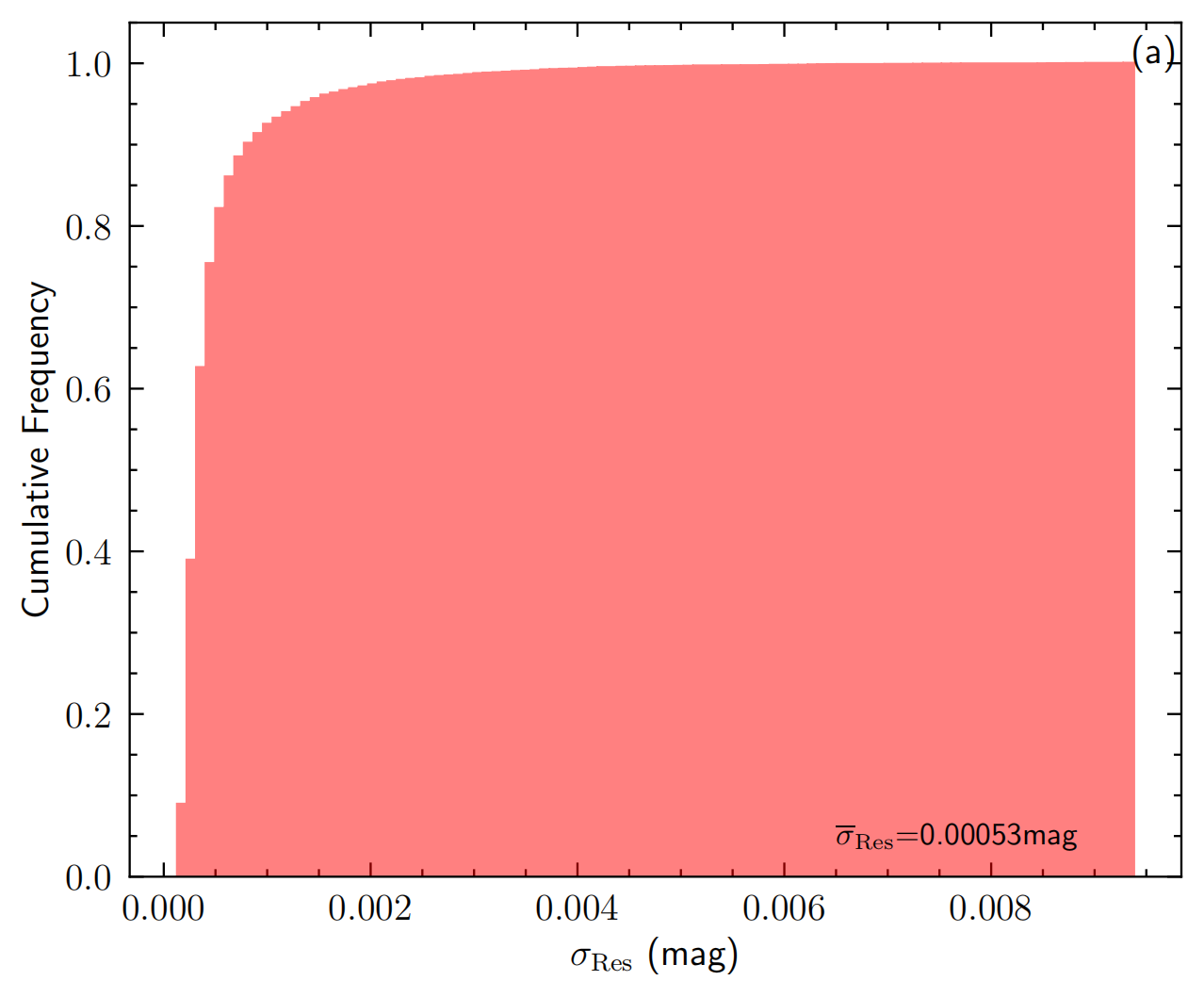}
  }
  \subfigure{
   \includegraphics[scale=0.44]{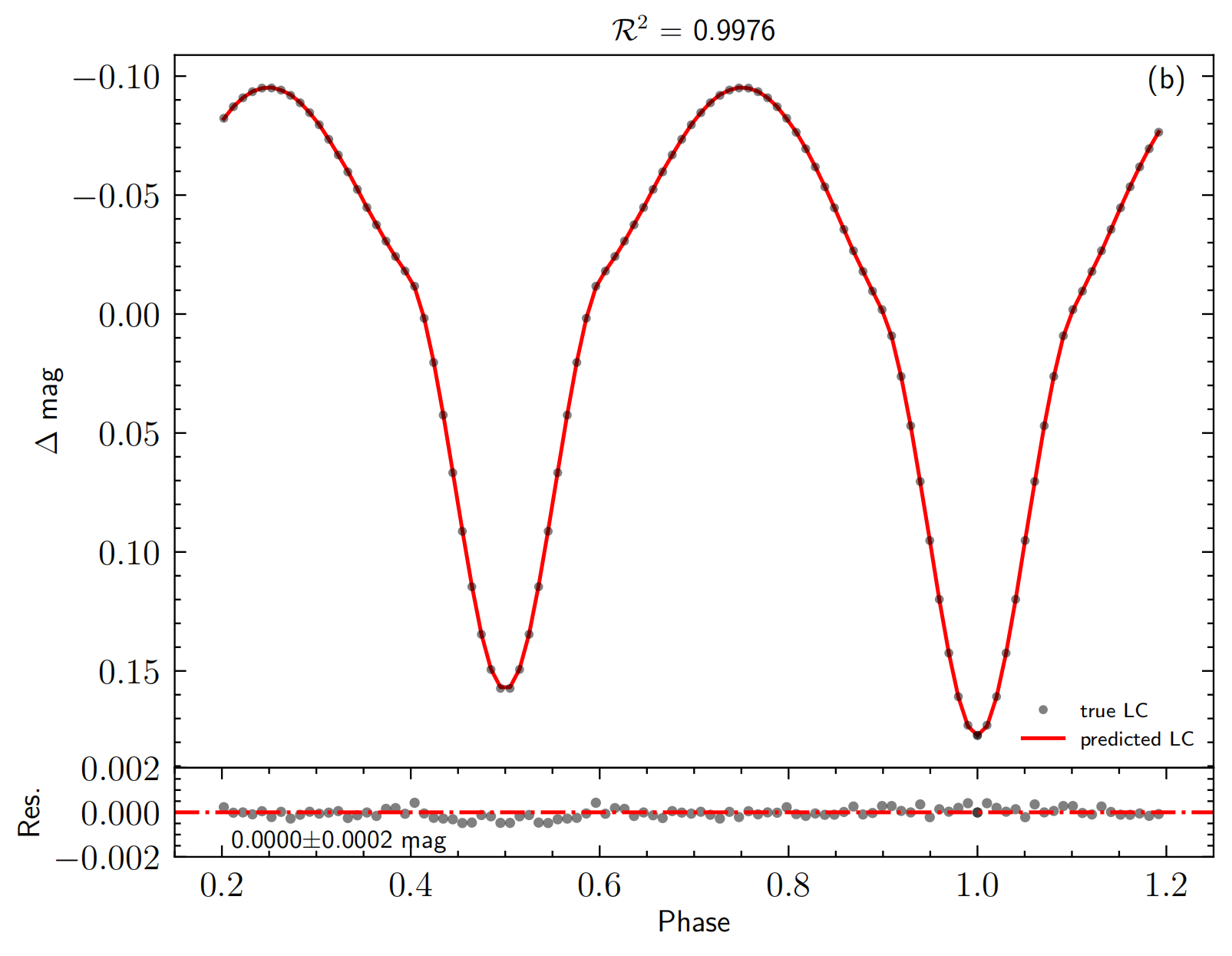}
  }
    \subfigure{
   \includegraphics[scale=0.5]{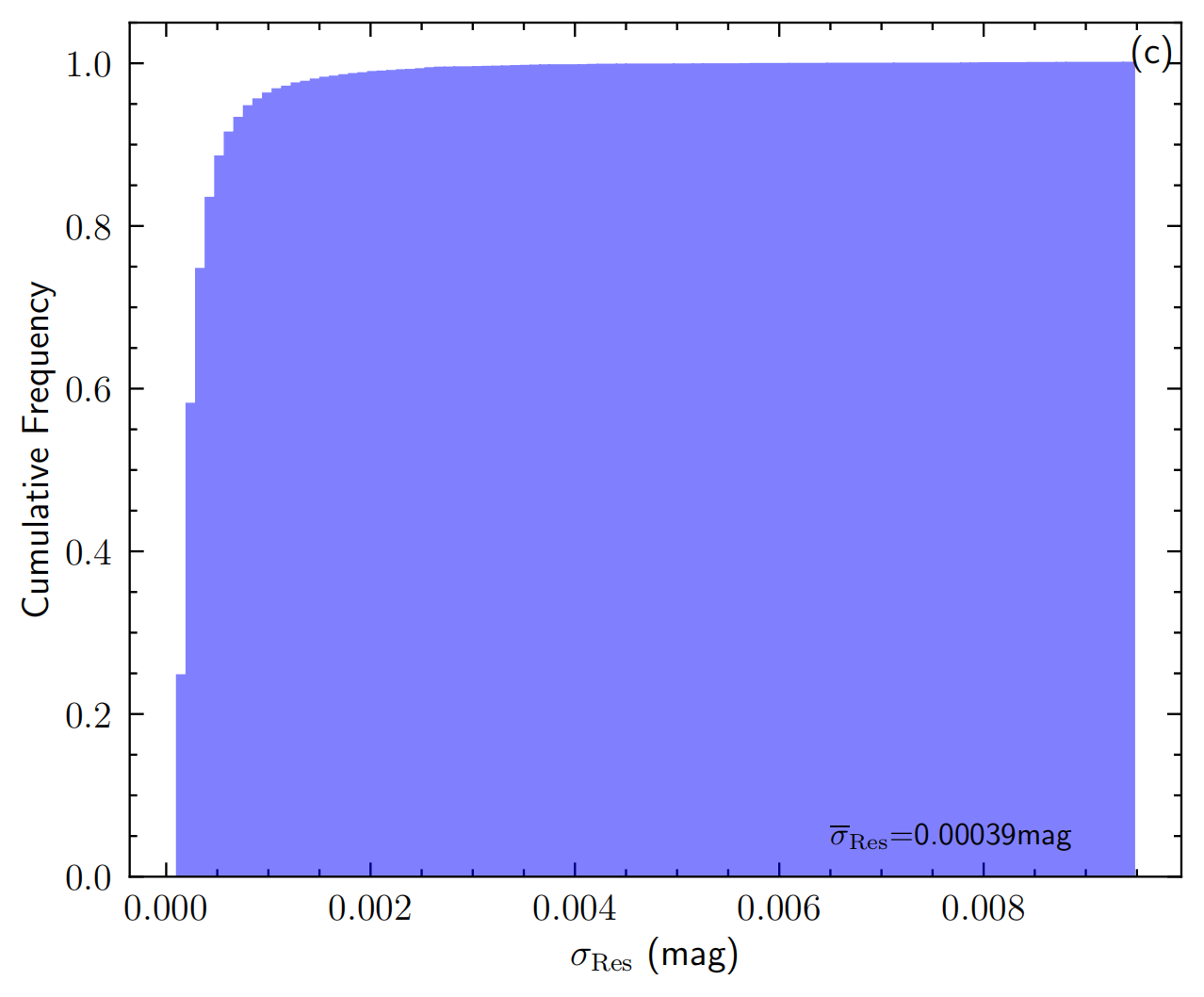}
  }
  \subfigure{
   \includegraphics[scale=0.44]{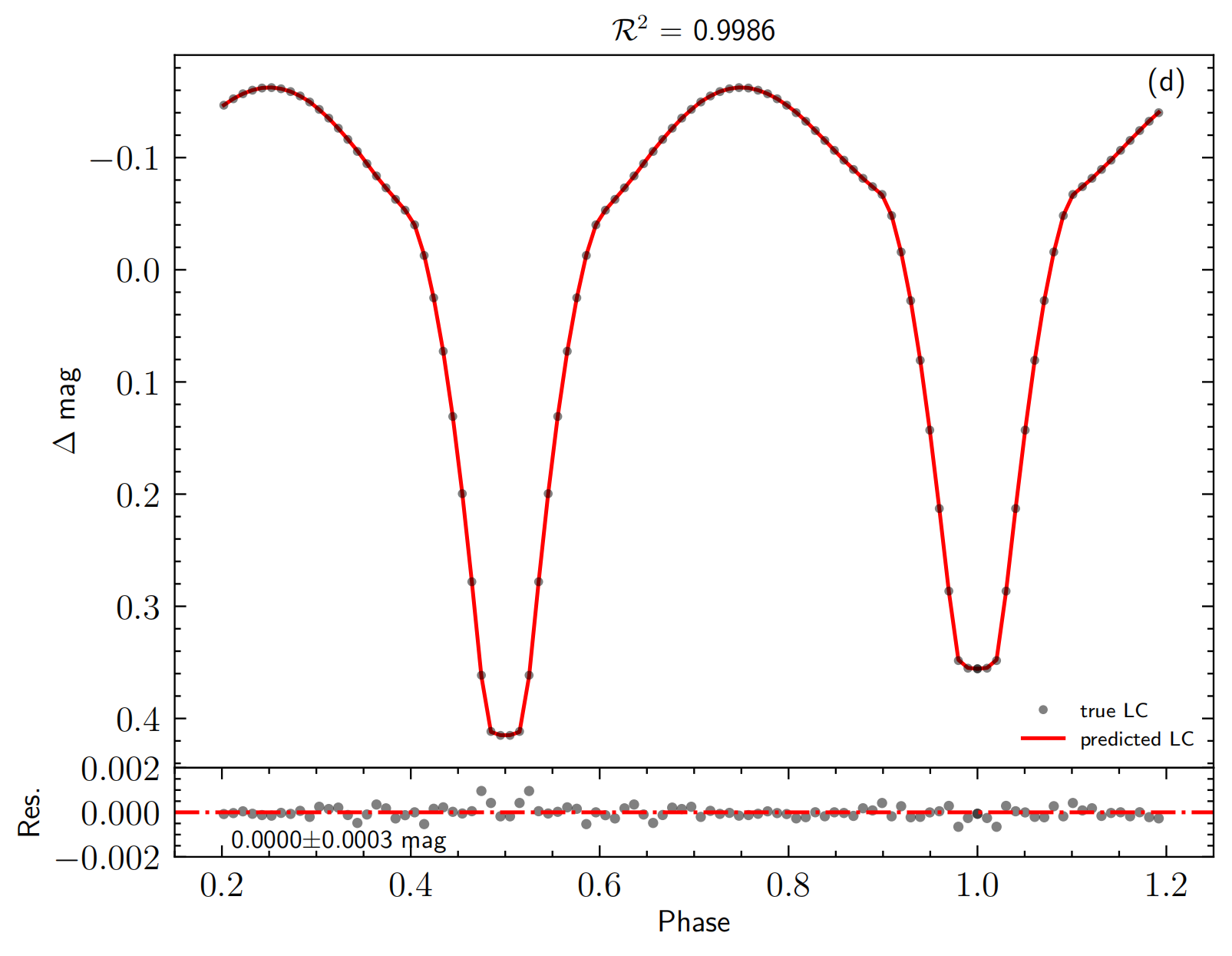}
  }
  \centering
  \caption{The upper plots display the results of the model with a more massive component filling its Roche lobe, while the bottom panels show the results of the model with a less massive component filling its Roche lobe. Panels (a) and (c) are cumulative histograms for standard deviations of the residuals ($\sigma$$\rm_{Res}$) between the predicted light curves and the simulated light curves. Panel (b) shows a direct comparison between the predicted light curve (red solid line) and the true light curve (black dots) for a target with the more massive component filling its Roche lobe, and the parameters are $T\rm_{1}$= 37768 K, $i$=68.97$^{\circ}$, $R\rm_{2}/a$=0.265, $q$=0.924, and $T\rm_{2}$/$T\rm_{1}$=0.892. Similarly, panel (d) presents a direct comparison for a target with the less massive component filling its Roche lobe, with the parameters being $T\rm_{1}$= 29179 K, $i$=89.90$^{\circ}$, $R\rm_{1}/a$=0.235, $q$=0.756, and $T\rm_{2}$/$T\rm_{1}$=1.118. In panels (b) and (d), the corresponding residuals between predictions and true values are shown at the bottom.}\label{fig:model_precision}
\end{figure*}

For model verification, we generate a test dataset consisting of 10,000 mock light curves by PHOEBE considering two scenarios. In Fig.\ref{fig:model_precision} (a) and (b), we present the evaluation of the model for a more massive component filling its Roche lobe. In panel (a), we depict the distribution of standard deviations for the residuals ($\sigma\rm _{Res}$) between the predictions and simulated light curves. It illustrates that the mean value of the residuals' standard deviation ($\overline{\sigma}\rm _{Res}$) is approximately 0.00053 \texttt{mag} for the model with a more massive component filling its Roche lobe. Additionally, in Fig.\ref{fig:model_precision} (b), a direct comparison is provided between the true light curve (black dots) and the predicted light curve (red solid line). Similarly, panels (c) and (d) exhibit the evaluation results of the model with a less massive component filling its Roche lobe. In Fig.\ref{fig:model_precision} (c), it shows that the mean value of the residuals' standard deviation ($\overline{\sigma}\rm _{Res}$) for this model is approximately 0.00039 \texttt{mag}. It can be seen in panels (b) and (d) that there are well-matched light curves between the predictions and mock data for the two models, with the R-squared value ($\mathcal{R}^{2}$) exceeding 0.99. The R-squared value is a statistical measure that represents the goodness of fit between the predicted and true values. And it is defined as:
\begin{equation}\label{eq:R2}
   \mathcal{R}^{2}=1-\frac{\sum^{n}_{i=1}(y_{i}-f(x_{i}))^{2}}{\sum^{n}_{i=1}(y_{i}- \overline{y})^{2}}
\end{equation}
in which, $f(x_{i})$ is the predicted light curve, and $y$ represents the true light curve. $n$ denotes the total number of data points in the light curve.

Fig.\ref{fig:model_precision} indicates that both models are able to accurately capture the characteristics and features of the light curve using the given input parameters. The predicted light curves closely resemble those generated by PHOEBE. Furthermore, the light curve fitting model constructed using this neural network significantly accelerates the process of fitting light curves for semi-detached binaries. Specifically, when comparing the two models to the PHOEBE package, they demonstrate superior efficiency in generating light curves. For example, when modeling a light curve with 100 data points, the light curve fitting model can be executed in just 5 ms on a CPU with 32GB of memory, operating at a frequency of 5.10GHz, such as the 12th Gen Intel$^\circledR$ Core$^\text{TM}$ i9-12900. In contrast, performing the same fitting process using the PHOEBE package under identical conditions would typically take around 4 seconds. As the number of data points increases, the slow speed of PHOEBE makes it almost useless for MCMC analysis. The substantial decrease in processing time by using neural network is particularly well-suited for analyzing extensive datasets.

\section{Photometric analysis based on Neural network} \label{sec: method}
\subsection{Method description}

\begin{figure*}[t]
  \centering
  \includegraphics[width=0.9\textwidth,height=0.55\textheight]{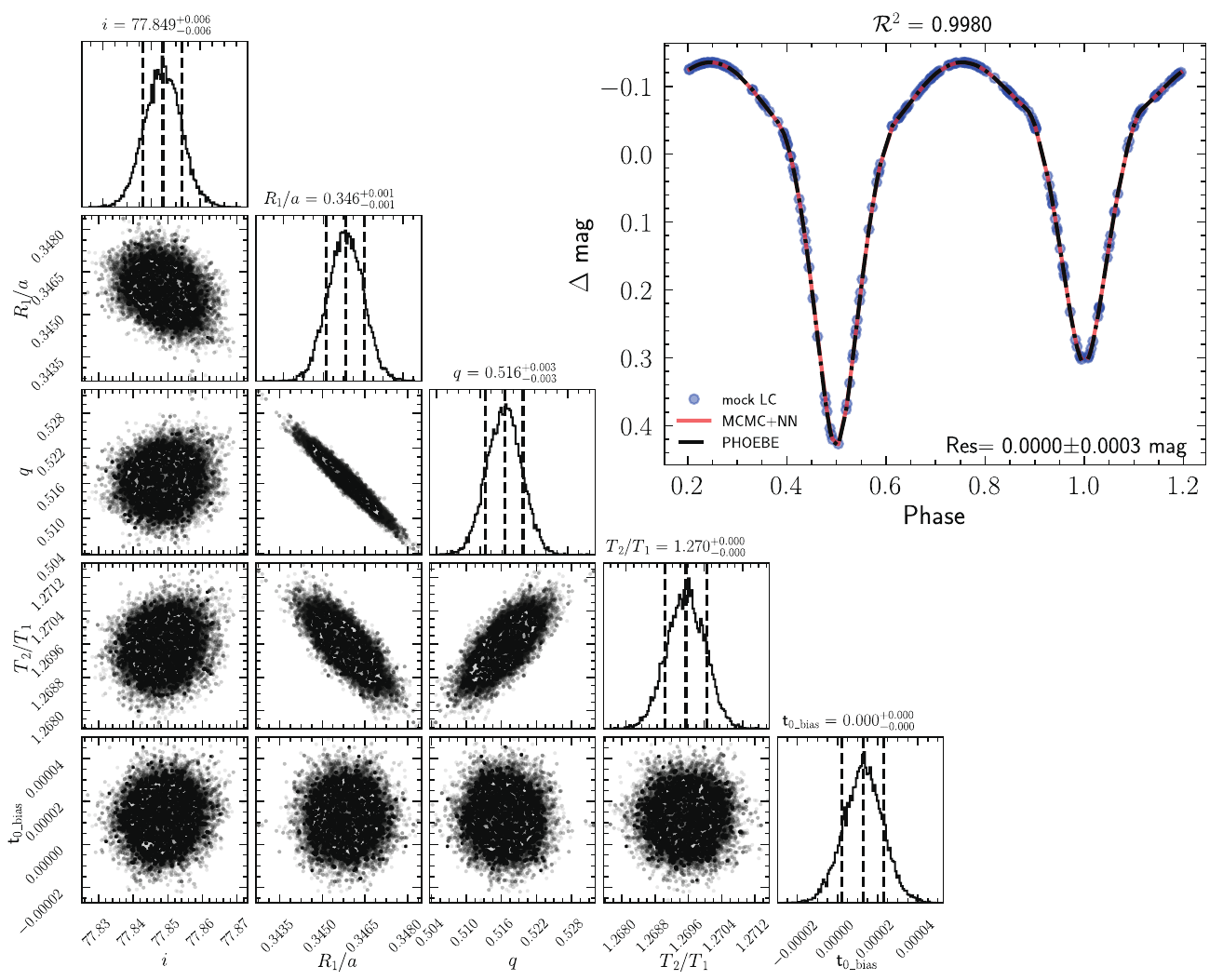}
  \centering
  \caption{The light curve fitting results obtained through our pipeline for a simulated target with the less massive component filling its Roche lobe. The true parameters are $T\rm_{1}$= 36552 K, $i$=77.72$^{\circ}$, $R\rm_{1}/a$=0.347, $q$=0.516, and $T\rm_{2}$/$T\rm_{1}$=1.268. The corner plot displays the distributions of measured parameters with the highest $\mathcal{R}^{2}$ of 0.9980. The corresponding light curves are presented in the upper right plot. In this plot, the blue dots represent the true points of the simulated target, the red line represents the reconstructed light curve using the parameters measured by our pipeline, and the black dashed line represents the light curve generated by PHOEBE based on the parameters measured from our pipeline.}\label{fig:example_final}
\end{figure*}

In our photometric analysis pipeline, we utilize EMCEE \citep{2013PASP..125..306F}\footnote{http://dan.iel.fm/emcee} as the framework for performing a MCMC fitting on light curves in two models (built-in Section \ref{sec:model}). Previously, to speed up the photometric solution process, observed light curves were often resampled along the phase axis. In this study, the models enable the rapid generation of light curves. Hence, in this work, we refrain from resampling, and instead employ interpolation to align the output light curve from the neural network with the phases of the observed data. To summarize, our photometric analysis progress is described as follows:

Initially, the light curves are folded into phases within the range of 0 to 1 using their known orbital periods, and normalized by Eq.\ref{Equ:normalized}. Subsequently, assuming the effective temperature of the primary star ($T\rm_{1}$) is obtained from spectroscopic or multiple-band observations, we set $i$, $R\rm_{2}/a$ (or $R\rm_{1}/a$), $q$, $T\rm_{2}$/$T\rm_{1}$ and t$\rm_{0\_bias}$ as free parameters with uniform priors, based on the initial values from model establishment. Where t$\rm_{0\_bias}$ represents the narrow deviation of the primary minimum potentially caused by measurement errors when we fold light curves.

An illustrative light curve demonstrating the secondary star filling its Roche lobe is used to show this photometric solution method, its parameters are $T\rm_{1}$= 36552 K, $i$=77.72$^{\circ}$, $R\rm_{1}/a$=0.347, $q$=0.516, and $T\rm_{2}$/$T\rm_{1}$=1.268. We have run 800 iterations of MCMC, with each free parameter being evaluated using 100 walkers. A single light curve might be generated by multiple parameter sets. Therefore, we employ DBSCAN (Density-based spatial clustering of applications with noise, \citealt{osti_421283}) to find the global maximum of the $\mathcal{R}^{2}$ among all other local solutions obtained from MCMC results. Finally, the parameters with the highest $\mathcal{R}^{2}$ values are selected as the final results.

\begin{figure*}[t]
  \centering
   \includegraphics[width=\textwidth]{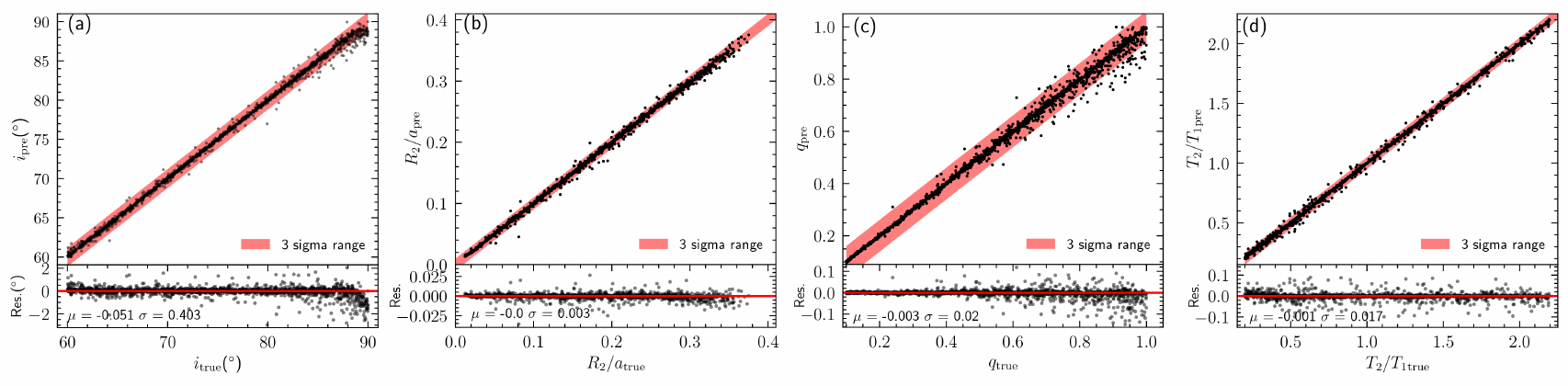}
  \centering
  \caption{The precision of parameter measurements for our pipeline (the model with a more massive component filling its Roche lobe). Panels (a) to (d) show the distributions of the discrepancies of inclination ($i$), relative radius of unfilled component ($R\rm_{2}/a$), mass ratio ($q$) and temperature ratio ($T\rm_{2}$/$T\rm_{1}$). The ordinate (\emph{y} axis) represents the measured values, while the abscissa (\emph{x} axis) is the true values. The mean value ($\mu$) and standard deviation ($\sigma$) of the residuals between measured values and true values are displayed at the bottom of the panels.}\label{fig:pri_paramaters_precision}
\end{figure*}

\begin{figure*}[t]
  \centering
   \includegraphics[width=\textwidth]{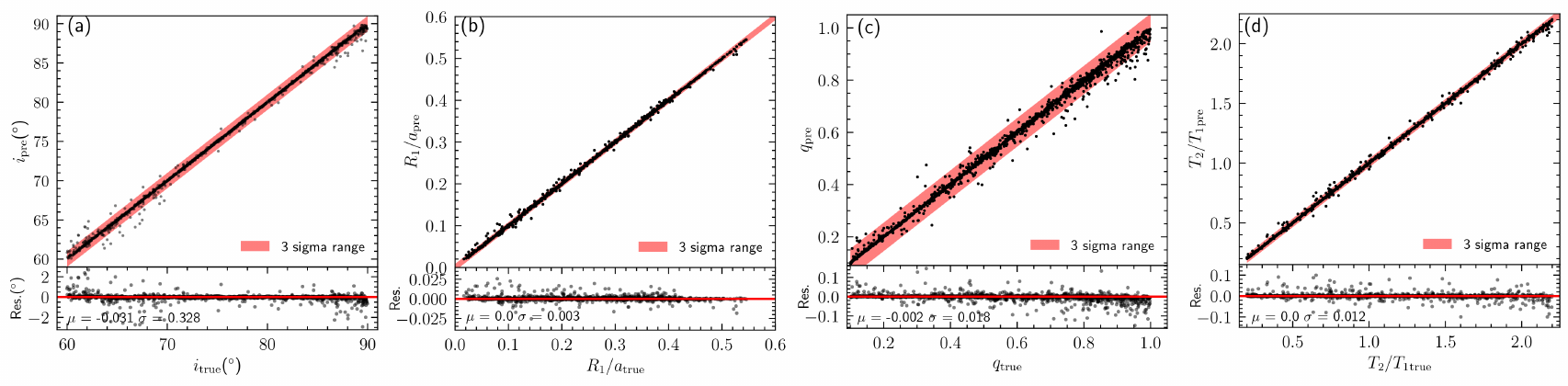}

  \centering
  \caption{The precision of parameter measurements for our pipeline (the model with a less massive component filling its Roche lobe). Panels (a) to (d) show the distributions of the discrepancies of inclination ($i$), relative radius of unfilled component ($R\rm_{1}/a$), mass ratio ($q$) and temperature ratio ($T\rm_{2}$/$T\rm_{1}$). The ordinate (\emph{y} axis) represents the measured values, while the abscissa (\emph{x} axis) is the true values. The mean value ($\mu$) and standard deviation ($\sigma$) of the residuals between measured values and true values are displayed at the bottom of the panels.}\label{fig:sec_paramaters_precision}
\end{figure*}

Fig.\ref{fig:example_final} displays the distributions of parameters with the highest $\mathcal{R}^{2}$. The corresponding light curves are shown in the upper right panel of Fig.\ref{fig:example_final}. In the plot, the blue dots represent the true points of the illustrative example, the red line depicts the reconstructed light curve using the parameters measured by this pipeline, and the black dashed line represents the light curve generated through PHOEBE based on the parameters. Fig.\ref{fig:example_final} demonstrates that the parameters measured by this method are identical to the true values, and the light curve reconstructed by the NN model closely matches the mocked light curve, as well as the one obtained through PHOEBE.

Subsequently, we use 1500 light curves to test the systematic performance and potential biases of this pipeline. Fig.\ref{fig:pri_paramaters_precision} shows the result of our pipeline on the model with a more massive component filling its Roche lobe, in panels (a) to (d), the differences between the parameters ($i$, $R\rm_{2}/a$, $q$ and $T\rm_{2}$/$T\rm_{1}$) obtained from our pipeline and true values are presented. As depicted in Fig.\ref{fig:pri_paramaters_precision}, the majority of samples in the mock dataset can be accurately measured their parameters by this pipeline with small deviations, and the corresponding standard deviations of the differences for the model with more massive component filling its Roche lobe in $i$, $R\rm_{2}/a$, $q$ and $T\rm_{2}$/$T\rm_{1}$ are 0.403$^{\circ}$, 0.003, 0.020, 0.017, respectively. Similarly, Fig.\ref{fig:sec_paramaters_precision} illustrates the evaluation result of our pipeline on the the model with a less massive component filling its Roche lobe. The corresponding standard deviations of differences for $i$, R$_{1}/a$, $q$ and $T\rm_{2}$/$T\rm_{1}$ are 0.328$^{\circ}$, 0.003, 0.018, 0.012, respectively.

\subsection{Method verification}

Here, we apply this pipeline to the observed light curve data for further validation. RT Per (TIC 385105755) was initially identified as an eclipsing binary by \citet{1904AN....166..155C}. It was further validated as a semi-detached binary system, with the lower-mass component filling its Roche lobe \citep{1996ApSS.243..275E}. Table \ref{tab:table1} summarizes the characteristics on RT Per. The effective temperature of the primary star ($T\rm_{1}$) for RT Per is required to set as fixed value. Gaia MSC\footnote{\scriptsize https://gea.esac.esa.int/archive/documentation/GDR3/Data\_ana\\lysis/chap\_cu8par/sec\_cu8par\_apsis/ssec\_cu8par\_apsis\_msc.html} (Multiple Star Classifier, \citet{2023AA...674A...1G}) has provided stellar parameters ($T\rm_{(1,2)}$, $\log g\rm_{(1,2)}$, [M/H], distance, $A\rm_{0}$ and $A\rm_{G}$) for all sources with \texttt{G}$\geq$18.25 \texttt{mag} from BP/RP spectra and parallaxes, assuming they are unresolved coeval binaries. Additionally, assuming they are single stars, Gaia DR3 also offers six homogeneous star samples (types OBA, FGKM, ultracool dwarfs (UCDs), solar analogues, carbon stars, and the Gaia spectrophotometric standard stars (SPSS)) with high-quality astrophysical parameters (golden sample) across the Hertzsprung-Russell (HR) diagram \citep{2023A&A...674A..39G}. For RT Per, the $T\rm_{1}$ values provided by Gaia MSC and the Gaia golden sample are $5921^{+186}_{-284}$ K and $6053^{+22}_{-22}$ K, respectively. These $T\rm_{1}$ values are utilized as fixed constants to fit the light curve. During the fitting process, we employ the pipeline on two NN models to determine the parameters sequentially, and the reproduced light curves from the two models are compared with the observational data using the $\mathcal{R}^{2}$ score. Finally, we select the parameters that yielded the highest $\mathcal{R}^{2}$ value as the final result for RT Per.

\begin{figure*}[ht]
  \centering
  \includegraphics[width=0.9\textwidth,height=0.55\textheight]{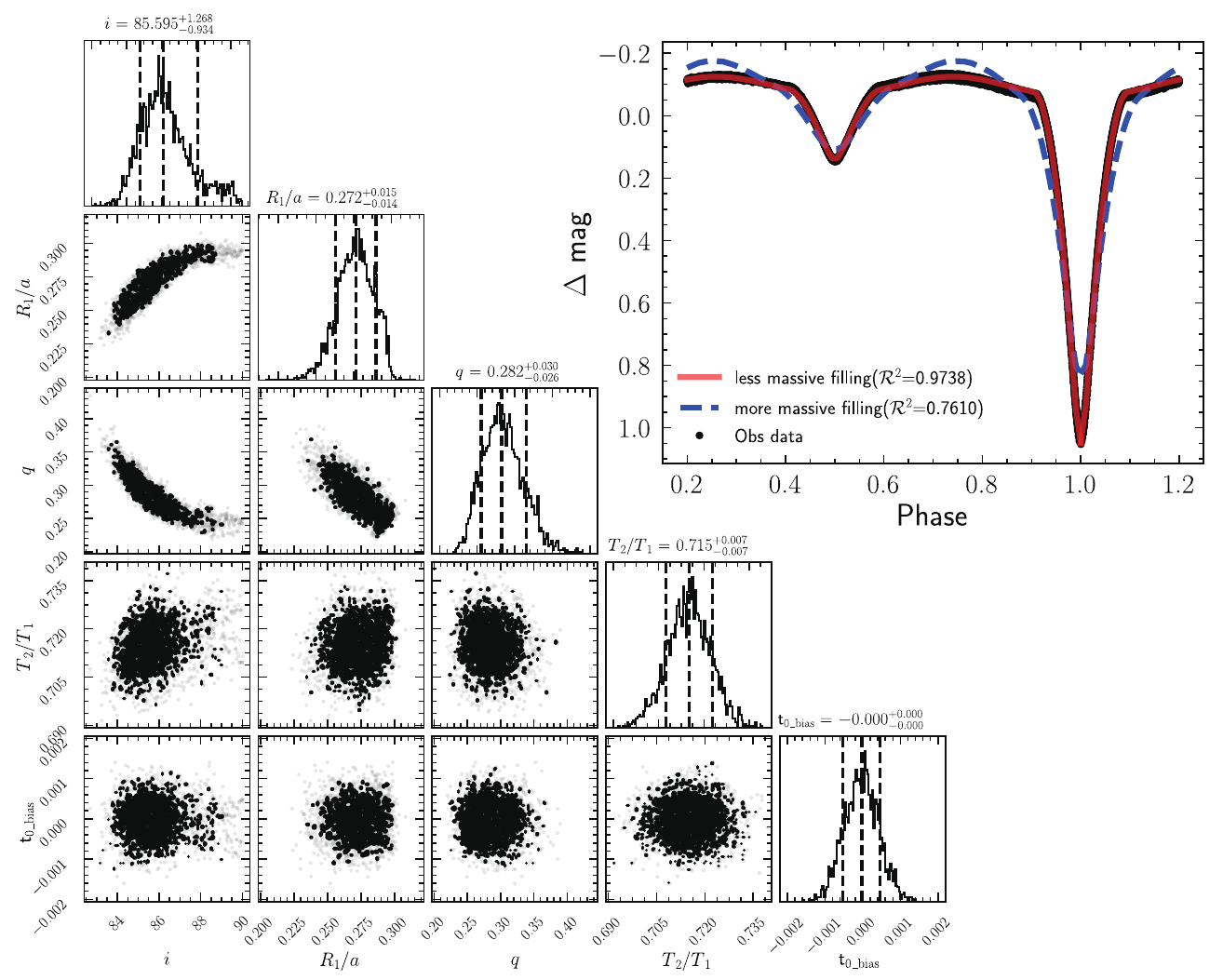}
  \centering
  \caption{Light curve fitting results for RT Per \citep{1904AN....166..155C} obtained by our pipeline with the priors from Gaia golden sample \citep{2023A&A...674A..39G}. The corner plot illustrates the distributions of these measured parameters with the highest $\mathcal{R}^{2}$=0.9748 for the model with less massive component filling its Roche lobe. In the upper right plot, observations from the TESS survey are represented by black dots, while the blue dashed and red solid lines depict reproduced light curves using models with a more massive and a less massive component filling their Roche lobes, respectively.}\label{fig:TIC_385105755_MCMC}
\end{figure*}

The fitting results for RT Per are presented in Table \ref{tab:table1}. As depicted in Table \ref{tab:table1}, the parameters ($i$, $R\rm _{1}/a$, $q$, and $T\rm _{2}$) obtained from Gaia MSC and golden sample show small differences compared to the existing literature. However, we notice relatively larger errors in the mass and radius when utilizing the MSC parameters. Furthermore, there is also a larger disparity between the parameters derived from the MSC and those reported in the existing literature.Fig.\ref{fig:TIC_385105755_MCMC} shows the light curve fitting results of RT Per with the priors from Gaia golden sample. In the upper right panel of Fig.\ref{fig:TIC_385105755_MCMC}, the comparison of the reproduced light curves and observations is presented. The black dots represent the observed light curve, and the blue dashed and red solid lines are the reproduced light curves from two models. And the higher $\mathcal{R}^{2}$ value of 0.9738 for the model with a less massive component filling its Roche lobe is selected as our final result, and the corner plot illustrates the distribution of these parameters. 

\begin{table*}[ht]
\footnotesize
\caption{Characteristics of RT Per}\label{tab:table1}
\centering
\begin{threeparttable}
\scalebox{0.9}{
\begin{tabular}{ccccccc}
\hline
\hline
  \multicolumn{1}{c}{Parameters} &
  \multicolumn{1}{c}{\citep{1979BASI....7..118S}} &
  \multicolumn{1}{c}{\citep{1990AcASn..31..140L}} &
  \multicolumn{1}{c}{\citep{1996ApSS.243..275E}} &
  \multicolumn{1}{c}{\citep{2001AJ....122.2686Q}} &
  \multicolumn{1}{c}{Our results}&
  \multicolumn{1}{c}{Our results} \\
  &  &  &   & &(Gaia MSC) &(Gaia golden sample)\\
\hline
\hline
$P$ (days) & -& 0.8494 & 0.84939889 &-&0.84941024&0.84941024 \\
$i$ ($^{\circ}$) & 85.2&-&85.801$\pm$0.127&-&85.423$^{+0.027}_{-0.025}$&85.595$^{+1.268}_{-0.934}$ \\
$q$=$M\rm _{2}$/$M\rm _{1}$ & 0.250& 0.282$\pm$ 0.004&0.281 $\pm$ 0.003 & 0.280 &0.286$^{+0.001}_{-0.001}$&0.282$^{+0.030}_{-0.026}$ \\
$T\rm _{1}$ (K) & - & -& 6100&F5V&5921$^{+186}_{-284}$ &6053$^{+22}_{-22}$\\
$T\rm _{2}$(K) & - & -& 4705$\pm$0.002&G0&4256$^{+1}_{-1}$&4326$^{+132}_{-135}$\\
$R\rm _{1}/a$ & - & - & 0.266$\pm$0.040&-&0.270$^{+0.000}_{-0.000}$&0.272$^{+0.015}_{-0.014}$ \\
$R\rm _{1}$($R_{\odot}$) & 1.500 & 1.197$\pm$0.009&1.132& 1.13&1.233$^{+0.135}_{-0.126}$& 1.148$^{+0.076}_{-0.064}$ \\
$R\rm _{2}$($R\rm _{\odot}$) &1.900&1.083$\pm$0.008& 1.085&1.09& 1.255$^{+0.138}_{-0.130}$& 1.165$^{+0.096}_{-0.089}$ \\
$M\rm _{1}$($M\rm _{\odot}$) & 1.200& 1.070$\pm$0.023& 1.068& 1.07& 1.343$^{+0.495}_{-0.376}$& 1.099$^{+0.290}_{-0.231}$ \\
$M\rm _{2}$($M\rm _{\odot}$) & 0.300&0.304$\pm$0.005&0.300&0.30&0.385$^{+0.141}_{-0.108}$& 0.309$^{+0.081}_{-0.065}$ \\
\hline
\end{tabular}}

\begin{tablenotes}

    \footnotesize
    \item[1]  The period we used is derived from \citet{2022ApJS..258...16P}. The final two columns are our results obtained using different priors, \\specifically Gaia MSC (Multiple Star Classifier,\citet{2023AA...674A...1G}) and Gaia golden sample of astrophysical \\parameters
    \citep{2023A&A...674A..39G}. For RT Per, the goodness-of-fit score \textit{logposterior\_msc}=201.

\end{tablenotes}

\end{threeparttable}

\end{table*}

\section{Photometric Analysis for TESS survey} \label{sec:tess}
\subsection{Target Selection}

To determine the properties of semi-detached binaries for the TESS survey, we combined the identified eclipsing binaries from \citet{2021AA...652A.120I} and \citet{2022ApJS..258...16P} with the effective temperatures from Gaia DR3. Initially, the samples consist of all EBs identified in the aforementioned studies. We then apply a rigorous selection process to refine the sample as follows:

First, we extract the complete information in TESS input catalog \citep{2019AJ....158..138S} for these EBs, we used the
latest version of the TIC (TIC v8.1). Second, we cross-match the TESS EBs with Gaia DR3 within 5 arcsec cone, and for cases where one star has multiple Gaia DR3 sources, we chose the source that exhibited the least deviation of magnitude in \texttt{G/BP/RP} band between TESS input catalog and Gaia DR3. Third, to ensure the accuracy of the parameters obtained from Gaia data, we also apply two quality filter parameter selection conditions on the Gaia DR3 dataset: (1) \texttt{parallax\_over\_error $>$ 5}; (2) \texttt{RUWE $<$ 1.4}. Such selection criteria were widely accepted practices and have been used in previous studies. For example, \citet{2019MNRAS.482.3831P,2019MNRAS.488.2892P} used a similar condition to search for extremely low-mass white dwarfs, and \citet{2023MNRAS.523.2369S} used Gaia DR3 parallaxes to calibrate preliminary period-luminosity relations of O-rich Mira variables, etc. In the next step, we review the light
curves for each star in the TESS survey.

\subsection{Light curve from TESS}

The TESS survey provides 2-minute and 30-minute cadence data on the public data releases of TESS-SPOC \citep{2016SPIE.9913E..3EJ}
and MIT QLP \citep{2020RNAAS...4..204H,2020RNAAS...4..206H} during its normal mission. The eclipsing binaries from \citet{2021AA...652A.120I} and \citet{2022ApJS..258...16P} are detected using 2-minute and 30-minute cadence data from the first and second year of TESS observations (sectors 1-26). Therefore, for our study, we will process both the 2-minute and 30-minute cadence data for the provided eclipsing binaries, if such data is available on the Mikulski Archive for Space Telescopes (MAST\footnote{https://archive.stsci.edu}).

Furthermore, the short-cadence (2-minute) light curve data obtained from the TESS survey includes \texttt{SAP} (Simple Aperture Photometry) flux \citep{2010SPIE.7740E..23T,2020ksci.rept....6M} and \texttt{PDCSAP} (Presearch Data Conditioning \texttt{SAP}) flux\citep{2012PASP..124.1000S,2014PASP..126..100S}. \texttt{SAP} flux is basically a background-corrected value, while the \texttt{PDCSAP} flux not only corrects for the background but also effectively preserves transits, eclipses, and the intrinsic characteristics of the stars by removing systematic stellar effects. And \texttt{PDCSAP} flux is typically preferred for scientific research in analyzing stellar variability and conducting studies on planetary transit events. For long-cadence (30-minute) light curve, \texttt{SAP} flux and \texttt{KSPSAP} flux are provided, where the \texttt{KSPSAP} flux represents the light curve from the optimal aperture\citep{2020RNAAS...4..204H,2020RNAAS...4..206H}. Therefore, in our study, when \texttt{PDCSAP} flux and \texttt{KSPSAP} flux are available, they are utilized for analysis. Then, we filter out the stars with few observations and involve data pre-processing, which includes removing invalid data observations and normalizing the light curve by Eq.\ref{Equ:normalized}.

\subsection{Photometric solution}
After conducting the above quality control screening, we obtained 2914 eclipsing binaries from the TESS survey. We next employ the pipeline proposed in this study to investigate the candidates of semi-detached binaries and extract their parameters. For each target, we sequentially employ the two models for these binaries. Finally, we select the candidates with an $\mathcal{R}^{2}$ score larger than 0.95 and exclude the candidates with a dispersion greater than 0.02 \texttt{mag} between the fitted light curves and the observed light curves. 


\subsection{Absolute parameters solution}
According to the light-curve solution, we can obtain the mass ratio, relative radii, effective temperatures, and luminosity ratio. Then, the absolute parameters (e.g. luminosity, mass, radii) are calculated.
For the purpose of measuring the luminosity of each star, the total bolometric magnitude is required to be entered into:
\begin{equation}\label{eq:mb}
	 	M_{\rm b\rm_{total}} = M'\mathrm{_{total}}-\textit{A}+\textit{BC},
\end{equation}
where $M'\mathrm{_{total}}$ is the absolute magnitude derived from:
\begin{equation}\label{eq:mg}
	 M'\rm _{total}=\textit{m}\rm _{total}+5-5\mathrm{log}_{10}(\textit{d})
\end{equation}
In Eq.~\ref{eq:mg}, $m\rm _{total}$ is the visual magnitude obtained from Gaia DR3. In this paper, the visual magnitudes provided by Gaia in three bands (\texttt{G, BP, RP}) are all used to calculate the absolute parameters. $d$ represents the distance in parsecs(\texttt{pc}) that obtained from \citet{2021AJ....161..147B} or Gaia MSC, where the distance estimation is based on color and magnitude priors. In Eq.~\ref{eq:mb}, $A$ represents the extinction in different band, which is calculated from $A_{\rm V}$ using the extinction coefficient provided by \cite{2019ApJ...877..116W}, and the $A_{\rm V}$ is given from the 3-D dust map \citep{2019ApJ...887...93G}. And $BC$ is the bolometric correction obtained from \cite{2019A&A...632A.105C}.
Then, the luminosity of each star can be calculated by:
\begin{equation}\label{eq:L12}
	 \rm log (\textit{L}_{1,2}/\textit{L}_{\odot}) = \frac{\textit{M}_{b (1,2)}-\textit{M}_{b\odot}}{-2.5}
\end{equation}
where $M\rm _{b\odot}$ is the absolute bolometric magnitude of the Sun that is taken as 4.73 \texttt{mag} 
\citep{2010AJ....140.1158T}. And the $M\rm _{b, (1,2)}$ is the separated bolometric magnitude for each component, which is calculated as:
\begin{equation}\label{eq:m12}
    M_{b (1,2)}=M_{\rm b_{total}}-2.5 \times \mathrm{log}_{10}(l_{(1,2)})
\end{equation}
where $l_{(1,2)}$ is a relative value of luminosity derived by:
\begin{equation}\label{eq:l12r}
\begin{aligned}
&l_{2}=1/(1+\frac{l_{1}}{l_{2}})\\
    &l_{1}=1-l_{2}\\
\end{aligned}
\end{equation}
in Eq.\ref{eq:l12r}, $\frac{l_{1}}{l_{2}}$ is derived by Steffan-Bolzman law, where $\frac{l_{1}}{l_{2}}$ = $(\frac{R_{1}}{R_{2}})^{2}$ $\times$ $(\frac{T_{1}}{T_{2}})^{4}$. As the $L_{(1,2)}$ is measured from bolometric magnitude (see Eq.\ref{eq:L12}), radii can be derived by the relation of $L_{(1,2)}=4\pi R^{2}_{(1,2)}\cdot \sigma T^{4}_{(1,2)}$. Then semi-major axis ($a$) is calculated by using relative radii ($R_{(1,2)}$/a). Finally, masses can be computed using the third Kepler's law (Eq.\ref{eq:kepler}) and mass ratio ($q$).:
\begin{equation}\label{eq:kepler}
  G(M_{1}+M_{2}) = \frac{4\pi ^{2} a^{3}}{P^{2}}
\end{equation}

\begin{figure*}[t]
  \centering
    \subfigure{
   \includegraphics[scale=0.43]{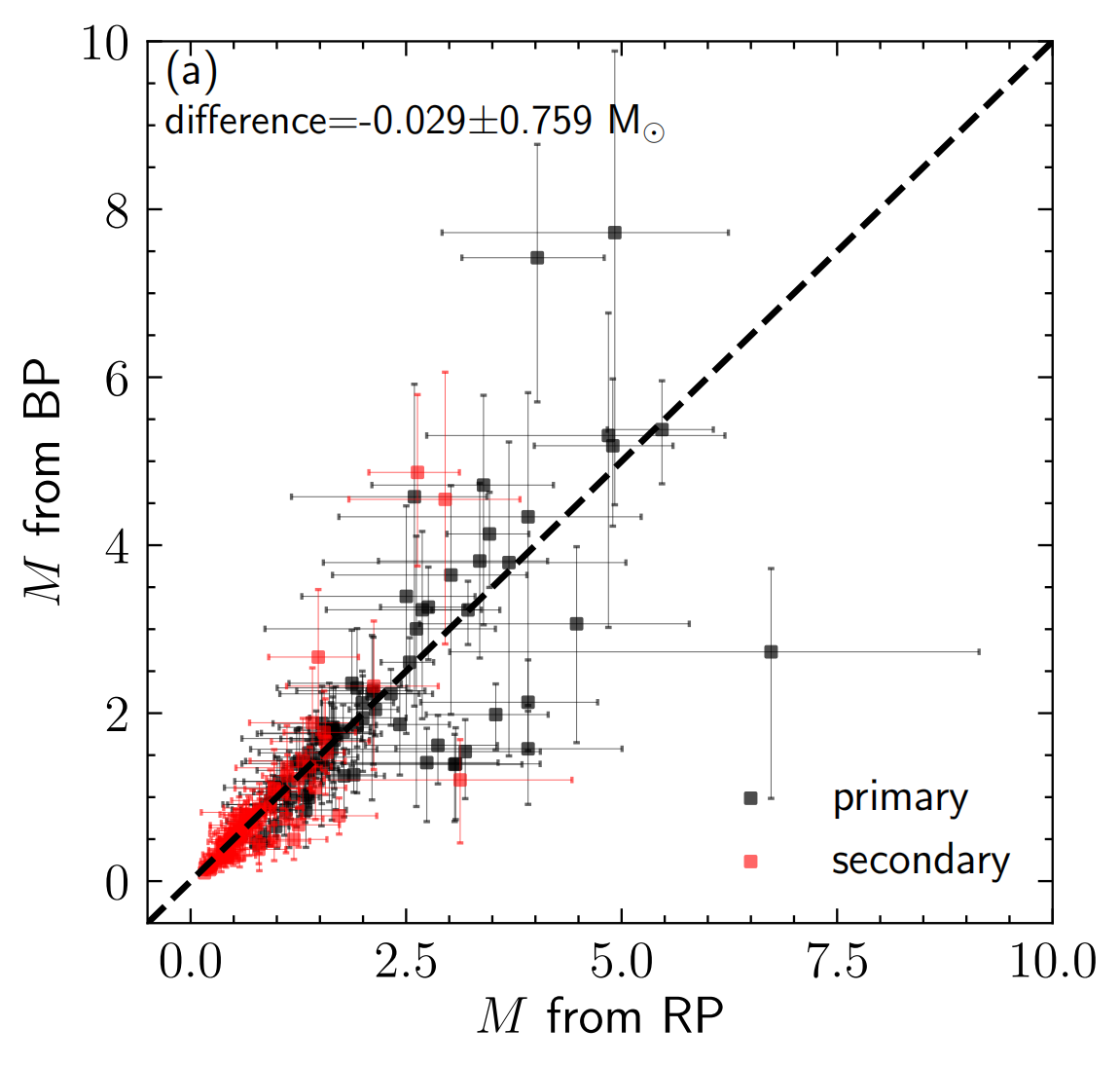}
  }
  \subfigure{
   \includegraphics[scale=0.43]{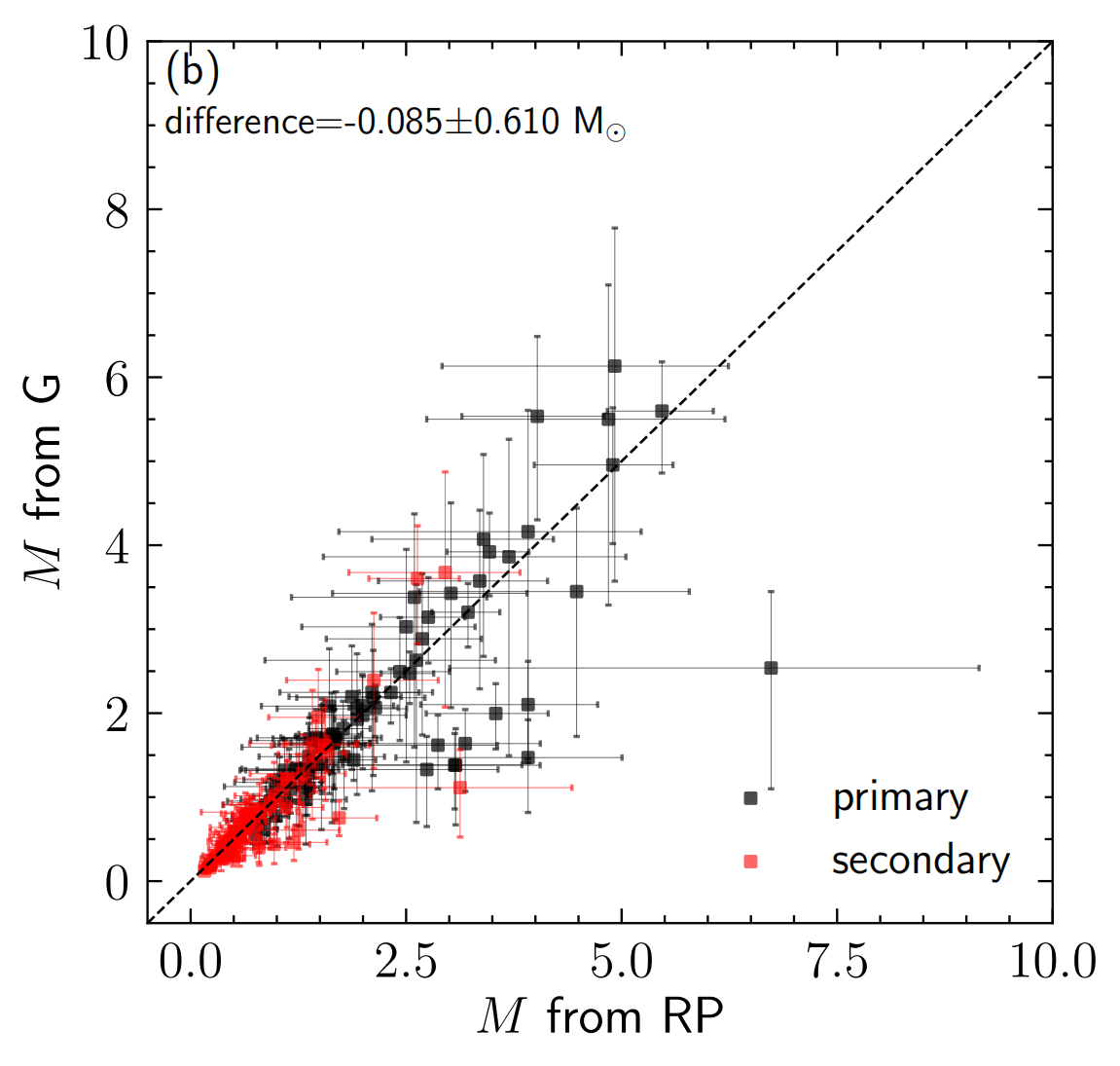}
  }
    \subfigure{
   \includegraphics[scale=0.43]{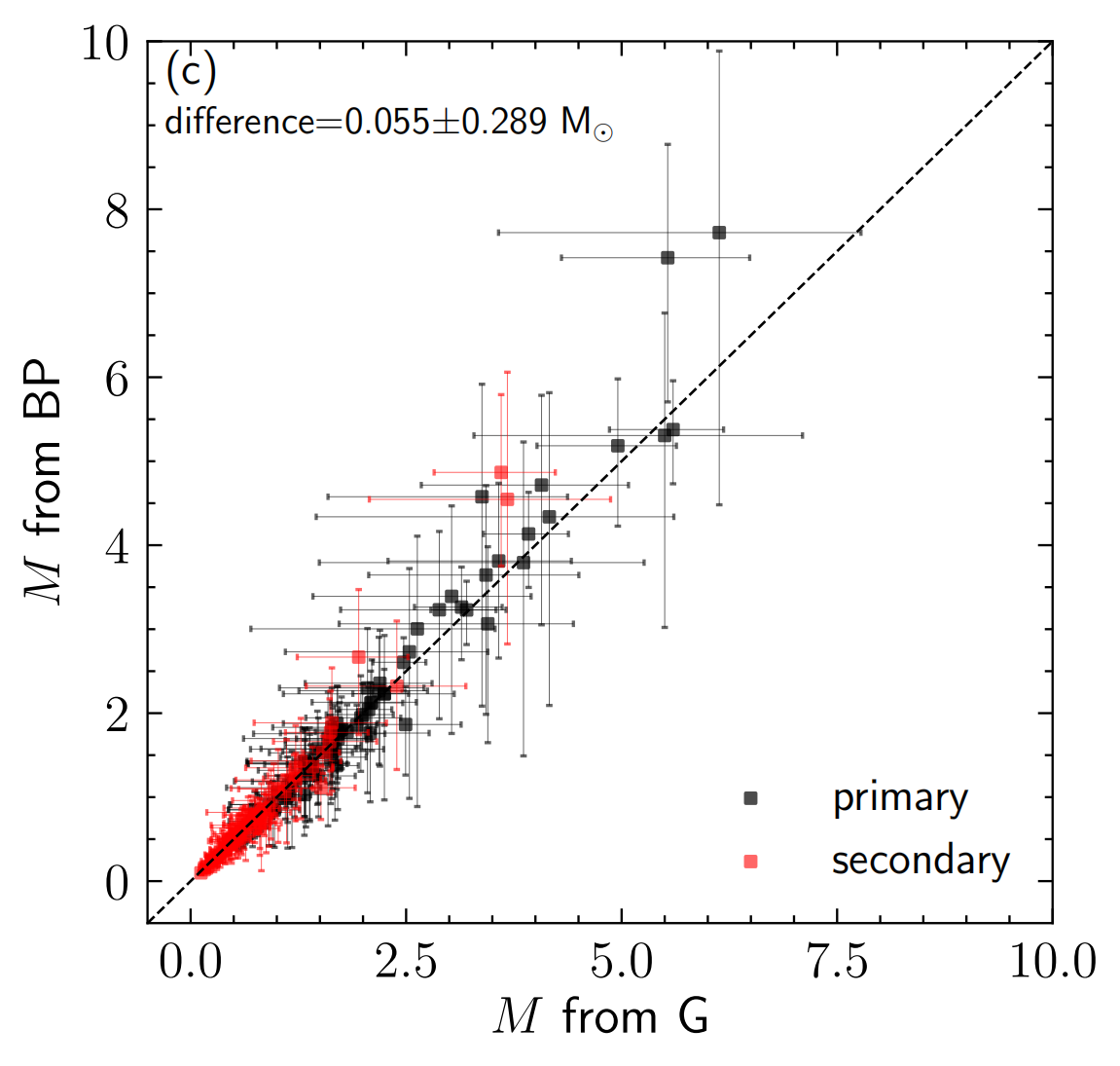}
  }
  \subfigure{
   \includegraphics[scale=0.43]{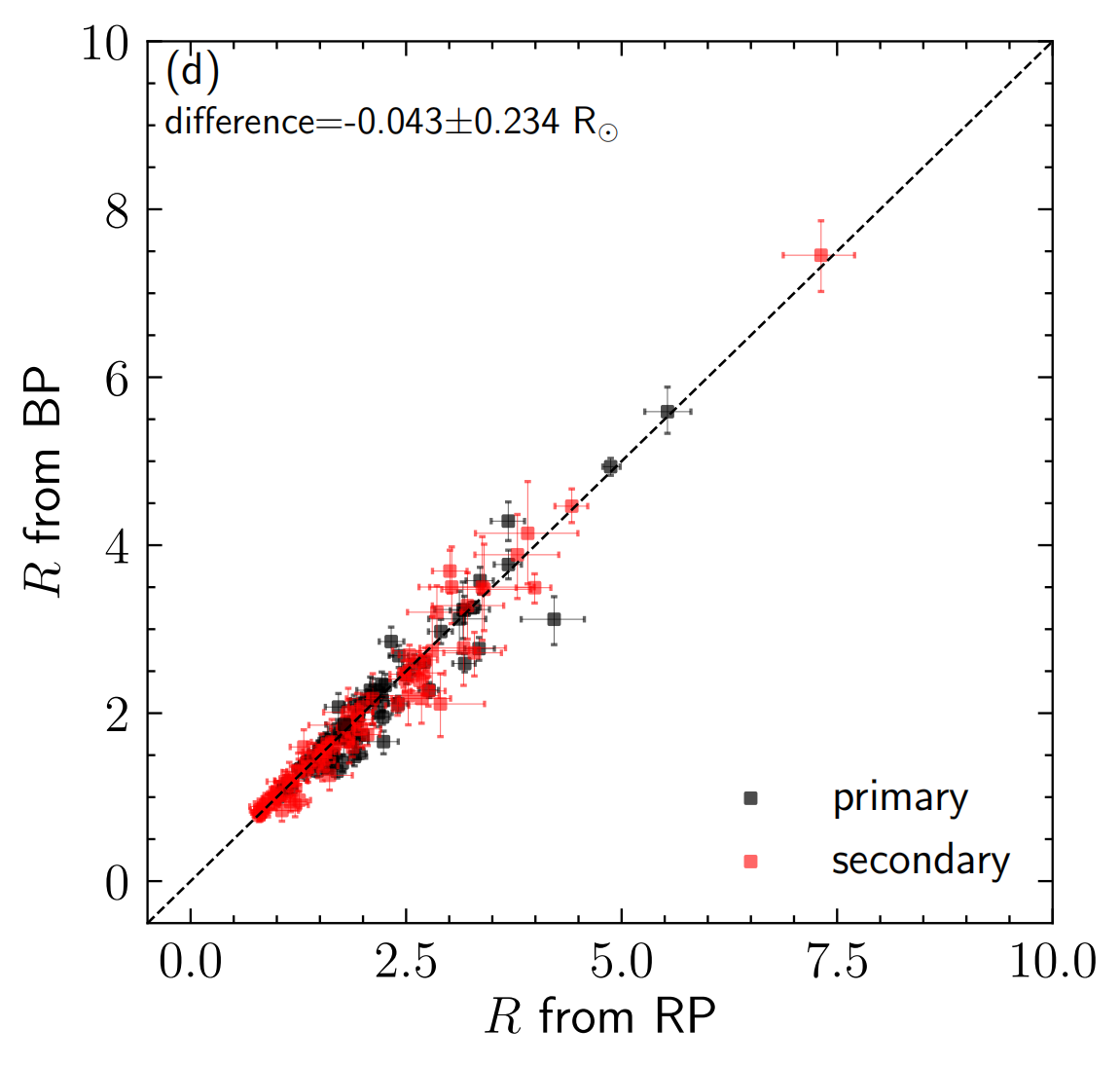}
  }
   \subfigure{
   \includegraphics[scale=0.43]{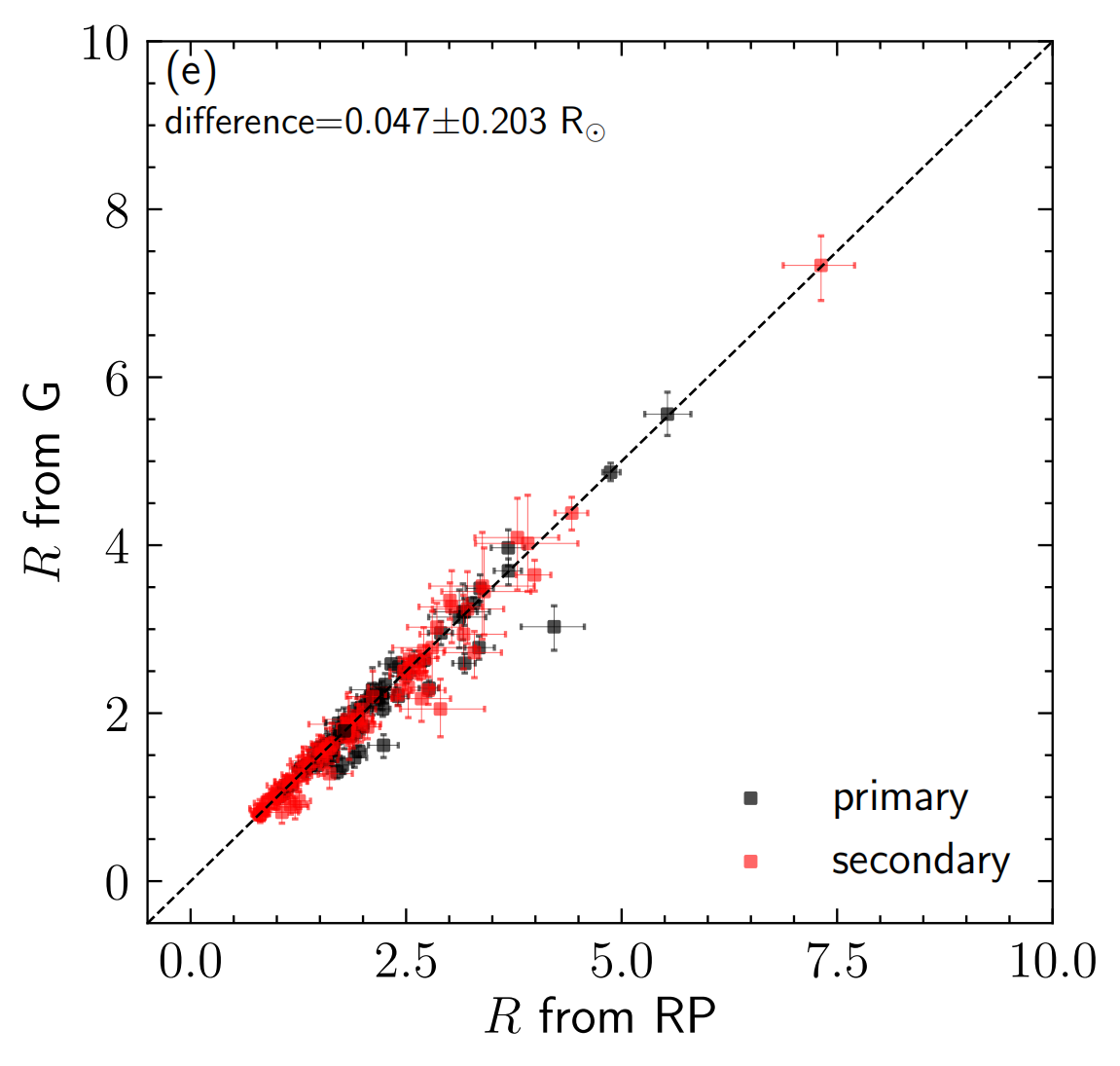}
  }
   \subfigure{
   \includegraphics[scale=0.43]{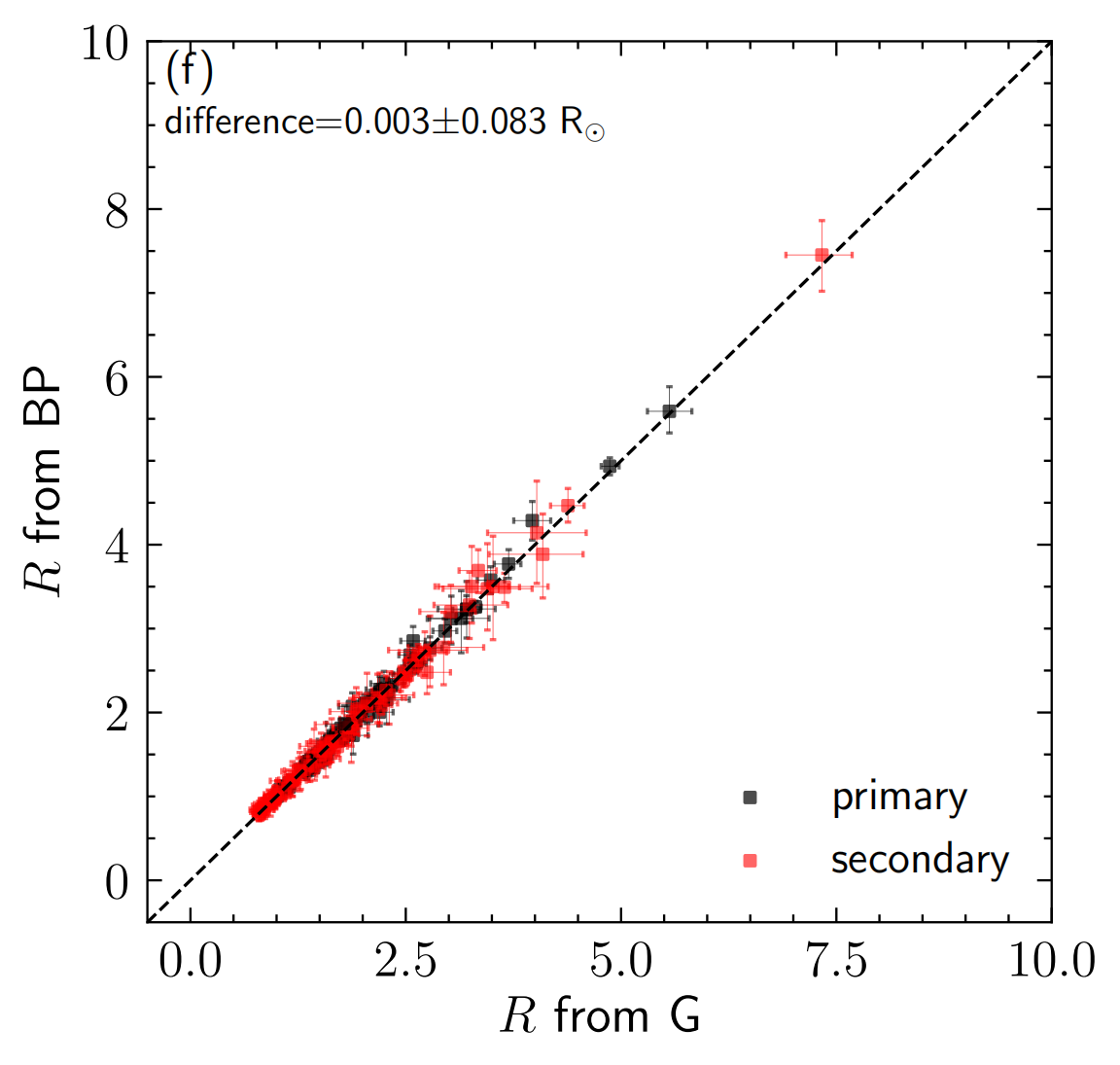}
  }
  \centering
  \caption{Comparison of mass (panel (a) to (c)) and radius (panel (d) to (f)) determined from Gaia \texttt{G-, BP-} and \texttt{RP-} bands using the parameters from Gaia golden sample. The black and red rectangles represent the primary and secondary components.}\label{fig:MR_compare}
\end{figure*}

\begin{figure*}[htb!]
  \centering
    \subfigure{
   \includegraphics[scale=0.65]{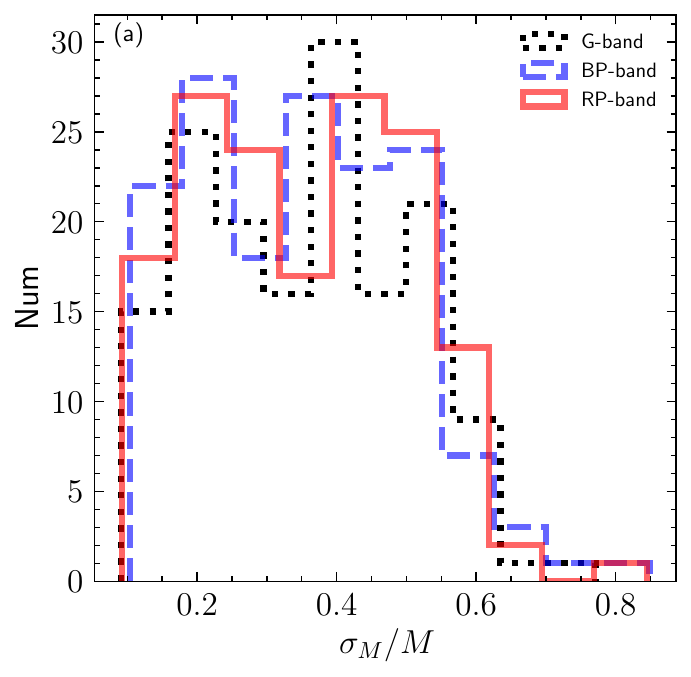}
  }
  \subfigure{
   \includegraphics[scale=0.65]{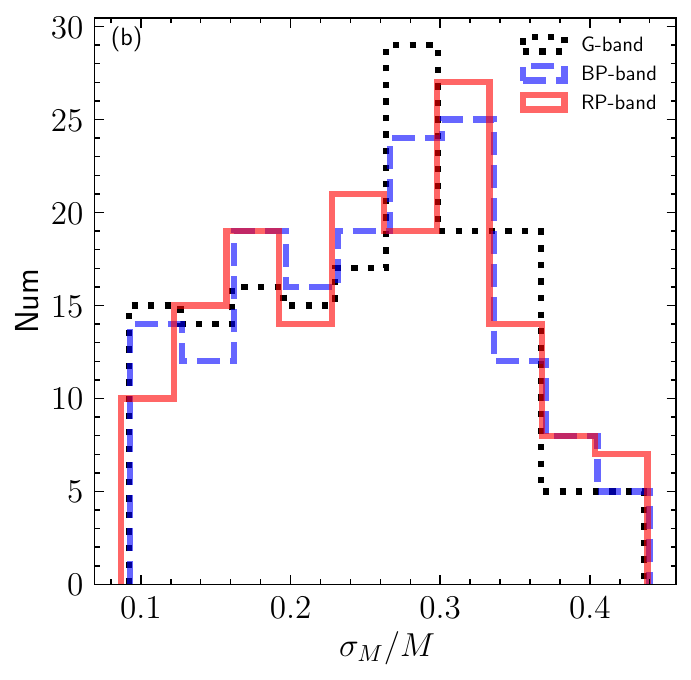}
  }\\
  \subfigure{
   \includegraphics[scale=0.65]{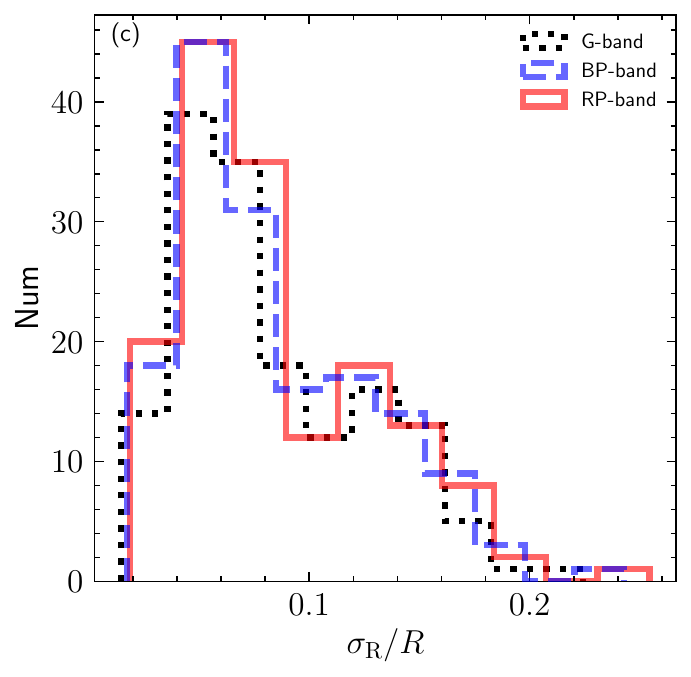}
  }
  \subfigure{
   \includegraphics[scale=0.65]{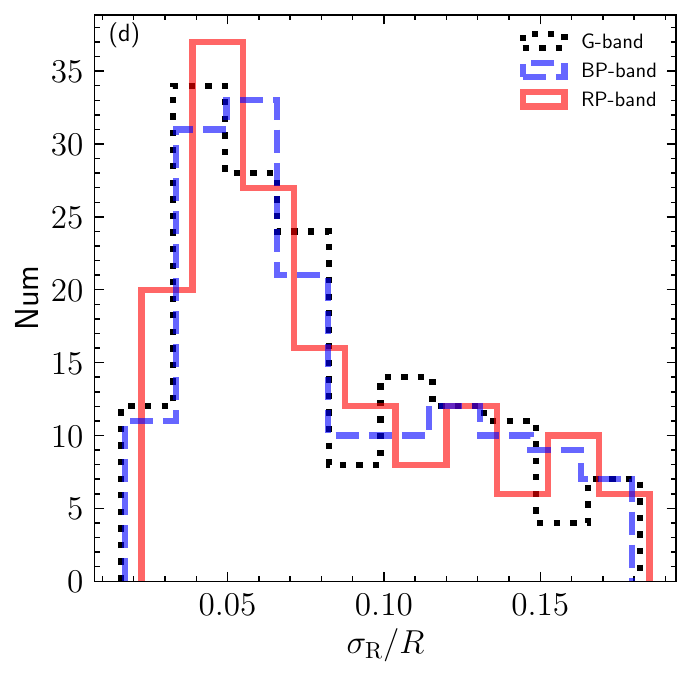}
  }
  \centering
  \caption{The distribution of relative uncertainties ($\sigma_{M}/M$ and $\sigma_{R}/R$) on masses ((a) and (b)) and radii ((c)and (d)) for our candidates that derived from Gaia three bands by using the parameters from Gaia golden sample. The black dotted, blue dashed, and red solid represent the relative uncertainties for Gaia’s
G-, BP- and RP-band, respectively.}\label{fig:MR_sec_precision}
\end{figure*}

\section{Results}\label{sec:result}


\subsection{Absolute parameters}

From the TESS EB catalogs \citep{2021AA...652A.120I,2022ApJS..258...16P}, we finally identified 77 semi-detached binary candidates, including 76 binaries with a low-mass component filling its Roche lobe and $1$ system with a more massive star filling its Roche lobe. In Appendix \ref{AppendixA}, we present their parameters computed using the priors from Gaia MSC and Gaia golden sample, respectively. Additionally, the upper limit for effective temperature in Gaia MSC is set at 8000K, and considering our sample includes systems with early-type stars\citep{2021AA...652A.120I}, in this section, we present the parameters calculated based on Gaia golden sample and distances from \citet{2021AJ....161..147B}. It's essential to note that our parameter solutions in this study are predicated on the assumption of these binaries being semi-detached systems.

We compare the differences and relative uncertainties of masses and radii ($\sigma_{M}/M$ or $\sigma_{R}/R$) for our candidates calculated from Gaia \texttt{G-, BP-} and \texttt{RP-}band in Fig.\ref{fig:MR_compare} and Fig.\ref{fig:MR_sec_precision}. In Fig.\ref{fig:MR_compare}, it can be observed that there are relatively larger discrepancies when comparing the mass and radius between the \texttt{RP-}band and \texttt{BP-}band (or \texttt{G-}band). In the Gaia mission, the \texttt{G-}band photometry demonstrates relatively higher precision, as it measures flux from the Image Parameter Determination (IPD) process, employing a complex model that incorporates extensive calibrations and employs a shape-based Point Spread Function (PSF). Conversely, \texttt{BP-} and \texttt{RP-}band photometry relies on the integration of low-resolution spectra\citep{2021A&A...649A...3R}. For stars showing larger mass and radius measurement discrepancies, their effective temperatures fall in the range of 6500 to 7500 K, which corresponds to early F-type or late A-type stars. These stars may also display variations in their spectral response in different bands.

Additionally, Fig.\ref{fig:MR_sec_precision} shows the distribution of relative uncertainties of masses (panel (a) and (b)) and radii (panel (c) and (d)). Due to the asymmetric nature of the prior distributions for effective temperature, distance etc., we generate 500 Monte Carlo samples from these asymmetric distributions to estimate their upper and lower levels $1\sigma$ confidence intervals (i.e., 16th and 84th percentiles). In Fig.\ref{fig:MR_sec_precision}, the histograms represented by black dotted lines, blue dashed lines, and red solid lines depict the relative uncertainties of masses and radii derived from Gaia's \texttt{G-, BP-} and \texttt{RP-}band with priors from Gaia golden sample. Panels (a) and (c) display the distribution of the upper $1\sigma$ confidence intervals for mass and radius, while panels (b) and (d) depict the distribution of the lower $1\sigma$ confidence intervals. As Fig.\ref{fig:MR_sec_precision} shows, the median relative uncertainties of the upper and lower $1\sigma$ confidence intervals for masses in the \texttt{G-, BP-} and \texttt{RP-}band are 36.4\% (upper) and 26.3\% (lower), 36.3\% (upper) and 25.8\% (lower), and 36.4\% (upper) and 25.5\% (lower), respectively. The larger measurement uncertainty of mass in our study is due to the mass being derived from the radius, which introduces error propagation and amplifies uncertainty in the calculation of mass. Subsequently, we compare these parameters with stellar models and samples from the literature.

\begin{figure*}[htb!]
  \centering
    \subfigure{
   \includegraphics[scale=0.6]{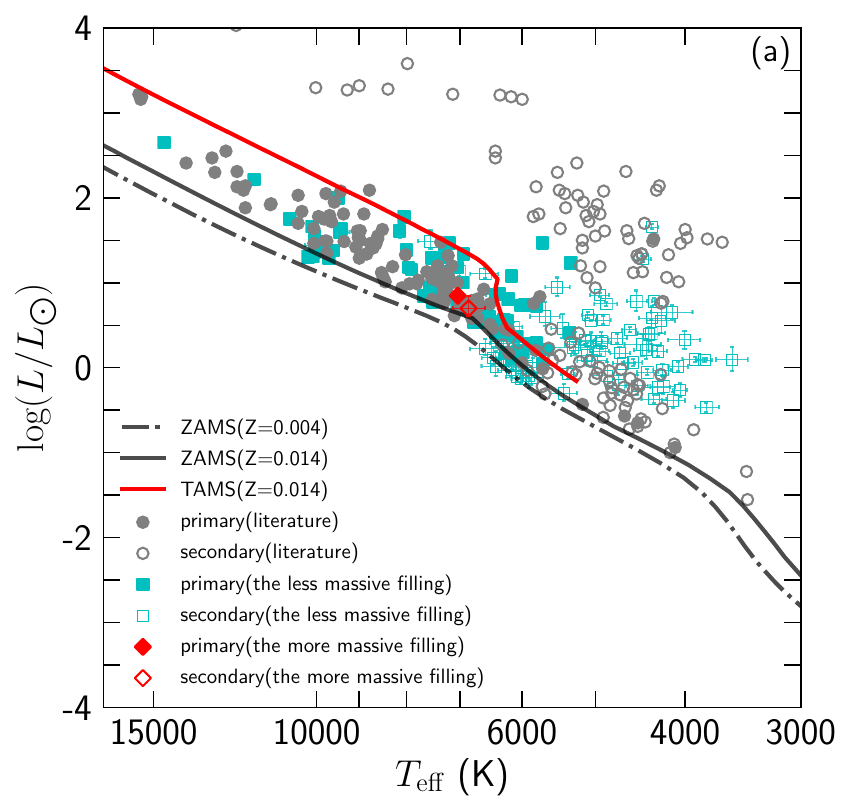}
  }
  \subfigure{
   \includegraphics[scale=0.6]{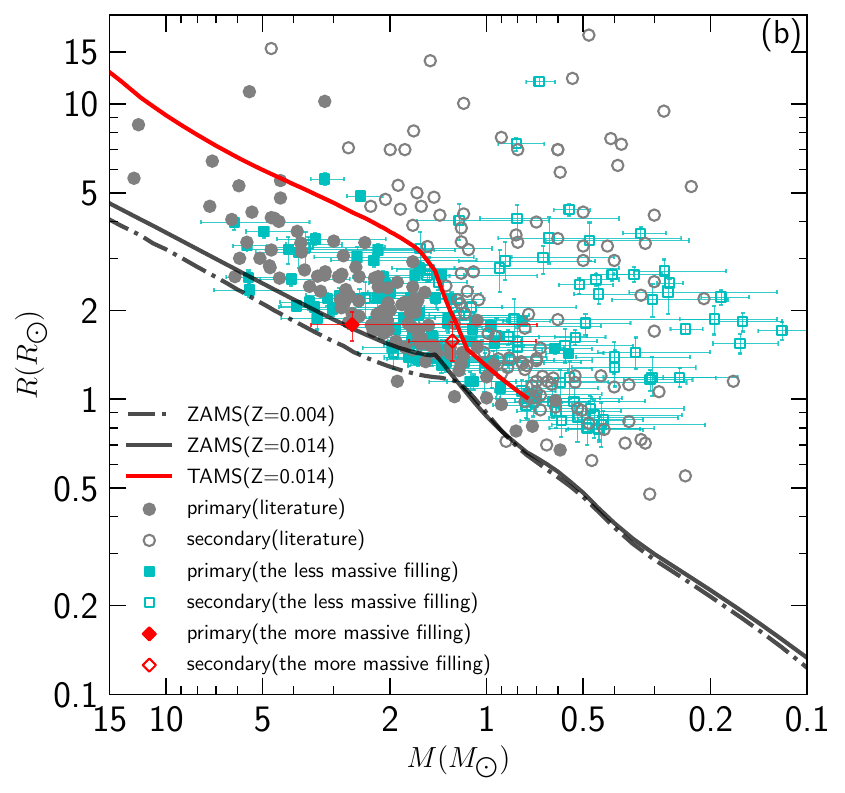}
  }
  \caption{The \teff-$L$ relation (panel (a)) and $M$-$R$ relation (panel (b)) of our semi-detached binary candidates and literature samples \citep{2006MNRAS.373..435I,2022RAA....22k5015M}. The black dash-dotted line and solid line represent the ZAMS (zero-age main sequence) lines with metallicities of $Z$=0.004 and $Z$=0.014 from the PARSEC model \citep{2012MNRAS.427..127B}, and the red line is the TAMS (terminal-age main sequence) line. The gray solid and hollow circles correspond to the primary and secondary components of the literature samples. The cyan solid and hollow rectangles represent the primary and secondary components of our semi-detached binary candidates with the less massive component filling its Roche lobe. The red solid and hollow diamonds represent our candidates with a more massive component filling its Roche lobe.}\label{fig:model_comparison}
\end{figure*}
\subsection{Distributions of SDs}
\subsubsection{\teff-$L$ and $M$-$R$}
Fig.\ref{fig:model_comparison} (a) compares our semi-detached candidates with the 50 Near-Contact Binaries \citep{2022RAA....22k5015M} and 61 Algol-type semi-detached binaries \citep{2006MNRAS.373..435I} on the \teff-$L$ relation. The black dash-dotted line and solid line represent the zero-age main sequence (ZAMS) lines with metallicities of $Z$=0.004 and $Z$=0.014 from the PARSEC model \citep{2012MNRAS.427..127B}, the red solid line is the terminal-age main sequence (TAMS) line from PARSEC model. The gray solid circles and hollow circles correspond to the primary and secondary components compiled from the literature. The cyan solid and hollow rectangles are the primary components and secondary components of our semi-detached binary candidates with a less massive component filling its Roche Lobe that identified from TESS survey. The red solid and hollow diamonds represent the our candidates with a more massive component filling its Roche lobe. As shown in Fig.\ref{fig:model_comparison} (a), the majority of our candidates are very close to Algol-type semi-detached systems, where the slow mass transfer is occurring \citep{2004A&A...419.1015M}. And the mass-accreting stars in these systems are typically the main sequence stars of spectral types B to F, while the donor stars are usually located in the region of the Hertzsprung-Russell diagram between the TAMS and the giants. Furthermore, we also include the samples with \teff\,$>$ 10000 K. Similarly, we locate our samples with masses and radii in Fig.\ref{fig:model_comparison} (b), it can be observed that the main sequence stars are in agreement with the $M$-$R$ relation, and the secondary components exhibit deviations from the main sequence and possess larger sizes compared to main-sequence stars with the same masses.

\begin{figure*}[t]
  \centering
    \subfigure{
   \includegraphics[scale=0.58]{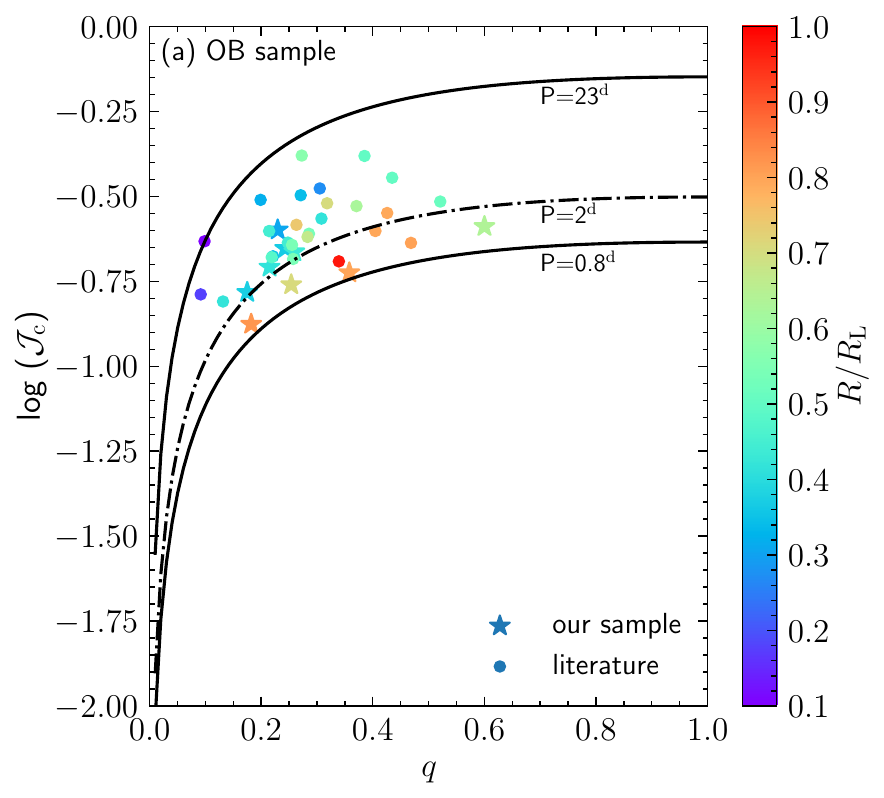}
  }
  \subfigure{
   \includegraphics[scale=0.58]{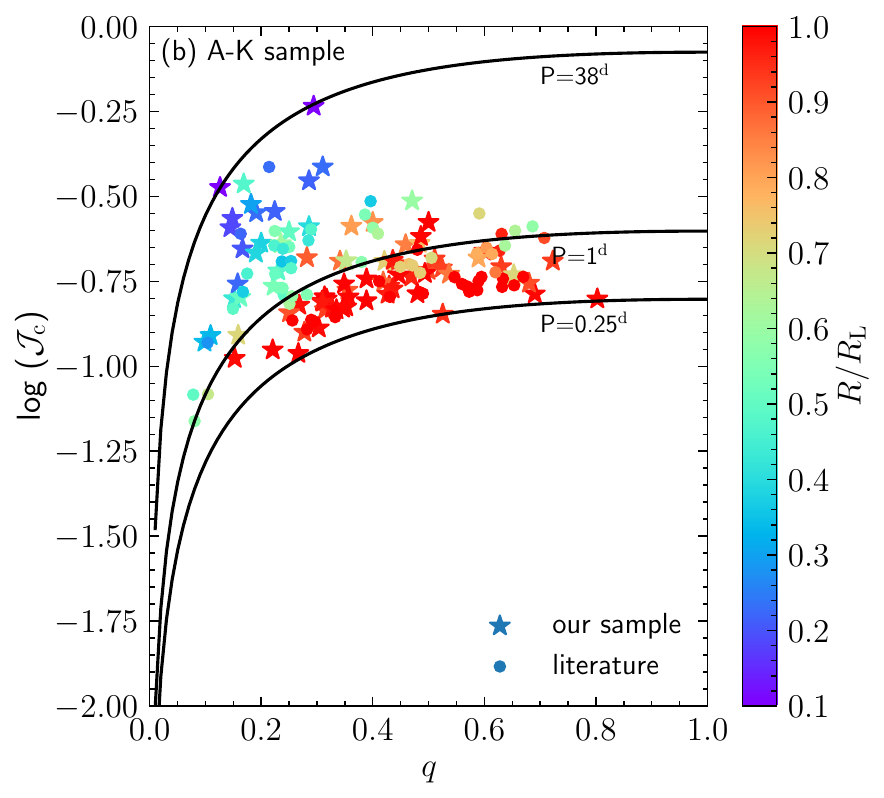}
  }
  \centering
  \caption{The relation between mass ratio ($q$=$M\mathrm{_{donor}}/M\mathrm{_{accretor}}$) with specific angular momentum ($\mathcal{J}_{c}$) for the semi-detached binaries that containing a giant (subgiant) donor. Panel (a) shows the relation of the semi-detached binaries with early-type accretors, and panel (b) is the relation of the semi-detached binaries with the late-type accretors. The color of the marker represents the filling fraction of the unfilled component ($R/R\rm_{L}$). The star markers are our semi-detached binary candidates identified from the TESS survey, and the dots represent the semi-detached binaries compiled from the literature \citep{2006MNRAS.373..435I,2022RAA....22k5015M}.}\label{fig:QJ}
\end{figure*}
\subsubsection{Specific angular momentum}
Fig.\ref{fig:QJ} shows the relation between mass ratio ($q$=$M\mathrm{_{donor}}/M\mathrm{_{accretor}}$) with specific angular momentum ($\mathcal{J}_{c}$) for SDs containing a giant (subgiant) donor, while $\mathcal{J}_{c}$ is defined as: $\mathcal{J}_{c}$=$q(1+q)^{-2}\cdot P^{1/3}$ \citep{1989AcASn..30..225Z}. In Fig.\ref{fig:QJ} (a), the relation of the SDs containing early-type accretors is shown, and panel (b) shows the relation of the SDs containing late-type accretors. In Fig.\ref{fig:QJ}, the star markers are our SD candidates identified from the TESS survey, and dots are the literature samples, and the color of the marker represents the filling fraction of accretors ($R/R\rm_{L}$). \citet{1989AcASn..30..225Z} presented a lower limit of the period for SDs with $P\mathrm{_{min}}$= 0.752 days for O-B stars and 0.248 days for A-F stars. Furthermore, they also pointed out that the presence of evolved SDs with extremely short periods is precluded due to the dynamic influence of the Roche critical surface. As the observations shown in Fig.\ref{fig:QJ}, the shortest periods of SDs in observation are 0.915 days for O-B stars and 0.2508 days for A-F stars. Compared to the theoretical simulations, the SDs containing an O-B accretor have not yet reached the theoretical lower limit of the period. Moreover, SDs with larger mass ratios and shorter periods tend to exhibit a larger filling fraction for the accretors. It also can be seen that our sample is consistent with the semi-detached binaries from previous studies, and it further validates the effectiveness of our pipeline.

\begin{figure*}[t]
  \centering
    \subfigure{
   \includegraphics[scale=0.6]{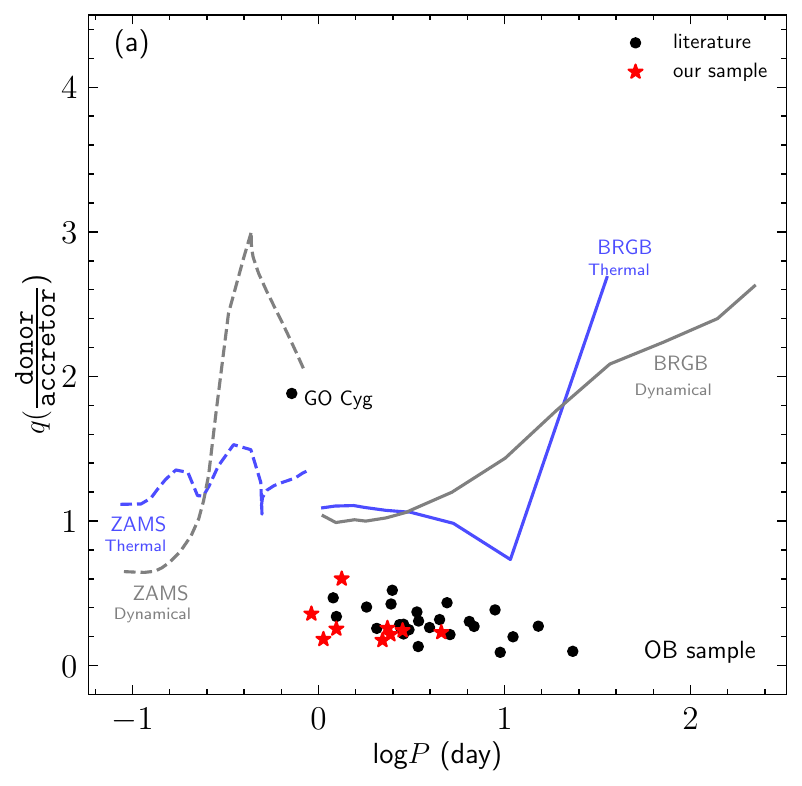}
  }
  \subfigure{
   \includegraphics[scale=0.6]{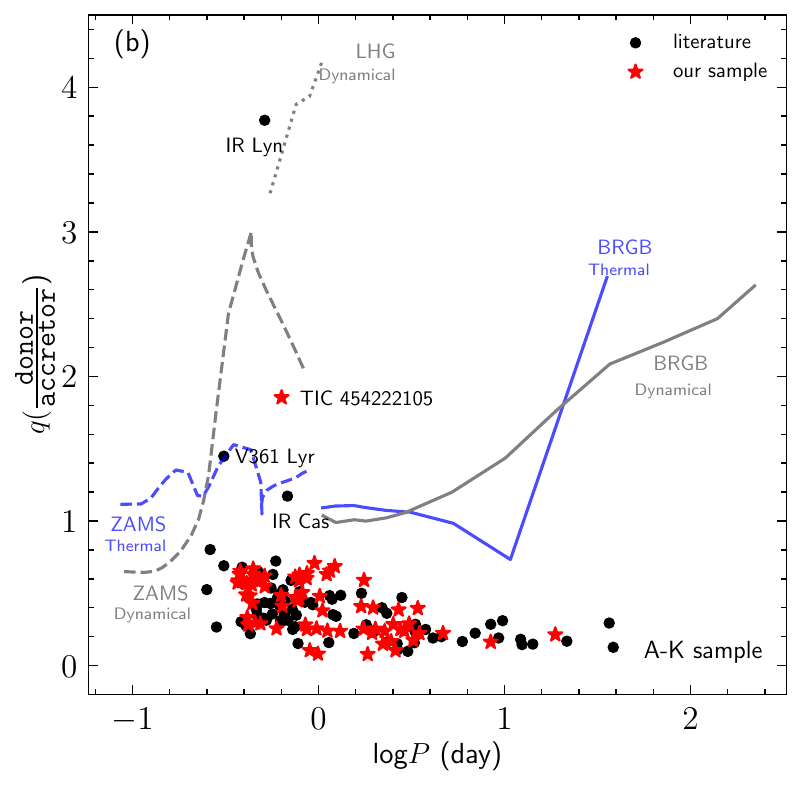}
  }
  \centering
  \caption{The distribution of mass ratio ($q$=$M\mathrm{_{donor}}/M\mathrm{_{accretor}}$) and orbital period for semi-detached binaries with early-type accretors (panel (a)) and late-type accretors (panel (b)). The colored lines represent the theoretical critical mass ratio limits for dynamical-timescale (gray lines) and thermal-timescale (blue lines) mass transfer at different evolutionary stages from \citet{2015ApJ...812...40G,2020ApJ...899..132G,2020ApJS..249....9G,2023ApJ...945....7G}, including the ZAMS in dashed lines, the LHG (late Hertzsprung–Russell gap) in dotted lines, and the BRGB (base of the red giant branch) in solid lines. The red star markers represent our semi-detached binary candidates, while the black dots correspond to the literature samples.}\label{fig:PQ}
\end{figure*}
\subsubsection{Mass transfer stage}
In Fig.\ref{fig:PQ}, we plot the relation between mass ratios ($q$=$M\mathrm{_{donor}}/M\mathrm{_{accretor}}$) and orbital periods of these SD samples. The red star markers and black dots represent our SD candidates and the literature samples, respectively. \citet{2015ApJ...812...40G,2020ApJ...899..132G,2020ApJS..249....9G,2023ApJ...945....7G} have proposed relationships between the critical mass ratio limits ($\widetilde{q}\mathrm{_{ad}}$) and orbital periods for stars of different masses. In Fig.\ref{fig:PQ}, we have selected the critical mass ratio limits for dynamical-timescale (gray lines) and thermal-timescale (blue lines) mass transfer at some significant evolutionary stages, including the ZAMS, the late Hertzsprung–Russell gap (LHG), and the base of the red giant branch (BRGB), represented by dashed, dotted, and solid lines, respectively. These limits form the foundation for discussing mass transfer phases in semi-detached binary systems. When a star with Roche lobe filling is positioned above the thermal-timescale limit but below the dynamical-timescale limit, it undergoes thermal-timescale mass transfer. Conversely, when a star is under the thermal-timescale limit, it experiences nuclear-timescale mass transfer. Furthermore, if a star is located above the dynamical-timescale limit, it will likely enter the phase of dynamical-timescale mass transfer. Depending on the structure of the donor, a binary system could enter the phase of the prompt (convective-dominated) or the delayed (radiative-dominated) dynamical-timescale transfer \citep{2015ApJ...812...40G,2020ApJ...899..132G}. And it is important to note that the criteria for mass transfer employed in this study are derived from the principles of conserved mass transfer. 

As Fig.\ref{fig:PQ} shows, we find the majority of these SD samples are located under the mass ratio limit of BRGB (solid blue lines) for thermal-timescale, which indicates that these samples are undergoing nuclear-timescale mass transfer. And we also know from Fig.\ref{fig:model_comparison} that these SD samples are Algol-type semi-detached systems, where the evolved, lower-mass star fills its Roche lobe. In addition, systems with the more massive component filling its Roche Lobe ($q>1$) are only observed in a few cases. These samples are undergoing rapid mass transfer, which is very short-lived and challenging to detect in observations. Among our candidates and the compiled literature samples, we only have $5$ such systems. All of these $5$ samples consist of two main-sequence stars, and they all have relatively short orbital periods ($P<1$ \texttt{days}). As shown in Fig.\ref{fig:PQ} (a) and (b), two of these samples (GO Cyg and TIC 454222105) are positioned below the dynamical-timescale limit and above the thermal-timescale limit. This indicates that these two samples are currently undergoing a thermal-timescale mass transfer. As the mass transfer progresses, if the initial mass ratios are large enough for such samples, the accretor may overfill its Roche lobe, then forming a contact binary. During the contact phase, if the more massive main-sequence star continues transferring material to the less massive companion, it can result in a reversal of the mass ratio. In other words, the current less massive component becomes larger and more evolved, while the current more massive main-sequence star is smaller and less evolved. In Fig.\ref{fig:PQ} (b), the two samples (V36 Lyr and IR Cas) with a main-sequence donor star, they are slightly under the limit of thermal-timescale in ZAMS, it suggests that they may undergo the nuclear timescale mass transfer. Moreover, an intriguing system (IR Lyn, \citet{2022RAA....22k5015M}) whose observation indicates it is located above the ZAMS line and below the TAMS line in the \teff-$L$ and $M$-$R$ diagrams. However, there is a discernible deviation from the established ZAMS criteria in Fig.\ref{fig:PQ} (b). And it is located above the limit of dynamical-timescale mass transfer for LHG. As depicted in Fig.\ref{fig:PQ} (b), under the assumption of conserved mass transfer, it shows that this system is nearing the threshold for entering the dynamical timescale of mass transfer. As the mass transfer continues, the radiative-dominated envelope of the primary may be transferred completely, and then a delayed dynamical-timescale mass transfer may occur in this system.

\section{Summary}\label{sec:Conclusion} 
In this paper, based on the MLP network with the MCMC method and DBSCAN clustering, we develop an efficient pipeline for identifying and deriving parameters of semi-detached binaries. In our pipeline, two light curve fitting models are established. They are the more massive component filling its Roche Lobe and the less massive component filling its Roche Lobe. We train the models by using the light curves generated by PHOEBE. The results demonstrate that the model with the more massive component filling its Roche Lobe provides residuals' standard deviations for inclination, relative radius, mass ratio, and temperature ratio of approximately 0.403$^\circ$, 0.003, 0.020, and 0.037, respectively. Similarly, the model with the less massive component filling its Roche Lobe yields standard deviations of residuals for these four parameters of approximately 0.328$^\circ$, 0.003, 0.018, and 0.0.012, respectively. Then, we apply this pipeline to the TESS survey to analyze semi-detached binaries and $77$ semi-detached binary candidates are identified. We also derive their Gaia-distance-dependent masses and radii with median relative uncertainties of $\sim$25\% (lower) and $\sim$36\% (upper), $\sim$6\% (lower) and $\sim$7\% (upper), respectively.

We also compare the distributions of our semi-detached binary candidates with the 111 compiled samples from previous studies. The \teff-$L$ and $M$-$R$ distributions demonstrate that our candidates show good agreement with the compiled samples and the PARSEC model. And the samples consist of configurations involving two main sequence stars, a main sequence star with a giant, or two giants. The majority of our candidates exhibit characteristics similar to Algol-type semi-detached systems. Based on a comparison with mass ratio limits and orbital periods, these Algol-type semi-detached systems are confirmed to undergo a nuclear-timescale mass transfer. Additionally, we highlight $5$ samples where the more massive component fills its Roche lobe. These samples exhibit a close proximity to the predicted mass ratio limits. The discovery of such samples holds great anticipation as they can serve as valuable constraints for models. Moreover, as an application, our pipeline allows for a significant reduction in the processing time for photometric analysis of semi-detached binaries. It proves highly applicable in the analysis of semi-detached binaries within the framework of big data. This pipeline has the potential for transfer learning across other photometric data, enabling its use in a broader range of semi-detached binary studies.

\section{acknowledgement}
We wish to thank the referee for his/her valuable comments and suggestions, which have helped us further improve this work. This work is supported by NSFC (grant No.12125303, 12288102, 12125303, 12173081, 12303106), the National Key R\&D Program of China (grant No.2021YFA1600401/ 2021YFA1600403), the key research program of frontier sciences, CAS, No.ZDBS-LY-7005, Yunnan Fundamental Research Projects (grant No.202101AV070001), the International Centre of Supernovae, Yunnan Key Laboratory (No.202302AN360001) and the Yunnan Revitalization Talent Support Program"—Science \& Technology Champion Project (No.202305AB350003). This work is also supported by the China Manned Space Project of No.CMS-CSST-2021-A10. We acknowledge the TESS mission provided by NASA's Science Mission Directorate. We also acknowledge the use of public TESS data from pipelines at the TESS Science Office and at the TESS Science Processing Operations Center. We thank the Gaia Data Processing and Analysis Consortium (DPAC) for their substantial contributions in producing and releasing high-quality data.

\bibliography{sample631}{}
\bibliographystyle{aasjournal}



\begin{appendices}

\section{Appendix A}\label{AppendixA}
\begin{table*}[htbp]
\tiny 
\caption{Absolute parameters of 77 semi-detached binaries from TESS survey (calculated based on $T\rm_{1}$ from Gaia golden sample \citep{2023A&A...674A..39G} and distance from \citet{2021AJ....161..147B}).}
\centering
\begin{tabular}{cccccccccccc}
\hline
\hline
  \multicolumn{1}{c}{Gaia DR3 Name} &
  \multicolumn{1}{c}{Period (d)} &
  \multicolumn{1}{c}{$T\mathrm{_{1}}$ (K)} &
  \multicolumn{1}{c}{$T\mathrm{_{2}}$ (K)} &
  \multicolumn{1}{c}{log$(L\mathrm{_{1}}$/$L_{\odot}$)} &
  \multicolumn{1}{c}{log$(L\mathrm{_{2}}$/$L_{\odot}$)} &
  \multicolumn{1}{c}{$R\mathrm{_{1}}$ ($R_{\odot}$)} &
  \multicolumn{1}{c}{$R\mathrm{_{2}}$ ($R_{\odot}$)} &
  \multicolumn{1}{c}{$M\mathrm{_{1}}$ ($M_{\odot}$)} &
  \multicolumn{1}{c}{$M\mathrm{_{2}}$ ($M_{\odot}$)}\\
\hline
\hline
4876695194334420096 & 3.067745211 & 7269 $^{+ 12 }_{- 14 }$& 4248 $^{+ 154 }_{- 149 }$& 1.343 $^{+ 0.029  }_{- 0.031  }$& 0.555 $^{+ 0.148  }_{- 0.175  }$& 2.955  $^{+ 0.140  }_{- 0.137  }$& 3.513  $^{+ 0.632  }_{- 0.639  }$& 2.248  $^{+ 1.172  }_{- 0.810  }$& 0.640  $^{+ 0.393  }_{- 0.279}$\\
2114453601444545024 & 0.993305002 & 6404 $^{+ 3 }_{- 2 }$& 6319 $^{+ 159 }_{- 145 }$& 0.860 $^{+ 0.017  }_{- 0.016  }$& 0.307 $^{+ 0.046  }_{- 0.056  }$& 2.188  $^{+ 0.113  }_{- 0.094  }$& 1.183  $^{+ 0.086  }_{- 0.092  }$& 3.143  $^{+ 0.546  }_{- 0.471  }$& 0.250  $^{+ 0.057  }_{- 0.053}$\\
2121872762311382912 & 0.86602103 & 7665 $^{+ 32 }_{- 26 }$& 4924 $^{+ 108 }_{- 113 }$& 0.845 $^{+ 0.020  }_{- 0.022  }$& 0.293 $^{+ 0.074  }_{- 0.077  }$& 1.494  $^{+ 0.077  }_{- 0.072  }$& 1.924  $^{+ 0.189  }_{- 0.177  }$& 1.975  $^{+ 0.638  }_{- 0.463  }$& 1.220  $^{+ 0.374  }_{- 0.289}$\\
699818436454276864 & 1.115976642 & 6668 $^{+ 30 }_{- 33 }$& 4224 $^{+ 137 }_{- 125 }$& 0.534 $^{+ 0.018  }_{- 0.019  }$& -0.221 $^{+ 0.091  }_{- 0.107  }$& 1.380  $^{+ 0.082  }_{- 0.072  }$& 1.442  $^{+ 0.181  }_{- 0.176  }$& 1.482  $^{+ 0.619  }_{- 0.428  }$& 0.343  $^{+ 0.142  }_{- 0.108}$\\
2436283197513649664 & 0.481424741 & 6771 $^{+ 11 }_{- 9 }$& 6575 $^{+ 252 }_{- 225 }$& 0.539 $^{+ 0.044  }_{- 0.043  }$& 0.223 $^{+ 0.073  }_{- 0.110  }$& 1.352  $^{+ 0.119  }_{- 0.106  }$& 0.983  $^{+ 0.125  }_{- 0.119  }$& 0.953  $^{+ 0.485  }_{- 0.273  }$& 0.534  $^{+ 0.213  }_{- 0.160}$\\
1333162251022354048 & 2.266959543 & 7553 $^{+ 30 }_{- 34 }$& 3812 $^{+ 76 }_{- 72 }$& 0.944 $^{+ 0.010  }_{- 0.009  }$& 0.091 $^{+ 0.025  }_{- 0.024  }$& 1.728  $^{+ 0.067  }_{- 0.062  }$& 2.542  $^{+ 0.123  }_{- 0.105  }$& 1.928  $^{+ 0.281  }_{- 0.227  }$& 0.456  $^{+ 0.070  }_{- 0.054}$\\
2526767987958943872 & 3.488532187 & 6942 $^{+ 44 }_{- 11 }$& 4347 $^{+ 139 }_{- 147 }$& 1.336 $^{+ 0.022  }_{- 0.026  }$& 0.586 $^{+ 0.116  }_{- 0.146  }$& 3.205  $^{+ 0.159  }_{- 0.160  }$& 3.448  $^{+ 0.516  }_{- 0.519  }$& 2.184  $^{+ 0.927  }_{- 0.564  }$& 0.478  $^{+ 0.244  }_{- 0.177}$\\
4601359272077948416 & 0.818090719 & 6691 $^{+ 19 }_{- 21 }$& 4668 $^{+ 120 }_{- 140 }$& 0.651 $^{+ 0.006  }_{- 0.006  }$& 0.054 $^{+ 0.018  }_{- 0.018  }$& 1.570  $^{+ 0.097  }_{- 0.077  }$& 1.629  $^{+ 0.101  }_{- 0.090  }$& 1.694  $^{+ 0.342  }_{- 0.252  }$& 0.857  $^{+ 0.171  }_{- 0.132}$\\
4597990196652263296 & 0.407603267 & 6679 $^{+ 12 }_{- 9 }$& 6451 $^{+ 184 }_{- 203 }$& 0.533 $^{+ 0.043  }_{- 0.041  }$& 0.118 $^{+ 0.089  }_{- 0.129  }$& 1.382  $^{+ 0.099  }_{- 0.092  }$& 0.911  $^{+ 0.114  }_{- 0.125  }$& 1.329  $^{+ 0.679  }_{- 0.394  }$& 0.607  $^{+ 0.226  }_{- 0.198}$\\
2882810400156716160 & 2.221301761 & 7055 $^{+ 34 }_{- 25 }$& 4410 $^{+ 134 }_{- 123 }$& 1.177 $^{+ 0.016  }_{- 0.018  }$& 0.209 $^{+ 0.101  }_{- 0.122  }$& 2.594  $^{+ 0.115  }_{- 0.114  }$& 2.173  $^{+ 0.271  }_{- 0.284  }$& 2.102  $^{+ 0.693  }_{- 0.516  }$& 0.304  $^{+ 0.127  }_{- 0.103}$\\
5180250215659097856 & 0.411306659 & 5905 $^{+ 8 }_{- 9 }$& 5844 $^{+ 134 }_{- 134 }$& 0.285 $^{+ 0.051  }_{- 0.052  }$& 0.002 $^{+ 0.081  }_{- 0.117  }$& 1.329  $^{+ 0.089  }_{- 0.098  }$& 0.977  $^{+ 0.103  }_{- 0.123  }$& 1.337  $^{+ 0.674  }_{- 0.389  }$& 0.715  $^{+ 0.235  }_{- 0.212}$\\
5179913528878335872 & 0.44528793 & 5637 $^{+ 18 }_{- 18 }$& 5357 $^{+ 90 }_{- 91 }$& 0.215 $^{+ 0.036  }_{- 0.032  }$& -0.072 $^{+ 0.056  }_{- 0.079  }$& 1.343  $^{+ 0.072  }_{- 0.065  }$& 1.066  $^{+ 0.075  }_{- 0.097  }$& 1.207  $^{+ 0.491  }_{- 0.271  }$& 0.778  $^{+ 0.168  }_{- 0.163}$\\
2905533010535981696 & 0.512831647 & 6079 $^{+ 12 }_{- 8 }$& 4122 $^{+ 113 }_{- 122 }$& 0.366 $^{+ 0.016  }_{- 0.019  }$& -0.386 $^{+ 0.090  }_{- 0.098  }$& 1.367  $^{+ 0.074  }_{- 0.067  }$& 1.254  $^{+ 0.162  }_{- 0.145  }$& 1.639  $^{+ 0.686  }_{- 0.407  }$& 0.964  $^{+ 0.406  }_{- 0.265}$\\
4653416302929032960 & 3.232081512 & 7316 $^{+ 17 }_{- 18 }$& 4379 $^{+ 131 }_{- 109 }$& 1.299 $^{+ 0.019  }_{- 0.017  }$& 0.395 $^{+ 0.079  }_{- 0.081  }$& 2.781  $^{+ 0.139  }_{- 0.138  }$& 2.721  $^{+ 0.298  }_{- 0.252  }$& 1.619  $^{+ 0.521  }_{- 0.358  }$& 0.279  $^{+ 0.100  }_{- 0.069}$\\
2032636398639228160 & 1.694822868 & 8134 $^{+ 24 }_{- 25 }$& 7535 $^{+ 233 }_{- 220 }$& 1.608 $^{+ 0.078  }_{- 0.088  }$& 1.488 $^{+ 0.092  }_{- 0.128  }$& 3.207  $^{+ 0.333  }_{- 0.332  }$& 3.240  $^{+ 0.412  }_{- 0.445  }$& 4.161  $^{+ 2.704  }_{- 1.445  }$& 1.613  $^{+ 0.652  }_{- 0.546}$\\
2990232754910113152 & 0.594445177 & 6213 $^{+ 42 }_{- 48 }$& 5938 $^{+ 186 }_{- 195 }$& 0.812 $^{+ 0.020  }_{- 0.043  }$& 0.153 $^{+ 0.150  }_{- 0.068  }$& 2.199  $^{+ 0.160  }_{- 0.154  }$& 1.139  $^{+ 0.212  }_{- 0.120  }$& 2.196  $^{+ 0.867  }_{- 0.605  }$& 0.591  $^{+ 0.353  }_{- 0.161}$\\
4889204372482883584 & 0.638486649 & 7325 $^{+ 22 }_{- 19 }$& 4722 $^{+ 155 }_{- 160 }$& 0.929 $^{+ 0.011  }_{- 0.012  }$& -0.202 $^{+ 0.124  }_{- 0.121  }$& 1.811  $^{+ 0.092  }_{- 0.095  }$& 1.185  $^{+ 0.183  }_{- 0.161  }$& 1.325  $^{+ 0.540  }_{- 0.376  }$& 0.551  $^{+ 0.270  }_{- 0.189}$\\
2457700412712178944 & 1.93975019 & 7278 $^{+ 24 }_{- 23 }$& 4281 $^{+ 100 }_{- 96 }$& 1.126 $^{+ 0.008  }_{- 0.009  }$& 0.193 $^{+ 0.061  }_{- 0.058  }$& 2.297  $^{+ 0.094  }_{- 0.089  }$& 2.273  $^{+ 0.168  }_{- 0.160  }$& 1.996  $^{+ 0.423  }_{- 0.355  }$& 0.447  $^{+ 0.104  }_{- 0.086}$\\
2172287874408422144 & 1.10574422 & 8045 $^{+ 54 }_{- 45 }$& 5500 $^{+ 162 }_{- 170 }$& 1.775 $^{+ 0.018  }_{- 0.019  }$& 0.947 $^{+ 0.102  }_{- 0.115  }$& 3.969  $^{+ 0.215  }_{- 0.214  }$& 3.266  $^{+ 0.427  }_{- 0.430  }$& 6.133  $^{+ 2.561  }_{- 1.644  }$& 3.675  $^{+ 1.602  }_{- 1.197}$\\
1351714684378550016 & 1.755696172 & 7508 $^{+ 43 }_{- 27 }$& 4353 $^{+ 148 }_{- 145 }$& 1.294 $^{+ 0.017  }_{- 0.021  }$& 0.405 $^{+ 0.126  }_{- 0.160  }$& 2.618  $^{+ 0.125  }_{- 0.121  }$& 2.781  $^{+ 0.481  }_{- 0.431  }$& 1.674  $^{+ 0.979  }_{- 0.565  }$& 0.911  $^{+ 0.524  }_{- 0.349}$\\
1035063002096359680 & 2.022459806 & 7955 $^{+ 28 }_{- 21 }$& 3560 $^{+ 146 }_{- 135 }$& 1.180 $^{+ 0.018  }_{- 0.017  }$& 0.096 $^{+ 0.130  }_{- 0.150  }$& 2.051  $^{+ 0.088  }_{- 0.077  }$& 2.940  $^{+ 0.413  }_{- 0.465  }$& 3.448  $^{+ 1.725  }_{- 0.993  }$& 0.877  $^{+ 0.418  }_{- 0.346}$\\
740582795692769408 & 0.376124124 & 6248 $^{+ 4 }_{- 5 }$& 6022 $^{+ 138 }_{- 135 }$& 0.171 $^{+ 0.038  }_{- 0.031  }$& -0.129 $^{+ 0.058  }_{- 0.096  }$& 1.042  $^{+ 0.073  }_{- 0.060  }$& 0.789  $^{+ 0.071  }_{- 0.088  }$& 0.743  $^{+ 0.305  }_{- 0.188  }$& 0.446  $^{+ 0.121  }_{- 0.119}$\\
5153975843420955136 & 1.01568807 & 6158 $^{+ 39 }_{- 33 }$& 5670 $^{+ 217 }_{- 204 }$& 1.077 $^{+ 0.045  }_{- 0.048  }$& 0.603 $^{+ 0.110  }_{- 0.167  }$& 3.029  $^{+ 0.280  }_{- 0.249  }$& 2.050  $^{+ 0.331  }_{- 0.355  }$& 2.538  $^{+ 1.439  }_{- 0.910  }$& 1.113  $^{+ 0.588  }_{- 0.452}$\\
4997907692641176192 & 0.413614378 & 6660 $^{+ 6 }_{- 5 }$& 6406 $^{+ 230 }_{- 211 }$& 0.582 $^{+ 0.037  }_{- 0.038  }$& 0.016 $^{+ 0.116  }_{- 0.178  }$& 1.469  $^{+ 0.113  }_{- 0.112  }$& 0.820  $^{+ 0.130  }_{- 0.149  }$& 1.380  $^{+ 0.710  }_{- 0.435  }$& 0.439  $^{+ 0.231  }_{- 0.188}$\\
4965590537640948992 & 2.434877225 & 10216 $^{+ 34 }_{- 50 }$& 4953 $^{+ 112 }_{- 117 }$& 1.303 $^{+ 0.010  }_{- 0.010  }$& 0.199 $^{+ 0.061  }_{- 0.068  }$& 1.431  $^{+ 0.044  }_{- 0.038  }$& 1.707  $^{+ 0.123  }_{- 0.116  }$& 0.556  $^{+ 0.124  }_{- 0.096  }$& 0.120  $^{+ 0.028  }_{- 0.022}$\\
2051947769857973120 & 0.513482922 & 6549 $^{+ 31 }_{- 32 }$& 4859 $^{+ 205 }_{- 201 }$& 0.711 $^{+ 0.022  }_{- 0.026  }$& -0.092 $^{+ 0.133  }_{- 0.161  }$& 1.747  $^{+ 0.115  }_{- 0.101  }$& 1.269  $^{+ 0.209  }_{- 0.213  }$& 2.052  $^{+ 1.021  }_{- 0.655  }$& 1.016  $^{+ 0.546  }_{- 0.398}$\\
4654488391189033344 & 0.418404117 & 6222 $^{+ 48 }_{- 55 }$& 5902 $^{+ 125 }_{- 127 }$& 0.554 $^{+ 0.019  }_{- 0.037  }$& -0.092 $^{+ 0.142  }_{- 0.060  }$& 1.617  $^{+ 0.081  }_{- 0.082  }$& 0.866  $^{+ 0.148  }_{- 0.074  }$& 1.698  $^{+ 0.622  }_{- 0.469  }$& 0.521  $^{+ 0.268  }_{- 0.118}$\\
5266269923742586496 & 0.437169155 & 6242 $^{+ 51 }_{- 56 }$& 6133 $^{+ 155 }_{- 151 }$& 0.426 $^{+ 0.048  }_{- 0.045  }$& 0.114 $^{+ 0.081  }_{- 0.118  }$& 1.399  $^{+ 0.107  }_{- 0.085  }$& 1.010  $^{+ 0.110  }_{- 0.140  }$& 1.321  $^{+ 0.666  }_{- 0.367  }$& 0.701  $^{+ 0.231  }_{- 0.233}$\\
4548008150397375872 & 0.387748803 & 6406 $^{+ 34 }_{- 36 }$& 6252 $^{+ 131 }_{- 123 }$& 0.410 $^{+ 0.041  }_{- 0.040  }$& 0.101 $^{+ 0.070  }_{- 0.100  }$& 1.301  $^{+ 0.078  }_{- 0.072  }$& 0.955  $^{+ 0.090  }_{- 0.107  }$& 1.383  $^{+ 0.524  }_{- 0.375  }$& 0.753  $^{+ 0.212  }_{- 0.201}$\\
5477873272970767360 & 0.896878797 & 6350 $^{+ 4 }_{- 3 }$& 6281 $^{+ 220 }_{- 242 }$& 0.874 $^{+ 0.033  }_{- 0.035  }$& 0.287 $^{+ 0.112  }_{- 0.147  }$& 2.247  $^{+ 0.153  }_{- 0.129  }$& 1.186  $^{+ 0.160  }_{- 0.199  }$& 2.885  $^{+ 1.145  }_{- 0.774  }$& 0.305  $^{+ 0.140  }_{- 0.127}$\\
5477159247544461312 & 2.59364758 & 7500 $^{+ 0 }_{- 0 }$& 4591 $^{+ 87 }_{- 76 }$& 1.082 $^{+ 0.009  }_{- 0.008  }$& 0.448 $^{+ 0.029  }_{- 0.030  }$& 2.068  $^{+ 0.070  }_{- 0.069  }$& 2.643  $^{+ 0.122  }_{- 0.111  }$& 3.921  $^{+ 0.524  }_{- 0.462  }$& 0.406  $^{+ 0.058  }_{- 0.049}$\\
5566666915748470912 & 1.049570677 & 7496 $^{+ 3 }_{- 7 }$& 4394 $^{+ 115 }_{- 120 }$& 0.768 $^{+ 0.007  }_{- 0.008  }$& -0.055 $^{+ 0.036  }_{- 0.037  }$& 1.434  $^{+ 0.077  }_{- 0.066  }$& 1.620  $^{+ 0.099  }_{- 0.092  }$& 1.392  $^{+ 0.271  }_{- 0.224  }$& 0.529  $^{+ 0.103  }_{- 0.085}$\\
2988525797467488256 & 0.91543371 & 10051 $^{+ 24 }_{- 22 }$& 5107 $^{+ 203 }_{- 182 }$& 1.551 $^{+ 0.008  }_{- 0.009  }$& 0.148 $^{+ 0.115  }_{- 0.142  }$& 1.963  $^{+ 0.066  }_{- 0.061  }$& 1.517  $^{+ 0.186  }_{- 0.214  }$& 1.649  $^{+ 0.550  }_{- 0.394  }$& 0.570  $^{+ 0.220  }_{- 0.195}$\\
938228841240985728 & 1.062648473 & 11681 $^{+ 38 }_{- 33 }$& 9781 $^{+ 518 }_{- 492 }$& 2.218 $^{+ 0.025  }_{- 0.037  }$& 1.443 $^{+ 0.176  }_{- 0.101  }$& 3.144  $^{+ 0.367  }_{- 0.323  }$& 1.868  $^{+ 0.419  }_{- 0.318  }$& 3.862  $^{+ 2.369  }_{- 1.399  }$& 0.821  $^{+ 0.633  }_{- 0.34}$\\
331337421011384448 & 0.753660737 & 6504 $^{+ 31 }_{- 29 }$& 4050 $^{+ 76 }_{- 67 }$& 0.596 $^{+ 0.013  }_{- 0.014  }$& -0.261 $^{+ 0.080  }_{- 0.084  }$& 1.563  $^{+ 0.023  }_{- 0.025  }$& 1.503  $^{+ 0.140  }_{- 0.127  }$& 1.711  $^{+ 0.483  }_{- 0.331  }$& 0.801  $^{+ 0.228  }_{- 0.18}$\\
2082658577035787392 & 4.573035676 & 10113 $^{+ 36 }_{- 36 }$& 4947 $^{+ 119 }_{- 116 }$& 1.661 $^{+ 0.007  }_{- 0.006  }$& 0.857 $^{+ 0.021  }_{- 0.022  }$& 2.202  $^{+ 0.107  }_{- 0.093  }$& 3.646  $^{+ 0.192  }_{- 0.175  }$& 1.443  $^{+ 0.243  }_{- 0.204  }$& 0.331  $^{+ 0.055  }_{- 0.045}$\\
2407526566105203072 & 0.862731672 & 7415 $^{+ 25 }_{- 25 }$& 5132 $^{+ 115 }_{- 126 }$& 0.986 $^{+ 0.013  }_{- 0.012  }$& 0.326 $^{+ 0.043  }_{- 0.044  }$& 1.884  $^{+ 0.092  }_{- 0.081  }$& 1.846  $^{+ 0.123  }_{- 0.116  }$& 1.815  $^{+ 0.368  }_{- 0.327  }$& 1.093  $^{+ 0.230  }_{- 0.188}$\\
2083160912108531456 & 0.628626478 & 5790 $^{+ 30 }_{- 22 }$& 3796 $^{+ 104 }_{- 119 }$& 0.293 $^{+ 0.009  }_{- 0.010  }$& -0.465 $^{+ 0.056  }_{- 0.053  }$& 1.384  $^{+ 0.088  }_{- 0.065  }$& 1.359  $^{+ 0.120  }_{- 0.109  }$& 1.751  $^{+ 0.470  }_{- 0.335  }$& 0.850  $^{+ 0.234  }_{- 0.182}$\\
1350220830329440256 & 0.787182881 & 6019 $^{+ 22 }_{- 20 }$& 4928 $^{+ 98 }_{- 88 }$& 0.730 $^{+ 0.033  }_{- 0.034  }$& 0.316 $^{+ 0.075  }_{- 0.100  }$& 2.131  $^{+ 0.118  }_{- 0.112  }$& 1.963  $^{+ 0.197  }_{- 0.212  }$& 3.574  $^{+ 1.284  }_{- 0.843  }$& 1.624  $^{+ 0.524  }_{- 0.435}$\\
1350891910379201280 & 0.486617826 & 6942 $^{+ 34 }_{- 18 }$& 6170 $^{+ 177 }_{- 176 }$& 0.749 $^{+ 0.019  }_{- 0.032  }$& 0.044 $^{+ 0.144  }_{- 0.069  }$& 1.641  $^{+ 0.092  }_{- 0.100  }$& 0.930  $^{+ 0.149  }_{- 0.097  }$& 1.482  $^{+ 0.569  }_{- 0.396  }$& 0.475  $^{+ 0.233  }_{- 0.133}$\\
1346411194338424704 & 1.305747766 & 7506 $^{+ 15 }_{- 20 }$& 4232 $^{+ 96 }_{- 90 }$& 1.007 $^{+ 0.008  }_{- 0.009  }$& -0.025 $^{+ 0.067  }_{- 0.064  }$& 1.882  $^{+ 0.070  }_{- 0.070  }$& 1.809  $^{+ 0.154  }_{- 0.136  }$& 2.090  $^{+ 0.463  }_{- 0.371  }$& 0.492  $^{+ 0.136  }_{- 0.100}$\\
2001818565159194496 & 2.206521406 & 10057 $^{+ 30 }_{- 172 }$& 5055 $^{+ 123 }_{- 136 }$& 1.355 $^{+ 0.010  }_{- 0.011  }$& 0.314 $^{+ 0.096  }_{- 0.101  }$& 1.568  $^{+ 0.049  }_{- 0.048  }$& 1.869  $^{+ 0.213  }_{- 0.196  }$& 1.126  $^{+ 0.343  }_{- 0.267  }$& 0.195  $^{+ 0.071  }_{- 0.055}$\\
1560572488648076800 & 4.669284406 & 5314 $^{+ 12 }_{- 11 }$& 4321 $^{+ 24 }_{- 21 }$& 1.232 $^{+ 0.018  }_{- 0.019  }$& 0.781 $^{+ 0.040  }_{- 0.039  }$& 4.869  $^{+ 0.100  }_{- 0.108  }$& 4.383  $^{+ 0.202  }_{- 0.188  }$& 2.472  $^{+ 0.358  }_{- 0.257  }$& 0.553  $^{+ 0.078  }_{- 0.065}$\\
4799811706322746752 & 1.83811605 & 9410 $^{+ 34 }_{- 34 }$& 6133 $^{+ 140 }_{- 137 }$& 1.631 $^{+ 0.011  }_{- 0.012  }$& 0.482 $^{+ 0.052  }_{- 0.056  }$& 2.470  $^{+ 0.112  }_{- 0.097  }$& 1.544  $^{+ 0.116  }_{- 0.120  }$& 2.065  $^{+ 0.342  }_{- 0.269  }$& 0.162  $^{+ 0.039  }_{- 0.035}$\\
4775507585905757440 & 0.785242682 & 7279 $^{+ 24 }_{- 23 }$& 5261 $^{+ 133 }_{- 131 }$& 0.861 $^{+ 0.033  }_{- 0.035  }$& 0.271 $^{+ 0.114  }_{- 0.145  }$& 1.697  $^{+ 0.097  }_{- 0.094  }$& 1.639  $^{+ 0.241  }_{- 0.242  }$& 1.707  $^{+ 0.977  }_{- 0.547  }$& 0.932  $^{+ 0.446  }_{- 0.332}$\\
5553742981197782400 & 2.894658261 & 9958 $^{+ 18 }_{- 17 }$& 4737 $^{+ 114 }_{- 115 }$& 1.478 $^{+ 0.012  }_{- 0.015  }$& 0.348 $^{+ 0.054  }_{- 0.052  }$& 1.842  $^{+ 0.055  }_{- 0.058  }$& 2.219  $^{+ 0.129  }_{- 0.118  }$& 0.775  $^{+ 0.142  }_{- 0.114  }$& 0.186  $^{+ 0.034  }_{- 0.028}$\\
1454617122923582976 & 0.366990139 & 5989 $^{+ 6 }_{- 5 }$& 5865 $^{+ 171 }_{- 174 }$& 0.182 $^{+ 0.051  }_{- 0.051  }$& -0.118 $^{+ 0.088  }_{- 0.128  }$& 1.146  $^{+ 0.095  }_{- 0.089  }$& 0.846  $^{+ 0.099  }_{- 0.125  }$& 1.126  $^{+ 0.709  }_{- 0.384  }$& 0.585  $^{+ 0.214  }_{- 0.200}$\\
4567617420587434624 & 0.370828898 & 5698 $^{+ 34 }_{- 24 }$& 5411 $^{+ 179 }_{- 180 }$& 0.052 $^{+ 0.041  }_{- 0.047  }$& -0.302 $^{+ 0.089  }_{- 0.110  }$& 1.091  $^{+ 0.094  }_{- 0.087  }$& 0.800  $^{+ 0.102  }_{- 0.105  }$& 0.907  $^{+ 0.457  }_{- 0.270  }$& 0.486  $^{+ 0.186  }_{- 0.151}$\\
2157168185074459648 & 2.83127773 & 10009 $^{+ 25 }_{- 29 }$& 5058 $^{+ 78 }_{- 72 }$& 1.430 $^{+ 0.007  }_{- 0.007  }$& 0.558 $^{+ 0.010  }_{- 0.009  }$& 1.726  $^{+ 0.052  }_{- 0.053  }$& 2.474  $^{+ 0.074  }_{- 0.078  }$& 1.108  $^{+ 0.101  }_{- 0.102  }$& 0.269  $^{+ 0.025  }_{- 0.025}$\\
4708499327219174912 & 0.865839827 & 6881 $^{+ 11 }_{- 10 }$& 4510 $^{+ 120 }_{- 120 }$& 0.685 $^{+ 0.015  }_{- 0.017  }$& -0.212 $^{+ 0.112  }_{- 0.129  }$& 1.547  $^{+ 0.068  }_{- 0.065  }$& 1.280  $^{+ 0.176  }_{- 0.184  }$& 1.638  $^{+ 0.570  }_{- 0.405  }$& 0.399  $^{+ 0.176  }_{- 0.144}$\\
3224767071968129152 & 1.746072739 & 9703 $^{+ 100 }_{- 136 }$& 4552 $^{+ 106 }_{- 103 }$& 1.283 $^{+ 0.012  }_{- 0.013  }$& 0.062 $^{+ 0.043  }_{- 0.039  }$& 1.550  $^{+ 0.047  }_{- 0.047  }$& 1.728  $^{+ 0.064  }_{- 0.067  }$& 0.945  $^{+ 0.106  }_{- 0.097  }$& 0.240  $^{+ 0.028  }_{- 0.027}$\\
5078057314303209984 & 0.456619343 & 6278 $^{+ 28 }_{- 21 }$& 6015 $^{+ 228 }_{- 253 }$& 0.569 $^{+ 0.025  }_{- 0.034  }$& -0.042 $^{+ 0.126  }_{- 0.080  }$& 1.618  $^{+ 0.146  }_{- 0.125  }$& 0.883  $^{+ 0.142  }_{- 0.110  }$& 1.471  $^{+ 0.654  }_{- 0.450  }$& 0.462  $^{+ 0.216  }_{- 0.149}$\\
3231583219428074752 & 0.950932477 & 5958 $^{+ 23 }_{- 28 }$& 4073 $^{+ 96 }_{- 100 }$& 0.739 $^{+ 0.011  }_{- 0.014  }$& 0.015 $^{+ 0.058  }_{- 0.056  }$& 2.198  $^{+ 0.109  }_{- 0.094  }$& 2.046  $^{+ 0.168  }_{- 0.152  }$& 1.728  $^{+ 0.415  }_{- 0.346  }$& 1.200  $^{+ 0.303  }_{- 0.236}$\\
5138409988587248896 & 0.790454125 & 7423 $^{+ 45 }_{- 46 }$& 5285 $^{+ 264 }_{- 274 }$& 0.991 $^{+ 0.014  }_{- 0.013  }$& 0.240 $^{+ 0.055  }_{- 0.059  }$& 1.889  $^{+ 0.211  }_{- 0.174  }$& 1.571  $^{+ 0.208  }_{- 0.168  }$& 1.320  $^{+ 0.553  }_{- 0.364  }$& 0.803  $^{+ 0.351  }_{- 0.224}$\\
2357884032723549184 & 0.445412171 & 6264 $^{+ 9 }_{- 9 }$& 6194 $^{+ 205 }_{- 222 }$& 0.398 $^{+ 0.052  }_{- 0.053  }$& 0.110 $^{+ 0.086  }_{- 0.117  }$& 1.338  $^{+ 0.123  }_{- 0.108  }$& 0.984  $^{+ 0.121  }_{- 0.128  }$& 1.176  $^{+ 0.668  }_{- 0.395  }$& 0.622  $^{+ 0.236  }_{- 0.193}$\\

\end{tabular}
\end{table*}

\begin{table*}[htbp]
\tiny 
\centering
\begin{tabular}{cccccccccccc}
\hline
\hline
  \multicolumn{1}{c}{Gaia DR3 Name} &
  \multicolumn{1}{c}{Period (d)} &
  \multicolumn{1}{c}{$T\mathrm{_{1}}$ (K)} &
  \multicolumn{1}{c}{$T\mathrm{_{2}}$ (K)} &
  \multicolumn{1}{c}{log$(L\mathrm{_{1}}$/$L_{\odot}$)} &
  \multicolumn{1}{c}{log$(L\mathrm{_{2}}$/$L_{\odot}$)} &
  \multicolumn{1}{c}{$R\mathrm{_{1}}$ ($R_{\odot}$)} &
  \multicolumn{1}{c}{$R\mathrm{_{2}}$ ($R_{\odot}$)} &
  \multicolumn{1}{c}{$M\mathrm{_{1}}$ ($M_{\odot}$)} &
  \multicolumn{1}{c}{$M\mathrm{_{2}}$ ($M_{\odot}$)}\\
\hline
\hline
3228811140158987392 & 2.455548648 & 7999 $^{+ 0 }_{- 3 }$& 4005 $^{+ 171 }_{- 149 }$& 1.385 $^{+ 0.020  }_{- 0.021  }$& 0.33 $^{+ 0.107  }_{- 0.097  }$& 2.550  $^{+ 0.130  }_{- 0.113  }$& 3.024  $^{+ 0.361  }_{- 0.284  }$& 4.071  $^{+ 1.395  }_{- 1.009  }$& 0.667  $^{+ 0.266  }_{- 0.170}$\\
2080482158185702784 & 1.332539798 & 10687 $^{+ 30 }_{- 31 }$& 6573 $^{+ 251 }_{- 208 }$& 1.751 $^{+ 0.016  }_{- 0.018  }$& 1.106 $^{+ 0.053  }_{- 0.062  }$& 2.195  $^{+ 0.150  }_{- 0.158  }$& 2.746  $^{+ 0.254  }_{- 0.273  }$& 2.494  $^{+ 0.819  }_{- 0.644  }$& 1.512  $^{+ 0.452  }_{- 0.397}$\\
4569841324592474240 & 0.749056683 & 7898 $^{+ 23 }_{- 21 }$& 5425 $^{+ 154 }_{- 161 }$& 1.159 $^{+ 0.029  }_{- 0.030  }$& 0.466 $^{+ 0.118  }_{- 0.162  }$& 2.020  $^{+ 0.112  }_{- 0.098  }$& 1.925  $^{+ 0.305  }_{- 0.320  }$& 3.028  $^{+ 1.609  }_{- 0.922  }$& 1.639  $^{+ 0.902  }_{- 0.633}$\\
5267716679183867264 & 2.758142599 & 9764 $^{+ 39 }_{- 32 }$& 4902 $^{+ 92 }_{- 89 }$& 1.754 $^{+ 0.007  }_{- 0.007  }$& 0.563 $^{+ 0.046  }_{- 0.047  }$& 2.632  $^{+ 0.074  }_{- 0.067  }$& 2.649  $^{+ 0.151  }_{- 0.152  }$& 1.387  $^{+ 0.208  }_{- 0.191  }$& 0.347  $^{+ 0.063  }_{- 0.054}$\\
5150422943394260224 & 0.975162921 & 7053 $^{+ 16 }_{- 15 }$& 4283 $^{+ 146 }_{- 134 }$& 0.745 $^{+ 0.011  }_{- 0.012  }$& -0.230 $^{+ 0.092  }_{- 0.101  }$& 1.578  $^{+ 0.092  }_{- 0.092  }$& 1.390  $^{+ 0.177  }_{- 0.169  }$& 1.626  $^{+ 0.585  }_{- 0.411  }$& 0.400  $^{+ 0.168  }_{- 0.126}$\\
5272868887594862464 & 2.694233969 & 7183 $^{+ 12 }_{- 10 }$& 4517 $^{+ 106 }_{- 117 }$& 1.466 $^{+ 0.024  }_{- 0.028  }$& 0.781 $^{+ 0.111  }_{- 0.128  }$& 3.486  $^{+ 0.156  }_{- 0.161  }$& 4.021  $^{+ 0.571  }_{- 0.574  }$& 3.427  $^{+ 1.361  }_{- 1.077  }$& 1.219  $^{+ 0.580  }_{- 0.431}$\\
1017991469167531136 & 1.222293931 & 7490 $^{+ 20 }_{- 17 }$& 5098 $^{+ 75 }_{- 76 }$& 0.977 $^{+ 0.017  }_{- 0.018  }$& 0.621 $^{+ 0.036  }_{- 0.036  }$& 1.826  $^{+ 0.059  }_{- 0.056  }$& 2.621  $^{+ 0.126  }_{- 0.118  }$& 2.246  $^{+ 0.364  }_{- 0.282  }$& 1.535  $^{+ 0.229  }_{- 0.192}$\\
4985326015445428608 & 0.379915237 & 6160 $^{+ 7 }_{- 6 }$& 6082 $^{+ 167 }_{- 183 }$& 0.122 $^{+ 0.036  }_{- 0.038  }$& -0.109 $^{+ 0.057  }_{- 0.072  }$& 1.007  $^{+ 0.080  }_{- 0.069  }$& 0.795  $^{+ 0.069  }_{- 0.079  }$& 0.719  $^{+ 0.275  }_{- 0.190  }$& 0.447  $^{+ 0.115  }_{- 0.109}$\\
3302503849723642496 & 0.443022083 & 5335 $^{+ 23 }_{- 21 }$& 5048 $^{+ 182 }_{- 192 }$& 0.409 $^{+ 0.040  }_{- 0.042  }$& 0.085 $^{+ 0.071  }_{- 0.099  }$& 1.880  $^{+ 0.158  }_{- 0.156  }$& 1.435  $^{+ 0.167  }_{- 0.163  }$& 3.380  $^{+ 1.786  }_{- 0.993  }$& 1.950  $^{+ 0.713  }_{- 0.571}$\\
5288762774849168640 & 18.75378863 & 5706 $^{+ 10 }_{- 9 }$& 4347 $^{+ 62 }_{- 73 }$& 1.471 $^{+ 0.039  }_{- 0.042  }$& 1.658 $^{+ 0.025  }_{- 0.029  }$& 5.561  $^{+ 0.255  }_{- 0.261  }$& 11.890  $^{+ 0.419  }_{- 0.402  }$& 3.202  $^{+ 0.415  }_{- 0.341  }$& 0.686  $^{+ 0.074  }_{- 0.066}$\\
2878255089122256128 & 2.521024161 & 9598 $^{+ 47 }_{- 68 }$& 4448 $^{+ 204 }_{- 202 }$& 1.378 $^{+ 0.026  }_{- 0.027  }$& 0.274 $^{+ 0.141  }_{- 0.122  }$& 1.771  $^{+ 0.085  }_{- 0.088  }$& 2.308  $^{+ 0.361  }_{- 0.286  }$& 0.973  $^{+ 0.533  }_{- 0.290  }$& 0.272  $^{+ 0.148  }_{- 0.087}$\\
233996763254722304 & 1.151619786 & 7052 $^{+ 14 }_{- 15 }$& 4865 $^{+ 148 }_{- 131 }$& 1.177 $^{+ 0.011  }_{- 0.012  }$& 0.754 $^{+ 0.024  }_{- 0.025  }$& 2.586  $^{+ 0.144  }_{- 0.141  }$& 3.342  $^{+ 0.223  }_{- 0.209  }$& 5.535  $^{+ 1.233  }_{- 0.951  }$& 3.603  $^{+ 0.778  }_{- 0.628}$\\
1364434900041227520 & 2.347587595 & 10092 $^{+ 91 }_{- 46 }$& 4708 $^{+ 109 }_{- 123 }$& 1.312 $^{+ 0.016  }_{- 0.017  }$& 0.174 $^{+ 0.062  }_{- 0.066  }$& 1.483  $^{+ 0.055  }_{- 0.055  }$& 1.836  $^{+ 0.141  }_{- 0.123  }$& 0.614  $^{+ 0.144  }_{- 0.100  }$& 0.159  $^{+ 0.039  }_{- 0.03}$\\
6630948129392624768 & 1.964594683 & 6950 $^{+ 11 }_{- 11 }$& 3902 $^{+ 123 }_{- 118 }$& 1.002 $^{+ 0.013  }_{- 0.013  }$& 0.098 $^{+ 0.031  }_{- 0.032  }$& 2.183  $^{+ 0.142  }_{- 0.115  }$& 2.441  $^{+ 0.164  }_{- 0.136  }$& 1.288  $^{+ 0.272  }_{- 0.211  }$& 0.514  $^{+ 0.111  }_{- 0.081}$\\
2055760055859798400 & 8.426641334 & 9483 $^{+ 21 }_{- 22 }$& 4447 $^{+ 101 }_{- 95 }$& 2.000 $^{+ 0.012  }_{- 0.011  }$& 1.280 $^{+ 0.036  }_{- 0.039  }$& 3.695  $^{+ 0.167  }_{- 0.142  }$& 7.332  $^{+ 0.418  }_{- 0.351  }$& 4.956  $^{+ 0.938  }_{- 0.680  }$& 0.808  $^{+ 0.147  }_{- 0.111}$\\
242962936981732608 & 0.84941024 & 6053 $^{+ 22 }_{- 22 }$& 4326 $^{+ 132 }_{- 135 }$& 0.204 $^{+ 0.012  }_{- 0.014  }$& -0.369 $^{+ 0.046  }_{- 0.048  }$& 1.148  $^{+ 0.076  }_{- 0.064  }$& 1.165  $^{+ 0.096  }_{- 0.089  }$& 1.099  $^{+ 0.290  }_{- 0.231  }$& 0.309  $^{+ 0.081  }_{- 0.065}$\\
5159619494871873024 & 3.417279372 & 5786 $^{+ 46 }_{- 29 }$& 4133 $^{+ 224 }_{- 207 }$& 0.723 $^{+ 0.066  }_{- 0.078  }$& 0.654 $^{+ 0.078  }_{- 0.082  }$& 2.278  $^{+ 0.298  }_{- 0.264  }$& 4.091  $^{+ 0.624  }_{- 0.468  }$& 2.084  $^{+ 1.042  }_{- 0.683  }$& 0.804  $^{+ 0.417  }_{- 0.247}$\\
4844494725003381632 & 0.426943756 & 6378 $^{+ 21 }_{- 14 }$& 6130 $^{+ 175 }_{- 165 }$& 0.541 $^{+ 0.042  }_{- 0.049  }$& 0.118 $^{+ 0.108  }_{- 0.137  }$& 1.530  $^{+ 0.110  }_{- 0.115  }$& 1.009  $^{+ 0.153  }_{- 0.143  }$& 1.594  $^{+ 0.980  }_{- 0.498  }$& 0.746  $^{+ 0.367  }_{- 0.25}$\\
5158131859934912640 & 0.440779116 & 6313 $^{+ 8 }_{- 7 }$& 6147 $^{+ 225 }_{- 223 }$& 0.409 $^{+ 0.006  }_{- 0.007  }$& -0.028 $^{+ 0.011  }_{- 0.012  }$& 1.338  $^{+ 0.097  }_{- 0.097  }$& 0.854  $^{+ 0.066  }_{- 0.061  }$& 1.010  $^{+ 0.245  }_{- 0.201  }$& 0.433  $^{+ 0.107  }_{- 0.085}$\\
1106402718122325760 & 0.771335966 & 9438 $^{+ 137 }_{- 222 }$& 6391 $^{+ 175 }_{- 174 }$& 1.594 $^{+ 0.019  }_{- 0.025  }$& 0.863 $^{+ 0.102  }_{- 0.124  }$& 2.342  $^{+ 0.130  }_{- 0.130  }$& 2.198  $^{+ 0.304  }_{- 0.302  }$& 5.500  $^{+ 2.213  }_{- 1.599  }$& 2.392  $^{+ 1.051  }_{- 0.801}$\\
211807141136171136$^{(a)}$ & 1.2471725 & 14612 $^{+ 131 }_{- 98 }$& 7015 $^{+ 49 }_{- 37 }$& 2.653 $^{+ 0.018  }_{- 0.018  }$& 1.134 $^{+ 0.055  }_{- 0.057  }$& 3.310  $^{+ 0.071  }_{- 0.066  }$& 2.496  $^{+ 0.157  }_{- 0.156  }$& 5.597  $^{+ 0.738  }_{- 0.586  }$& 1.416  $^{+ 0.265  }_{- 0.234}$\\
3234434046920438656 $^{(*)}$& 0.632792879 & 7221 $^{+ 352 }_{- 462 }$& 6584 $^{+ 273 }_{- 275 }$& 0.847 $^{+ 0.085  }_{- 0.078 }$& 0.699 $^{+ 0.087  }_{- 0.146  }$& 1.789  $^{+ 0.212  }_{- 0.190 }$& 1.573  $^{+ 0.224  }_{- 0.247  }$& 2.630  $^{+ 1.931  }_{- 0.902  }$& 1.279  $^{+ 0.577  }_{- 0.463}$\\
\hline
\end{tabular}

\begin{tablenotes}
    \footnotesize
    \item[1] (a): Period is extracted from \citet{2021AA...652A.120I}, while other periods are obtained from \citet{2022ApJS..258...16P}. \\(*): The target with the more massive component filling its Roche lobe.
\end{tablenotes}

\end{table*}

\begin{table*}[htbp]
\tiny 
\caption{Absolute parameters of 77 semi-detached binaries from TESS survey (calculated based on $T\rm_{1}$ and distance from Gaia MSC \citep{2023AA...674A...1G}).}
\centering
\begin{tabular}{cccccccccccc}
\hline
\hline
  \multicolumn{1}{c}{Gaia DR3 Name} &
  \multicolumn{1}{c}{Period (d)} &
  \multicolumn{1}{c}{$T\mathrm{_{1}}$ (K)} &
  \multicolumn{1}{c}{$T\mathrm{_{2}}$ (K)} &
  \multicolumn{1}{c}{log$(L\mathrm{_{1}}$/$L_{\odot}$)} &
  \multicolumn{1}{c}{log$(L\mathrm{_{2}}$/$L_{\odot}$)} &
  \multicolumn{1}{c}{$R\mathrm{_{1}}$ ($R_{\odot}$)} &
  \multicolumn{1}{c}{$R\mathrm{_{2}}$ ($R_{\odot}$)} &
  \multicolumn{1}{c}{$M\mathrm{_{1}}$ ($M_{\odot}$)} &
  \multicolumn{1}{c}{$M\mathrm{_{2}}$ ($M_{\odot}$)}\\
\hline
\hline
4876695194334420096 & 3.067745211 & 7973 $^{+ 27 }_{- 328 }$& 4572 $^{+ 4 }_{- 4 }$& 1.215 $^{+ 0.246  }_{- 0.292  }$& 0.340 $^{+ 0.257  }_{- 0.314  }$& 2.134  $^{+ 0.713  }_{- 0.649  }$& 2.359  $^{+ 0.780  }_{- 0.718  }$& 0.804  $^{+ 1.090  }_{- 0.533  }$& 0.198  $^{+ 0.269  }_{- 0.132}$\\
2114453601444545024 & 0.993305002 & 6581 $^{+ 221 }_{- 220 }$& 6481 $^{+ 2 }_{- 2 }$& 0.852 $^{+ 0.066  }_{- 0.083  }$& 0.346 $^{+ 0.102  }_{- 0.096  }$& 2.047  $^{+ 0.207  }_{- 0.195  }$& 1.185  $^{+ 0.121  }_{- 0.113  }$& 2.500  $^{+ 0.845  }_{- 0.647  }$& 0.250  $^{+ 0.084  }_{- 0.065}$\\
2121872762311382912 & 0.86602103 & 7216 $^{+ 83 }_{- 100 }$& 4738 $^{+ 1 }_{- 1 }$& 0.922 $^{+ 0.116  }_{- 0.139  }$& 0.404 $^{+ 0.120  }_{- 0.152  }$& 1.854  $^{+ 0.258  }_{- 0.264  }$& 2.354  $^{+ 0.327  }_{- 0.340  }$& 3.672  $^{+ 1.736  }_{- 1.376  }$& 2.251  $^{+ 1.078  }_{- 0.841}$\\
699818436454276864 & 1.115976642 & 6663 $^{+ 66 }_{- 84 }$& 4238 $^{+ 3 }_{- 3 }$& 0.597 $^{+ 0.049  }_{- 0.058  }$& -0.155 $^{+ 0.084  }_{- 0.096  }$& 1.494  $^{+ 0.086  }_{- 0.103  }$& 1.559  $^{+ 0.093  }_{- 0.107  }$& 1.838  $^{+ 0.342  }_{- 0.352  }$& 0.434  $^{+ 0.082  }_{- 0.083}$\\
2436283197513649664 & 0.481424741 & 7030 $^{+ 453 }_{- 397 }$& 6766 $^{+ 13 }_{- 8 }$& 0.595 $^{+ 0.101  }_{- 0.114  }$& 0.061 $^{+ 0.136  }_{- 0.173  }$& 1.351  $^{+ 0.229  }_{- 0.208  }$& 0.769  $^{+ 0.127  }_{- 0.120  }$& 0.803  $^{+ 0.465  }_{- 0.327  }$& 0.273  $^{+ 0.155  }_{- 0.110}$\\
1333162251022354048 & 2.266959543 & 7932 $^{+ 68 }_{- 488 }$& 3934 $^{+ 1 }_{- 1 }$& 0.929 $^{+ 0.075  }_{- 0.076  }$& 0.044 $^{+ 0.103  }_{- 0.103  }$& 1.548  $^{+ 0.180  }_{- 0.166  }$& 2.265  $^{+ 0.263  }_{- 0.245  }$& 1.380  $^{+ 0.539  }_{- 0.401  }$& 0.323  $^{+ 0.126  }_{- 0.094}$\\
2526767987958943872 & 3.488532187 & 6800 $^{+ 7 }_{- 5 }$& 4311 $^{+ 5 }_{- 3 }$& 1.074 $^{+ 0.014  }_{- 0.016  }$& 0.311 $^{+ 0.081  }_{- 0.079  }$& 2.478  $^{+ 0.040  }_{- 0.045  }$& 2.568  $^{+ 0.044  }_{- 0.048  }$& 0.982  $^{+ 0.050  }_{- 0.052  }$& 0.200  $^{+ 0.010  }_{- 0.011}$\\
4601359272077948416 & 0.818090719 & 6807 $^{+ 302 }_{- 74 }$& 4719 $^{+ 1 }_{- 1 }$& 0.655 $^{+ 0.036  }_{- 0.040  }$& 0.047 $^{+ 0.067  }_{- 0.069  }$& 1.524  $^{+ 0.100  }_{- 0.091  }$& 1.582  $^{+ 0.102  }_{- 0.094  }$& 1.556  $^{+ 0.325  }_{- 0.262  }$& 0.785  $^{+ 0.163  }_{- 0.132}$\\
4597990196652263296 & 0.407603267 & 7007 $^{+ 117 }_{- 56 }$& 6691 $^{+ 5 }_{- 6 }$& 0.677 $^{+ 0.032  }_{- 0.029  }$& -0.014 $^{+ 0.067  }_{- 0.079  }$& 1.476  $^{+ 0.059  }_{- 0.059  }$& 0.730  $^{+ 0.034  }_{- 0.034  }$& 1.323  $^{+ 0.199  }_{- 0.160  }$& 0.332  $^{+ 0.048  }_{- 0.043}$\\
2882810400156716160 & 2.221301761 & 6997 $^{+ 128 }_{- 61 }$& 4443 $^{+ 13 }_{- 4 }$& 1.165 $^{+ 0.112  }_{- 0.126  }$& 0.106 $^{+ 0.137  }_{- 0.147  }$& 2.598  $^{+ 0.360  }_{- 0.358  }$& 1.912  $^{+ 0.268  }_{- 0.263  }$& 1.935  $^{+ 0.912  }_{- 0.697  }$& 0.209  $^{+ 0.100  }_{- 0.075}$\\
5180250215659097856 & 0.411306659 & 6099 $^{+ 73 }_{- 125 }$& 6052 $^{+ 7 }_{- 5 }$& 0.182 $^{+ 0.055  }_{- 0.055  }$& 0.103 $^{+ 0.066  }_{- 0.054  }$& 1.106  $^{+ 0.069  }_{- 0.074  }$& 1.032  $^{+ 0.063  }_{- 0.069  }$& 0.869  $^{+ 0.184  }_{- 0.161  }$& 0.788  $^{+ 0.157  }_{- 0.147}$\\
5179913528878335872 & 0.44528793 & 5817 $^{+ 481 }_{- 455 }$& 5505 $^{+ 2 }_{- 2 }$& 0.209 $^{+ 0.059  }_{- 0.070  }$& -0.134 $^{+ 0.121  }_{- 0.117  }$& 1.250  $^{+ 0.176  }_{- 0.150  }$& 0.941  $^{+ 0.130  }_{- 0.113  }$& 0.925  $^{+ 0.447  }_{- 0.288  }$& 0.547  $^{+ 0.260  }_{- 0.174}$\\
2905533010535981696 & 0.512831647 & 6301 $^{+ 134 }_{- 83 }$& 4202 $^{+ 3 }_{- 3 }$& 0.384 $^{+ 0.041  }_{- 0.039  }$& -0.473 $^{+ 0.076  }_{- 0.087  }$& 1.306  $^{+ 0.076  }_{- 0.068  }$& 1.091  $^{+ 0.067  }_{- 0.056  }$& 1.299  $^{+ 0.245  }_{- 0.191  }$& 0.655  $^{+ 0.130  }_{- 0.097}$\\
4653416302929032960 & 3.232081512 & 7501 $^{+ 339 }_{- 575 }$& 4459 $^{+ 3 }_{- 3 }$& 1.219 $^{+ 0.101  }_{- 0.104  }$& 0.303 $^{+ 0.149  }_{- 0.146  }$& 2.421  $^{+ 0.407  }_{- 0.357  }$& 2.383  $^{+ 0.399  }_{- 0.355  }$& 1.083  $^{+ 0.645  }_{- 0.412  }$& 0.188  $^{+ 0.111  }_{- 0.072}$\\
2032636398639228160 & 1.694822868 & 7600 $^{+ 6 }_{- 9 }$& 7046 $^{+ 4 }_{- 4 }$& 1.283 $^{+ 0.039  }_{- 0.047  }$& 1.130 $^{+ 0.055  }_{- 0.064  }$& 2.526  $^{+ 0.116  }_{- 0.132  }$& 2.456  $^{+ 0.113  }_{- 0.125  }$& 1.905  $^{+ 0.275  }_{- 0.279  }$& 0.709  $^{+ 0.102  }_{- 0.102}$\\
2990232754910113152 & 0.594445177 & 6231 $^{+ 347 }_{- 251 }$& 5942 $^{+ 2 }_{- 2 }$& 0.783 $^{+ 0.088  }_{- 0.086  }$& 0.082 $^{+ 0.125  }_{- 0.133  }$& 2.126  $^{+ 0.289  }_{- 0.287  }$& 1.048  $^{+ 0.143  }_{- 0.141  }$& 1.860  $^{+ 0.868  }_{- 0.657  }$& 0.462  $^{+ 0.216  }_{- 0.163}$\\
4889204372482883584 & 0.638486649 & 7610 $^{+ 277 }_{- 205 }$& 4872 $^{+ 4 }_{- 3 }$& 0.926 $^{+ 0.058  }_{- 0.062  }$& -0.287 $^{+ 0.111  }_{- 0.126  }$& 1.670  $^{+ 0.180  }_{- 0.160  }$& 0.997  $^{+ 0.108  }_{- 0.096  }$& 0.997  $^{+ 0.361  }_{- 0.260  }$& 0.338  $^{+ 0.122  }_{- 0.088}$\\
2457700412712178944 & 1.93975019 & 7916 $^{+ 84 }_{- 879 }$& 4504 $^{+ 3 }_{- 3 }$& 1.141 $^{+ 0.057  }_{- 0.059  }$& 0.146 $^{+ 0.129  }_{- 0.117  }$& 1.963  $^{+ 0.281  }_{- 0.218  }$& 1.956  $^{+ 0.279  }_{- 0.219  }$& 1.269  $^{+ 0.622  }_{- 0.384  }$& 0.284  $^{+ 0.140  }_{- 0.085}$\\
2172287874408422144 & 1.10574422 & 6365 $^{+ 1057 }_{- 393 }$& 4746 $^{+ 10 }_{- 6 }$& 1.605 $^{+ 0.136  }_{- 0.172  }$& 1.057 $^{+ 0.204  }_{- 0.249  }$& 5.134  $^{+ 1.329  }_{- 1.119  }$& 4.928  $^{+ 1.300  }_{- 1.109  }$& 14.603  $^{+ 14.475  }_{- 7.977  }$& 12.051  $^{+ 12.166  }_{- 6.387}$\\
1351714684378550016 & 1.755696172 & 7819 $^{+ 181 }_{- 540 }$& 4459 $^{+ 4 }_{- 4 }$& 1.255 $^{+ 0.075  }_{- 0.085  }$& 0.363 $^{+ 0.131  }_{- 0.144  }$& 2.311  $^{+ 0.296  }_{- 0.280  }$& 2.552  $^{+ 0.331  }_{- 0.313  }$& 1.156  $^{+ 0.508  }_{- 0.367  }$& 0.700  $^{+ 0.308  }_{- 0.227}$\\
1035063002096359680 & 2.022459806 & 7762 $^{+ 238 }_{- 504 }$& 3568 $^{+ 4 }_{- 5 }$& 1.417 $^{+ 0.129  }_{- 0.139  }$& 0.365 $^{+ 0.169  }_{- 0.189  }$& 2.825  $^{+ 0.557  }_{- 0.520  }$& 3.773  $^{+ 0.740  }_{- 0.695  }$& 8.264  $^{+ 5.825  }_{- 3.803  }$& 1.877  $^{+ 1.338  }_{- 0.857}$\\
740582795692769408 & 0.376124124 & 6319 $^{+ 189 }_{- 115 }$& 6125 $^{+ 3 }_{- 2 }$& 0.106 $^{+ 0.046  }_{- 0.051  }$& 0.001 $^{+ 0.051  }_{- 0.060  }$& 0.942  $^{+ 0.058  }_{- 0.056  }$& 0.887  $^{+ 0.056  }_{- 0.051  }$& 0.597  $^{+ 0.119  }_{- 0.099  }$& 0.590  $^{+ 0.117  }_{- 0.097}$\\
5153975843420955136 & 1.01568807 & 5804 $^{+ 578 }_{- 392 }$& 5352 $^{+ 4 }_{- 9 }$& 1.101 $^{+ 0.116  }_{- 0.131  }$& 0.414 $^{+ 0.195  }_{- 0.220  }$& 3.486  $^{+ 0.787  }_{- 0.613  }$& 1.861  $^{+ 0.478  }_{- 0.370  }$& 3.265  $^{+ 2.811  }_{- 1.551  }$& 0.878  $^{+ 0.853  }_{- 0.428}$\\
4997907692641176192 & 0.413614378 & 7002 $^{+ 29 }_{- 25 }$& 6640 $^{+ 5 }_{- 15 }$& 0.632 $^{+ 0.032  }_{- 0.035  }$& -0.027 $^{+ 0.101  }_{- 0.122  }$& 1.403  $^{+ 0.055  }_{- 0.054  }$& 0.732  $^{+ 0.068  }_{- 0.073  }$& 1.123  $^{+ 0.294  }_{- 0.200  }$& 0.321  $^{+ 0.090  }_{- 0.079}$\\
4965590537640948992 & 2.434877225 & 7966 $^{+ 34 }_{- 362 }$& 4268 $^{+ 3 }_{- 9 }$& 1.387 $^{+ 0.053  }_{- 0.061  }$& 0.453 $^{+ 0.099  }_{- 0.103  }$& 2.591  $^{+ 0.210  }_{- 0.198  }$& 3.072  $^{+ 0.247  }_{- 0.248  }$& 3.222  $^{+ 0.828  }_{- 0.712  }$& 0.701  $^{+ 0.184  }_{- 0.156}$\\
2051947769857973120 & 0.513482922 & 7076 $^{+ 295 }_{- 373 }$& 5128 $^{+ 8 }_{- 8 }$& 0.669 $^{+ 0.064  }_{- 0.067  }$& 0.087 $^{+ 0.130  }_{- 0.133  }$& 1.441  $^{+ 0.164  }_{- 0.147  }$& 1.395  $^{+ 0.165  }_{- 0.137  }$& 1.292  $^{+ 0.505  }_{- 0.352  }$& 1.239  $^{+ 0.493  }_{- 0.332}$\\
4654488391189033344 & 0.418404117 & 6346 $^{+ 209 }_{- 411 }$& 6008 $^{+ 2 }_{- 3 }$& 0.557 $^{+ 0.067  }_{- 0.075  }$& -0.124 $^{+ 0.122  }_{- 0.117  }$& 1.581  $^{+ 0.194  }_{- 0.190  }$& 0.803  $^{+ 0.099  }_{- 0.096  }$& 1.521  $^{+ 0.635  }_{- 0.485  }$& 0.418  $^{+ 0.174  }_{- 0.133}$\\
5266269923742586496 & 0.437169155 & 6378 $^{+ 251 }_{- 237 }$& 6244 $^{+ 3 }_{- 3 }$& 0.491 $^{+ 0.056  }_{- 0.063  }$& -0.044 $^{+ 0.098  }_{- 0.099  }$& 1.438  $^{+ 0.149  }_{- 0.131  }$& 0.809  $^{+ 0.086  }_{- 0.075  }$& 1.201  $^{+ 0.414  }_{- 0.295  }$& 0.386  $^{+ 0.137  }_{- 0.098}$\\
4548008150397375872 & 0.387748803 & 6666 $^{+ 271 }_{- 357 }$& 6482 $^{+ 3 }_{- 7 }$& 0.466 $^{+ 0.053  }_{- 0.057  }$& -0.019 $^{+ 0.091  }_{- 0.102  }$& 1.280  $^{+ 0.133  }_{- 0.124  }$& 0.772  $^{+ 0.087  }_{- 0.078  }$& 1.131  $^{+ 0.393  }_{- 0.313  }$& 0.420  $^{+ 0.156  }_{- 0.112}$\\
5477873272970767360 & 0.896878797 & 6498 $^{+ 216 }_{- 354 }$& 6429 $^{+ 5 }_{- 5 }$& 0.859 $^{+ 0.080  }_{- 0.098  }$& 0.256 $^{+ 0.138  }_{- 0.123  }$& 2.124  $^{+ 0.278  }_{- 0.270  }$& 1.090  $^{+ 0.142  }_{- 0.138  }$& 2.384  $^{+ 1.057  }_{- 0.795  }$& 0.239  $^{+ 0.106  }_{- 0.080}$\\
5477159247544461312 & 2.59364758 & 7943 $^{+ 57 }_{- 504 }$& 4780 $^{+ 1 }_{- 1 }$& 1.287 $^{+ 0.076  }_{- 0.081  }$& 0.616 $^{+ 0.097  }_{- 0.094  }$& 2.317  $^{+ 0.249  }_{- 0.245  }$& 2.977  $^{+ 0.320  }_{- 0.314  }$& 5.540  $^{+ 1.980  }_{- 1.574  }$& 0.580  $^{+ 0.208  }_{- 0.165}$\\
5566666915748470912 & 1.049570677 & 7211 $^{+ 138 }_{- 87 }$& 4296 $^{+ 2 }_{- 2 }$& 0.802 $^{+ 0.034  }_{- 0.036  }$& 0.004 $^{+ 0.076  }_{- 0.079  }$& 1.608  $^{+ 0.083  }_{- 0.076  }$& 1.808  $^{+ 0.092  }_{- 0.086  }$& 1.948  $^{+ 0.315  }_{- 0.265  }$& 0.737  $^{+ 0.119  }_{- 0.100}$\\
2988525797467488256 & 0.91543371 & 7474 $^{+ 360 }_{- 709 }$& 4329 $^{+ 4 }_{- 4 }$& 1.631 $^{+ 0.031  }_{- 0.030  }$& 0.541 $^{+ 0.173  }_{- 0.164  }$& 3.898  $^{+ 0.658  }_{- 0.533  }$& 3.320  $^{+ 0.560  }_{- 0.441  }$& 13.099  $^{+ 7.780  }_{- 4.608  }$& 5.886  $^{+ 3.475  }_{- 2.052}$\\
938228841240985728 & 1.062648473 & 7600 $^{+ 2 }_{- 5 }$& 6526 $^{+ 69 }_{- 46 }$& 1.746 $^{+ 0.087  }_{- 0.088  }$& 0.995 $^{+ 0.189  }_{- 0.243  }$& 4.308  $^{+ 0.455  }_{- 0.418  }$& 2.493  $^{+ 0.382  }_{- 0.364  }$& 8.953  $^{+ 3.742  }_{- 2.755  }$& 1.968  $^{+ 1.007  }_{- 0.730}$\\
331337421011384448 & 0.753660737 & 6481 $^{+ 144 }_{- 231 }$& 4037 $^{+ 2 }_{- 2 }$& 0.595 $^{+ 0.065  }_{- 0.062  }$& -0.268 $^{+ 0.087  }_{- 0.088  }$& 1.573  $^{+ 0.142  }_{- 0.136  }$& 1.497  $^{+ 0.134  }_{- 0.127  }$& 1.707  $^{+ 0.494  }_{- 0.403  }$& 0.792  $^{+ 0.232  }_{- 0.185}$\\
2082658577035787392 & 4.573035676 & 7828 $^{+ 172 }_{- 992 }$& 4223 $^{+ 1 }_{- 1 }$& 1.528 $^{+ 0.099  }_{- 0.111  }$& 0.908 $^{+ 0.160  }_{- 0.162  }$& 3.147  $^{+ 0.543  }_{- 0.514  }$& 5.287  $^{+ 0.912  }_{- 0.863  }$& 4.215  $^{+ 2.588  }_{- 1.742  }$& 1.007  $^{+ 0.616  }_{- 0.417}$\\
2407526566105203072 & 0.862731672 & 7401 $^{+ 599 }_{- 185 }$& 5116 $^{+ 2 }_{- 2 }$& 1.012 $^{+ 0.326  }_{- 0.489  }$& 0.345 $^{+ 0.354  }_{- 0.471  }$& 1.944  $^{+ 0.939  }_{- 0.831  }$& 1.896  $^{+ 0.923  }_{- 0.809  }$& 1.974  $^{+ 4.482  }_{- 1.602  }$& 1.188  $^{+ 2.715  }_{- 0.964}$\\
2083160912108531456 & 0.628626478 & 6228 $^{+ 135 }_{- 36 }$& 3993 $^{+ 3 }_{- 3 }$& 0.297 $^{+ 0.038  }_{- 0.039  }$& -0.500 $^{+ 0.074  }_{- 0.074  }$& 1.209  $^{+ 0.065  }_{- 0.057  }$& 1.179  $^{+ 0.064  }_{- 0.057  }$& 1.155  $^{+ 0.198  }_{- 0.158  }$& 0.554  $^{+ 0.095  }_{- 0.076}$\\
1350220830329440256 & 0.787182881 & 6104 $^{+ 195 }_{- 77 }$& 4982 $^{+ 1 }_{- 1 }$& 0.678 $^{+ 0.119  }_{- 0.151  }$& 0.169 $^{+ 0.140  }_{- 0.147  }$& 1.947  $^{+ 0.311  }_{- 0.311  }$& 1.629  $^{+ 0.260  }_{- 0.259  }$& 2.453  $^{+ 1.378  }_{- 0.988  }$& 0.955  $^{+ 0.533  }_{- 0.388}$\\
1350891910379201280 & 0.486617826 & 7377 $^{+ 284 }_{- 734 }$& 6515 $^{+ 5 }_{- 4 }$& 0.775 $^{+ 0.069  }_{- 0.080  }$& 0.011 $^{+ 0.143  }_{- 0.135  }$& 1.495  $^{+ 0.243  }_{- 0.203  }$& 0.794  $^{+ 0.129  }_{- 0.108  }$& 1.079  $^{+ 0.621  }_{- 0.382  }$& 0.298  $^{+ 0.170  }_{- 0.106}$\\
1346411194338424704 & 1.305747766 & 7430 $^{+ 337 }_{- 249 }$& 4211 $^{+ 3 }_{- 5 }$& 1.133 $^{+ 0.018  }_{- 0.018  }$& 0.105 $^{+ 0.088  }_{- 0.098  }$& 2.220  $^{+ 0.193  }_{- 0.162  }$& 2.116  $^{+ 0.187  }_{- 0.179  }$& 3.416  $^{+ 0.991  }_{- 0.786  }$& 0.793  $^{+ 0.231  }_{- 0.184}$\\
2001818565159194496 & 2.206521406 & 7622 $^{+ 378 }_{- 2880 }$& 4255 $^{+ 3 }_{- 3 }$& 1.36 $^{+ 0.371  }_{- 0.455  }$& 0.537 $^{+ 0.519  }_{- 0.583  }$& 2.816  $^{+ 2.218  }_{- 1.374  }$& 3.452  $^{+ 2.706  }_{- 1.690  }$& 6.651  $^{+ 30.894  }_{- 5.765  }$& 1.223  $^{+ 5.717  }_{- 1.060}$\\
1560572488648076800 & 4.669284406 & 5393 $^{+ 415 }_{- 118 }$& 4376 $^{+ 1 }_{- 1 }$& 1.153 $^{+ 0.242  }_{- 0.353  }$& 0.692 $^{+ 0.243  }_{- 0.353  }$& 4.350  $^{+ 1.433  }_{- 1.452  }$& 3.857  $^{+ 1.267  }_{- 1.292  }$& 1.743  $^{+ 2.334  }_{- 1.229  }$& 0.378  $^{+ 0.507  }_{- 0.267}$\\
4799811706322746752 & 1.83811605 & 7599 $^{+ 5 }_{- 6 }$& 5232 $^{+ 89 }_{- 42 }$& 1.576 $^{+ 0.015  }_{- 0.019  }$& 0.591 $^{+ 0.145  }_{- 0.155  }$& 3.540  $^{+ 0.063  }_{- 0.078  }$& 2.383  $^{+ 0.411  }_{- 0.316  }$& 5.289  $^{+ 1.521  }_{- 1.236  }$& 0.591  $^{+ 0.340  }_{- 0.202}$\\
4775507585905757440 & 0.785242682 & 7345 $^{+ 476 }_{- 824 }$& 5308 $^{+ 4 }_{- 4 }$& 0.862 $^{+ 0.097  }_{- 0.108  }$& 0.408 $^{+ 0.160  }_{- 0.157  }$& 1.639  $^{+ 0.325  }_{- 0.253  }$& 1.879  $^{+ 0.372  }_{- 0.290  }$& 1.740  $^{+ 1.283  }_{- 0.682  }$& 1.344  $^{+ 0.977  }_{- 0.528}$\\
5553742981197782400 & 2.894658261 & 7985 $^{+ 15 }_{- 199 }$& 4168 $^{+ 4 }_{- 3 }$& 1.51 $^{+ 0.453  }_{- 0.593  }$& 0.527 $^{+ 0.478  }_{- 0.581  }$& 2.977  $^{+ 2.057  }_{- 1.466  }$& 3.551  $^{+ 2.449  }_{- 1.751  }$& 3.176  $^{+ 12.192  }_{- 2.762  }$& 0.761  $^{+ 2.912  }_{- 0.662}$\\
1454617122923582976 & 0.366990139 & 5991 $^{+ 85 }_{- 49 }$& 5872 $^{+ 3 }_{- 4 }$& 0.121 $^{+ 0.052  }_{- 0.059  }$& 0.052 $^{+ 0.057  }_{- 0.060  }$& 1.066  $^{+ 0.066  }_{- 0.065  }$& 1.023  $^{+ 0.063  }_{- 0.063  }$& 0.951  $^{+ 0.189  }_{- 0.164  }$& 0.948  $^{+ 0.187  }_{- 0.164}$\\
4567617420587434624 & 0.370828898 & 5786 $^{+ 213 }_{- 167 }$& 5493 $^{+ 8 }_{- 13 }$& -0.015 $^{+ 0.062  }_{- 0.069  }$& -0.244 $^{+ 0.084  }_{- 0.108  }$& 0.979  $^{+ 0.080  }_{- 0.080  }$& 0.833  $^{+ 0.074  }_{- 0.081  }$& 0.717  $^{+ 0.235  }_{- 0.189  }$& 0.531  $^{+ 0.149  }_{- 0.134}$\\
2157168185074459648 & 2.83127773 & 7201 $^{+ 15 }_{- 8 }$& 4131 $^{+ 2 }_{- 1 }$& 1.443 $^{+ 0.026  }_{- 0.025  }$& 0.792 $^{+ 0.041  }_{- 0.037  }$& 3.383  $^{+ 0.103  }_{- 0.099  }$& 4.856  $^{+ 0.150  }_{- 0.139  }$& 8.178  $^{+ 0.772  }_{- 0.686  }$& 2.031  $^{+ 0.194  }_{- 0.170}$\\
4708499327219174912 & 0.865839827 & 6929 $^{+ 28 }_{- 33 }$& 4552 $^{+ 2 }_{- 2 }$& 0.75 $^{+ 0.016  }_{- 0.017  }$& -0.174 $^{+ 0.069  }_{- 0.070  }$& 1.645  $^{+ 0.036  }_{- 0.032  }$& 1.306  $^{+ 0.029  }_{- 0.026  }$& 1.866  $^{+ 0.125  }_{- 0.111  }$& 0.425  $^{+ 0.029  }_{- 0.025}$\\
3224767071968129152 & 1.746072739 & 7600 $^{+ 3 }_{- 5 }$& 3958 $^{+ 3 }_{- 2 }$& 1.591 $^{+ 0.010  }_{- 0.010  }$& 0.548 $^{+ 0.069  }_{- 0.074  }$& 3.600  $^{+ 0.039  }_{- 0.042  }$& 3.982  $^{+ 0.043  }_{- 0.045  }$& 11.402  $^{+ 0.372  }_{- 0.397  }$& 2.936  $^{+ 0.096  }_{- 0.100}$\\
5078057314303209984 & 0.456619343 & 6493 $^{+ 232 }_{- 428 }$& 6218 $^{+ 3 }_{- 3 }$& 0.562 $^{+ 0.086  }_{- 0.105  }$& -0.093 $^{+ 0.137  }_{- 0.144  }$& 1.509  $^{+ 0.215  }_{- 0.211  }$& 0.778  $^{+ 0.111  }_{- 0.108  }$& 1.129  $^{+ 0.552  }_{- 0.410  }$& 0.318  $^{+ 0.156  }_{- 0.115}$\\
3231583219428074752 & 0.950932477 & 5774 $^{+ 236 }_{- 336 }$& 3975 $^{+ 2 }_{- 2 }$& 0.712 $^{+ 0.060  }_{- 0.058  }$& 0.009 $^{+ 0.108  }_{- 0.111  }$& 2.276  $^{+ 0.281  }_{- 0.241  }$& 2.129  $^{+ 0.265  }_{- 0.228  }$& 1.877  $^{+ 0.786  }_{- 0.537  }$& 1.349  $^{+ 0.569  }_{- 0.388}$\\
5138409988587248896 & 0.790454125 & 7948 $^{+ 52 }_{- 782 }$& 5480 $^{+ 2 }_{- 2 }$& 0.997 $^{+ 0.108  }_{- 0.122  }$& 0.162 $^{+ 0.165  }_{- 0.171  }$& 1.662  $^{+ 0.292  }_{- 0.247  }$& 1.327  $^{+ 0.233  }_{- 0.197  }$& 0.880  $^{+ 0.549  }_{- 0.336  }$& 0.490  $^{+ 0.306  }_{- 0.187}$\\
2357884032723549184 & 0.445412171 & 6403 $^{+ 138 }_{- 73 }$& 6291 $^{+ 5 }_{- 5 }$& 0.52 $^{+ 0.099  }_{- 0.117  }$& -0.061 $^{+ 0.123  }_{- 0.132  }$& 1.476  $^{+ 0.179  }_{- 0.188  }$& 0.792  $^{+ 0.096  }_{- 0.106  }$& 1.226  $^{+ 0.503  }_{- 0.408  }$& 0.353  $^{+ 0.146  }_{- 0.122}$\\

\end{tabular}
\end{table*}

\begin{table*}[htbp]
\tiny
\centering
\begin{tabular}{cccccccccccc}
\hline
\hline
  \multicolumn{1}{c}{Gaia DR3 Name} &
  \multicolumn{1}{c}{Period (d)} &
  \multicolumn{1}{c}{$T\mathrm{_{1}}$ (K)} &
  \multicolumn{1}{c}{$T\mathrm{_{2}}$ (K)} &
  \multicolumn{1}{c}{log$(L\mathrm{_{1}}$/$L_{\odot}$)} &
  \multicolumn{1}{c}{log$(L\mathrm{_{2}}$/$L_{\odot}$)} &
  \multicolumn{1}{c}{$R\mathrm{_{1}}$ ($R_{\odot}$)} &
  \multicolumn{1}{c}{$R\mathrm{_{2}}$ ($R_{\odot}$)} &
  \multicolumn{1}{c}{$M\mathrm{_{1}}$ ($M_{\odot}$)} &
  \multicolumn{1}{c}{$M\mathrm{_{2}}$ ($M_{\odot}$)}\\
\hline
\hline
3228811140158987392 & 2.455548648 & 7572 $^{+ 197 }_{- 215 }$& 3910 $^{+ 5 }_{- 5 }$& 1.665 $^{+ 0.056  }_{- 0.052  }$& 0.661 $^{+ 0.120  }_{- 0.119  }$& 3.962  $^{+ 0.341  }_{- 0.313  }$& 4.663  $^{+ 0.401  }_{- 0.371  }$& 14.884  $^{+ 4.137  }_{- 3.314  }$& 2.447  $^{+ 0.687  }_{- 0.539}$\\
2080482158185702784 & 1.332539798 & 7601 $^{+ 9 }_{- 6 }$& 5162 $^{+ 2 }_{- 2 }$& 1.617 $^{+ 0.026  }_{- 0.029  }$& 1.101 $^{+ 0.079  }_{- 0.079  }$& 3.712  $^{+ 0.110  }_{- 0.122  }$& 4.483  $^{+ 0.132  }_{- 0.143  }$& 11.123  $^{+ 1.039  }_{- 1.042  }$& 6.602  $^{+ 0.600  }_{- 0.616}$\\
4569841324592474240 & 0.749056683 & 7795 $^{+ 182 }_{- 2014 }$& 5386 $^{+ 5 }_{- 5 }$& 1.324 $^{+ 0.062  }_{- 0.082  }$& 0.691 $^{+ 0.212  }_{- 0.205  }$& 2.540  $^{+ 0.629  }_{- 0.505  }$& 2.553  $^{+ 0.639  }_{- 0.502  }$& 5.924  $^{+ 5.665  }_{- 2.887  }$& 3.807  $^{+ 3.660  }_{- 1.835}$\\
5267716679183867264 & 2.758142599 & 6999 $^{+ 212 }_{- 258 }$& 4034 $^{+ 2 }_{- 2 }$& 1.644 $^{+ 0.070  }_{- 0.078  }$& 0.702 $^{+ 0.099  }_{- 0.120  }$& 4.511  $^{+ 0.516  }_{- 0.500  }$& 4.577  $^{+ 0.519  }_{- 0.510  }$& 6.909  $^{+ 2.638  }_{- 2.064  }$& 1.787  $^{+ 0.679  }_{- 0.534}$\\
5150422943394260224 & 0.975162921 & 7137 $^{+ 64 }_{- 91 }$& 4329 $^{+ 3 }_{- 4 }$& 0.798 $^{+ 0.034  }_{- 0.033  }$& -0.278 $^{+ 0.089  }_{- 0.091  }$& 1.636  $^{+ 0.074  }_{- 0.078  }$& 1.293  $^{+ 0.058  }_{- 0.058  }$& 1.589  $^{+ 0.229  }_{- 0.211  }$& 0.327  $^{+ 0.046  }_{- 0.042}$\\
5272868887594862464 & 2.694233969 & 6200 $^{+ 13 }_{- 36 }$& 4072 $^{+ 2 }_{- 2 }$& 1.324 $^{+ 0.117  }_{- 0.154  }$& 0.710 $^{+ 0.145  }_{- 0.157  }$& 3.982  $^{+ 0.566  }_{- 0.658  }$& 4.580  $^{+ 0.671  }_{- 0.756  }$& 4.888  $^{+ 2.421  }_{- 2.054  }$& 1.818  $^{+ 0.923  }_{- 0.759}$\\
1017991469167531136 & 1.222293931 & 7597 $^{+ 293 }_{- 211 }$& 5147 $^{+ 2 }_{- 2 }$& 1.007 $^{+ 0.171  }_{- 0.200  }$& 0.658 $^{+ 0.170  }_{- 0.206  }$& 1.845  $^{+ 0.402  }_{- 0.404  }$& 2.699  $^{+ 0.582  }_{- 0.591  }$& 2.399  $^{+ 1.920  }_{- 1.257  }$& 1.671  $^{+ 1.331  }_{- 0.876}$\\
4985326015445428608 & 0.379915237 & 6119 $^{+ 36 }_{- 39 }$& 6017 $^{+ 1 }_{- 7 }$& 0.212 $^{+ 0.047  }_{- 0.041  }$& -0.173 $^{+ 0.061  }_{- 0.068  }$& 1.136  $^{+ 0.061  }_{- 0.054  }$& 0.750  $^{+ 0.047  }_{- 0.042  }$& 0.908  $^{+ 0.187  }_{- 0.137  }$& 0.394  $^{+ 0.075  }_{- 0.060}$\\
3302503849723642496 & 0.443022083 & 5009 $^{+ 2306 }_{- 774 }$& 4765 $^{+ 3 }_{- 2 }$& 0.332 $^{+ 0.198  }_{- 0.284  }$& 0.143 $^{+ 0.256  }_{- 0.339  }$& 1.791  $^{+ 0.608  }_{- 0.568  }$& 1.731  $^{+ 0.594  }_{- 0.547  }$& 3.210  $^{+ 4.523  }_{- 2.184  }$& 3.176  $^{+ 4.480  }_{- 2.166}$\\
5288762774849168640 & 18.75378863 & 4802 $^{+ 76 }_{- 63 }$& 3792 $^{+ 2 }_{- 2 }$& 1.032 $^{+ 0.173  }_{- 0.234  }$& 1.247 $^{+ 0.178  }_{- 0.219  }$& 4.739  $^{+ 1.065  }_{- 1.090  }$& 9.725  $^{+ 2.198  }_{- 2.244  }$& 1.790  $^{+ 1.502  }_{- 0.975  }$& 0.376  $^{+ 0.317  }_{- 0.205}$\\
2878255089122256128 & 2.521024161 & 7256 $^{+ 744 }_{- 659 }$& 3819 $^{+ 7 }_{- 7 }$& 1.394 $^{+ 0.300  }_{- 0.476  }$& 0.492 $^{+ 0.372  }_{- 0.500  }$& 3.177  $^{+ 1.553  }_{- 1.401  }$& 3.990  $^{+ 1.971  }_{- 1.756  }$& 5.228  $^{+ 12.105  }_{- 4.315  }$& 1.413  $^{+ 3.295  }_{- 1.165}$\\
233996763254722304 & 1.151619786 & 5504 $^{+ 165 }_{- 119 }$& 4024 $^{+ 0 }_{- 0 }$& 1.115 $^{+ 0.145  }_{- 0.166  }$& 0.798 $^{+ 0.161  }_{- 0.174  }$& 3.973  $^{+ 0.732  }_{- 0.732  }$& 5.153  $^{+ 0.949  }_{- 0.949  }$& 19.893  $^{+ 13.137  }_{- 9.093  }$& 13.200  $^{+ 8.717  }_{- 6.034}$\\
1364434900041227520 & 2.347587595 & 7605 $^{+ 133 }_{- 104 }$& 3997 $^{+ 4 }_{- 4 }$& 1.288 $^{+ 0.064  }_{- 0.066  }$& 0.347 $^{+ 0.097  }_{- 0.107  }$& 2.531  $^{+ 0.226  }_{- 0.194  }$& 3.097  $^{+ 0.276  }_{- 0.237  }$& 2.927  $^{+ 0.846  }_{- 0.629  }$& 0.763  $^{+ 0.223  }_{- 0.162}$\\
6630948129392624768 & 1.964594683 & 6401 $^{+ 54 }_{- 15 }$& 3707 $^{+ 3 }_{- 3 }$& 0.907 $^{+ 0.408  }_{- 0.146  }$& 0.078 $^{+ 0.404  }_{- 0.167  }$& 2.309  $^{+ 1.380  }_{- 0.355  }$& 2.581  $^{+ 1.542  }_{- 0.398  }$& 1.515  $^{+ 4.660  }_{- 0.598  }$& 0.608  $^{+ 1.869  }_{- 0.240}$\\
2055760055859798400 & 8.426641334 & 7610 $^{+ 390 }_{- 272 }$& 3929 $^{+ 2 }_{- 2 }$& 1.684 $^{+ 0.099  }_{- 0.095  }$& 1.130 $^{+ 0.132  }_{- 0.128  }$& 4.026  $^{+ 0.611  }_{- 0.534  }$& 7.948  $^{+ 1.205  }_{- 1.055  }$& 6.359  $^{+ 3.360  }_{- 2.211  }$& 1.030  $^{+ 0.544  }_{- 0.358}$\\
242962936981732608 & 0.84941024 & 5921 $^{+ 186 }_{- 284 }$& 4256 $^{+ 1 }_{- 1 }$& 0.219 $^{+ 0.077  }_{- 0.074  }$& -0.337 $^{+ 0.103  }_{- 0.111  }$& 1.223  $^{+ 0.135  }_{- 0.126  }$& 1.255  $^{+ 0.138  }_{- 0.130  }$& 1.343  $^{+ 0.495  }_{- 0.376  }$& 0.385  $^{+ 0.141  }_{- 0.108}$\\
5159619494871873024 & 3.417279372 & 4723 $^{+ 304 }_{- 186 }$& 3558 $^{+ 3 }_{- 3 }$& 0.456 $^{+ 0.128  }_{- 0.136  }$& 0.523 $^{+ 0.124  }_{- 0.128  }$& 2.526  $^{+ 0.392  }_{- 0.379  }$& 4.313  $^{+ 0.671  }_{- 0.638  }$& 2.449  $^{+ 1.296  }_{- 0.932  }$& 0.942  $^{+ 0.510  }_{- 0.360}$\\
4844494725003381632 & 0.426943756 & 6749 $^{+ 174 }_{- 272 }$& 6404 $^{+ 4 }_{- 15 }$& 0.605 $^{+ 0.043  }_{- 0.045  }$& -0.132 $^{+ 0.112  }_{- 0.127  }$& 1.469  $^{+ 0.120  }_{- 0.113  }$& 0.696  $^{+ 0.076  }_{- 0.081  }$& 1.140  $^{+ 0.351  }_{- 0.279  }$& 0.263  $^{+ 0.093  }_{- 0.079}$\\
5158131859934912640 & 0.440779116 & 6451 $^{+ 146 }_{- 121 }$& 6273 $^{+ 1 }_{- 1 }$& 0.409 $^{+ 0.035  }_{- 0.035  }$& -0.028 $^{+ 0.062  }_{- 0.071  }$& 1.284  $^{+ 0.067  }_{- 0.064  }$& 0.818  $^{+ 0.043  }_{- 0.041  }$& 0.886  $^{+ 0.146  }_{- 0.126  }$& 0.381  $^{+ 0.063  }_{- 0.054}$\\
1106402718122325760 & 0.771335966 & 7596 $^{+ 31 }_{- 27 }$& 5453 $^{+ 1 }_{- 1 }$& 1.51 $^{+ 0.028  }_{- 0.024  }$& 0.943 $^{+ 0.063  }_{- 0.073  }$& 3.288  $^{+ 0.115  }_{- 0.099  }$& 3.303  $^{+ 0.117  }_{- 0.101  }$& 14.922  $^{+ 1.619  }_{- 1.321  }$& 7.986  $^{+ 0.887  }_{- 0.711}$\\
211807141136171136$^{(a)}$& 1.2471725  & 7600 $^{+ 2 }_{- 4 }$& 4674 $^{+ 5 }_{- 21 }$& 1.905 $^{+ 0.008  }_{- 0.008  }$& 1.017 $^{+ 0.038  }_{- 0.034  }$& 5.170  $^{+ 0.047  }_{- 0.045  }$& 4.916  $^{+ 0.212  }_{- 0.191  }$& 24.561  $^{+ 2.485  }_{- 2.098  }$& 10.353  $^{+ 1.317  }_{- 1.104}$\\
3234434046920438656 $^{(*)}$& 0.632792879 & 7221 $^{+ 352 }_{- 462 }$& 7043 $^{+ 14 }_{- 115 }$& 0.806 $^{+ 0.161  }_{- 0.188  }$& 0.688 $^{+ 0.170  }_{- 0.214  }$& 1.618  $^{+ 0.341  }_{- 0.322  }$& 1.480  $^{+ 0.317  }_{- 0.319  }$& 2.131  $^{+ 1.894  }_{- 1.045  }$& 1.068  $^{+ 0.841  }_{- 0.537}$\\
\hline
\end{tabular}

\begin{tablenotes}
    \footnotesize
    \item[1]  (a): Period is extracted from \citet{2021AA...652A.120I}, while other periods are obtained from \citet{2022ApJS..258...16P}. \\(*): The target with the more massive component filling its Roche lobe.
\end{tablenotes}

\end{table*}
\end{appendices}

\end{CJK}
\end{document}